\documentclass[prb,reprint]{revtex4-2}
    \setcitestyle{super}

	\usepackage{lmodern}			
	\usepackage[latin1]{inputenc}
	\usepackage[T1]{fontenc}

	\usepackage{graphicx}     
	\usepackage{amsmath}      
	\usepackage{tabularx}		  
	\usepackage{xcolor}			  

	\usepackage{epstopdf}			
	\usepackage{hyperref}   	
	\usepackage[all]{hypcap} 	

	\usepackage{blindtext}
	
	\usepackage{todonotes}
	\usepackage{tikz}
	\usepackage{subcaption}		
	
	\newcolumntype{C}[1]{>{\centering\arraybackslash}p{#1}}		

	\hypersetup{pdfstartview={XYZ null null 0.93}}
	\hypersetup{pdfpagemode=UseNone}

\newcommand{\esub}{\epsilon_{\text{sub}}}

\newcommand{\psid}{\Psi_{\text{1-2}}}

\begin{document}

\linespread{1.2}


\title{Lattice gas study of thin film growth scenarios and transitions between them: Role of substrate}

\author{E. Empting}
\email{eelco.empting@uni-tuebingen.de}
\affiliation{Institut f\"ur Angewandte Physik, Universit\"at T\"ubingen, Auf der Morgenstelle 10, 72076 T\"ubingen, Germany} 


\author{M. Klopotek}
\affiliation{Institut f\"ur Angewandte Physik, Universit\"at T\"ubingen, Auf der Morgenstelle 10, 72076 T\"ubingen, Germany} 

\author{A. Hinderhofer}
\affiliation{Institut f\"ur Angewandte Physik, Universit\"at T\"ubingen, Auf der Morgenstelle 10, 72076 T\"ubingen, Germany} 


\author{F. Schreiber}
\affiliation{Institut f\"ur Angewandte Physik, Universit\"at T\"ubingen, Auf der Morgenstelle 10, 72076 T\"ubingen, Germany} 

\author{M. Oettel}
\affiliation{Institut f\"ur Angewandte Physik, Universit\"at T\"ubingen, Auf der Morgenstelle 10, 72076 T\"ubingen, Germany} 


\begin{abstract}
\noindent

Thin film growth is investigated in two types of lattice gas models where substrate and film particles are different, 
expressed by unequal interaction energy parameters. The first is of solid--on--solid type, whereas the second additionally incorporates
desorption, diffusion in the gas phase above the film and re-adsorption at the film (appropriate for growth in colloidal
systems). In both models, the difference between particle--substrate and particle--particle interactions plays a central role for the evolution of the film morphology at intermediate times.
The models exhibit a dynamic layering transition which occurs at generally lower substrate attraction strengths than the equilibrium layering transition. 
A second, flattening transition is found where initial island growth transforms to layer--by--layer growth at intermediate deposition times.
Combined with the known roughening behavior in such models for very large deposition times, we present four global growth scenarios, charting out the possible types of roughness evolution.

\end{abstract}


\maketitle


\section{Introduction}
The evolution of structure in thin film growth is a topic of broad interest,
both from the perspective of non-equilibrium statistical mechanics as well as from
an applied point of view where certain properties such as e.g. a smooth film
with minimal roughness may be desirable. For homoepitaxial growth, quantitative and 
detailed insight has been reached. \cite{evans_thiel_bartelt_2006}
For growth of a substance on a substrate consisting of a different material the phenomenology is broader.
If both the substrate and the film material are crystalline with different equilibrium
lattice parameters, one speaks of genuine heteroepitaxial growth. However, substrate or film may be 
amorphous in which case heteroepitaxial effects (such as residual stresses) are absent, but, 
effects of different surface energies are still present. 
The experimental systems
of interest encompass metal or semiconductor growth (which is in most cases heteroepitaxial and 
in which the film particles can be considered
as isotropic) \cite{Shchukin_1999_RevModPhys} as well as the growth of organic semiconductors on varying substrates
where the film molecules are mostly anisotropic. \cite{Witte_2004_JMaterRes,Schreiber04}
For organic semiconductor growth, genuine heteroepitaxy may occur (e.g. for growth of Pentacene on C60 crystal layers)
but, on the other hand, may also be absent (e.g when using amorphous silica substrates, which are of enormous practical importance and probably the most popular material).

 Typically, one distinguishes between
layer--by--layer (LBL) growth, Vollmer--Weber/island (ISL) growth and Stranski--Krastanov
(SK) growth which is characterized by initial LBL growth changing to ISL growth. \cite{venables_2006}
The latter is often  characterized  as a transition from 2D growth to 3D growth (see Ref. \citenum{Krause_2004} for an early experimental and simulation study with organic molecules).    
These growth modes can be distinguished using the film roughness $\sigma$ (root-mean-square (RMS) deviation
from the average film height $\bar h$) as an observable, which is easily accessible in both experiment and simulations.
LBL growth is characterized by oscillations in $\sigma (\bar h)$, and ISL growth is reflected in a quick
rise in $\sigma (\bar h)$, stronger than in statistical (Poisson) growth ($\sigma (\bar h) \propto {\bar h}^{1/2}$).
SK growth shows initial roughness oscillations, which subsequently change to a monotonic rise of $\sigma$ with $\bar h$. 
The latter may be described with a power law, $\sigma (\bar h) \propto  \bar h^\beta$ where $\beta$ is the roughening exponent.\cite{Krug_2002_PhysicaA}  

In explaining the occurrence of LBL vs.\ ISL growth or near--equilibrium conditions, it is common practice to invoke equilibrium surface free energies
(interface tensions),
in particular the ratio $r=(\gamma_\text{sv}-\gamma_\text{sf})/\gamma$.
Here, $\gamma$ is the interface tension between film and vapor/vacuum, $ \gamma_\text{sv}$ the one  between 
substrate and vapor/vacuum
and $\gamma_\text{sf}$ the one between substrate and film (see e.g. Ref.~\citenum{Burke_2009_JPhysCondensMatter}).
If $r=1$, equilibrium wetting occurs
(the free energy is lowest when a thick film is inserted between substrate and vapor/vacuum), which is understood as the condition for LBL growth. If $|r|<1$, partial wetting occurs with droplets of film material  appearing, whose contact angle $\theta_\text{Y}=\arccos(r)$ varies between 0 and 180 degrees.
This is understood as a condition for ISL growth. (The case $r=-1$ is not relevant to film growth
as it refers to equilibrium drying, the formation of a thick film of vapor between substrate and film material.) 
By the use of surface free energies, this argument for the distinction between LBL and ISL growth is an 
equilibrium one.
However, in a real system kinetics will enter the picture, and may invalidate the use of equilibrium arguments. We will address this point by studying growth in a simple lattice model where the equilibrium properties are known and can be used to gauge the results for growth.

Computational studies of film growth generally invoke one of two widely used simulation methods:
Kinetic Monte Carlo (KMC) and molecular dynamics (MD). KMC simulations can be lattice--based or 
off--lattice, \cite{biehl_2005} and are characterized by local, discrete `moves' of particles with no explicit memory of past ones. 
Each `move' occurs at a rate that adheres to the instantaneous energy (barrier) encountered at the current configuration.
The simplicity of especially lattice--based KMC simulations allows to study large systems and thick 
films.\cite{Assis_2015,Reis_2020}
On the other hand, MD simulations of growth incorporate the full particle dynamics, 
but studying multilayer growth for reasonably large systems is only practically possible for isotropic particles (see e.g. Refs. \citenum{gilmore_1991, xie_2014}). All--atom simulations of organic semiconductor growth can model 
faithfully the growth process of a specific molecule. Examples are Pentacene (PEN) growth on C$_{60}$ (Refs. \citenum{muccioli_2011,Fu_2013_AdvMater}), reversely the C$_{60}$ growth on PEN\cite{clancy_2016} or silica\cite{muccioli_2018} and 6T
monolayer growth on SiO$_\text{x}$ (Ref. \citenum{muccioli_2019}).
These models are, however, generically limited by the small number of particles for computational reasons. 
Therefore, lattice--based KMC simulations are more suited to study the multilayer regime and to explore the parameter space more thoroughly.  

In this paper, we investigate dynamic transitions between different growth modes using KMC simulations. We both quantify these transitions as well as map out the conditions under which they occur. We adopt the well-known lattice gas model for particles with nearest-neighbor interactions living on a simple cubic (SC) lattice.
Particles at the substrate and in the film are treated differently by means of their interaction energies, 
i.e.\ the interaction strength between two film particles is different from the interaction strength
between a film and a substrate particle.
However, in our modeling genuine heteroepitaxy is absent since the lattices of the substrate and the film are 
assumed to be equivalent. The modeling should thus actually correspond nicely to the growth on amorphous substrates such as the very commonly used oxidized silica wafers.
Both a solid--on--solid (SOS) model and
a second, more general model are investigated. Within both, deposition of particles and diffusion occurs at the film surface. However, in the second model, arbitrary desorption and re-adsorption of particles can occur, along with diffusion in a gas phase hovering above the film. This model is less restrictive in local transport than the SOS counterpart (its catalog of possible local `moves' is broader), and is 
novel (to our knowledge) in the KMC literature for thin-film growth.
We call this version of the model the Colloidal Growth Model (CGM) since it is describing a typical
setup for the growth of colloidal crystalline films.
Colloidal particles are immersed in a solvent, generally rendering their bulk and surface dynamics purely diffusive.
Inspired by sedimentation-diffusion,\cite{dhont_1996} deposition on top of a substrate proceeds by drift--diffusion.

As a main result, we identify two dynamic transitions in both models:
(i) between ISL and LBL growth (''dynamic layering transition'') via a novel order parameter that quantifies the difference in coverages of the first and 
second layer and (ii) a ''flattening transition'' of the ISL growth mode back to (near--)LBL growth at intermediate 
times --- after a total deposition of a few monolayers --- the onset of which can be shown to be the moment the first layer becomes completely filled. The possible occurrence of these transitions
gives rise to certain global growth scenarios (characterized by the roughness evolution with time) when observing the film evolution over long deposition time-scales.   

The paper is structured as follows: Sec. II introduces lattice--based KMC simulations, in particular the 
two variants studied here. In Sec. III the two dynamic transitions mentioned above are characterized, and additionally
the global growth scenarios are discussed.
In Sec. IV we relate our observations to
existing experimental results for thin film growth, and in Sec. V we provide some conclusions and an outlook.

\section{KMC lattice simulations}
\begin{figure}[t]
	\centering
	\includegraphics[width=0.49\linewidth]{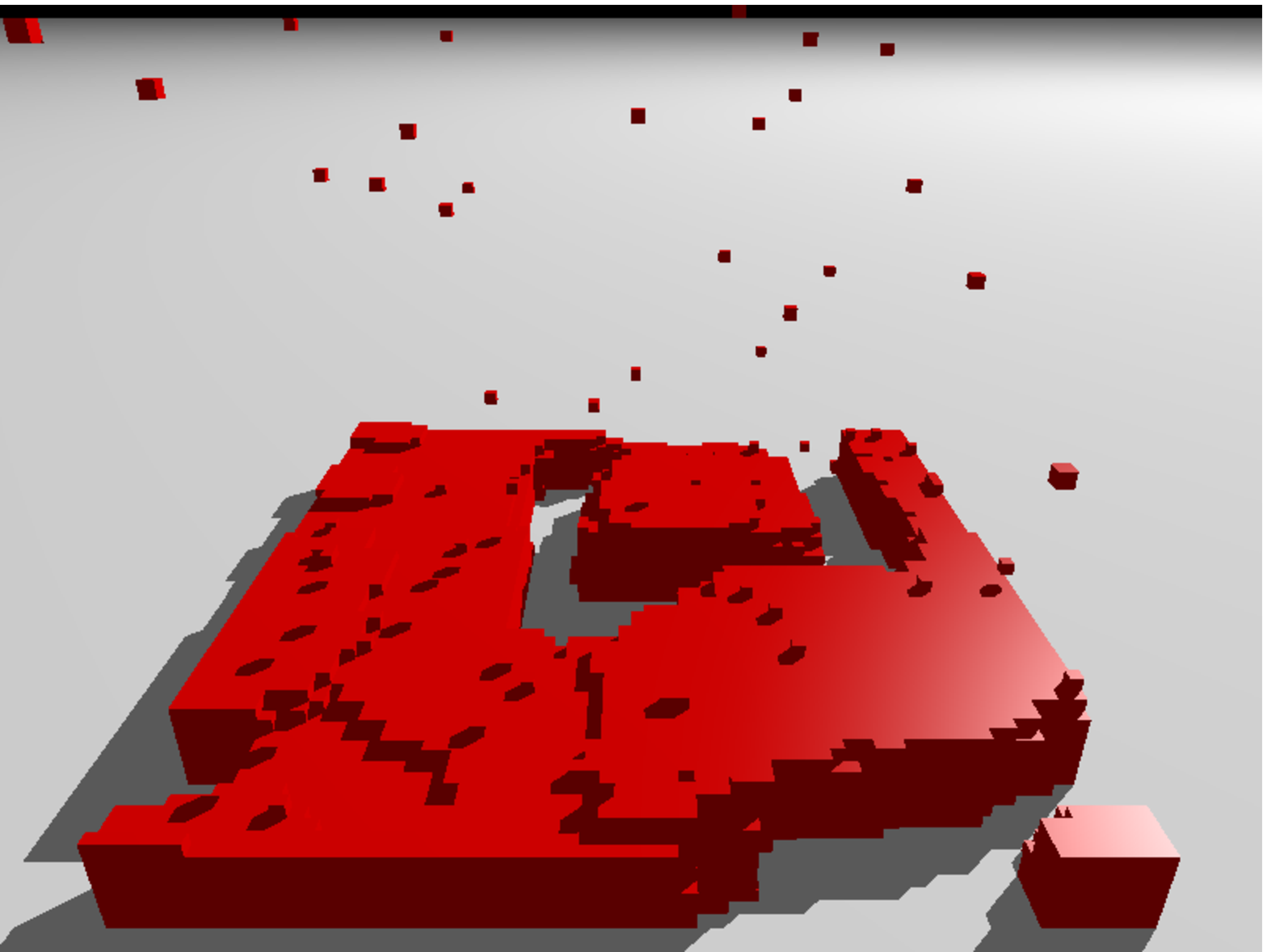}
	\includegraphics[width=0.49\linewidth]{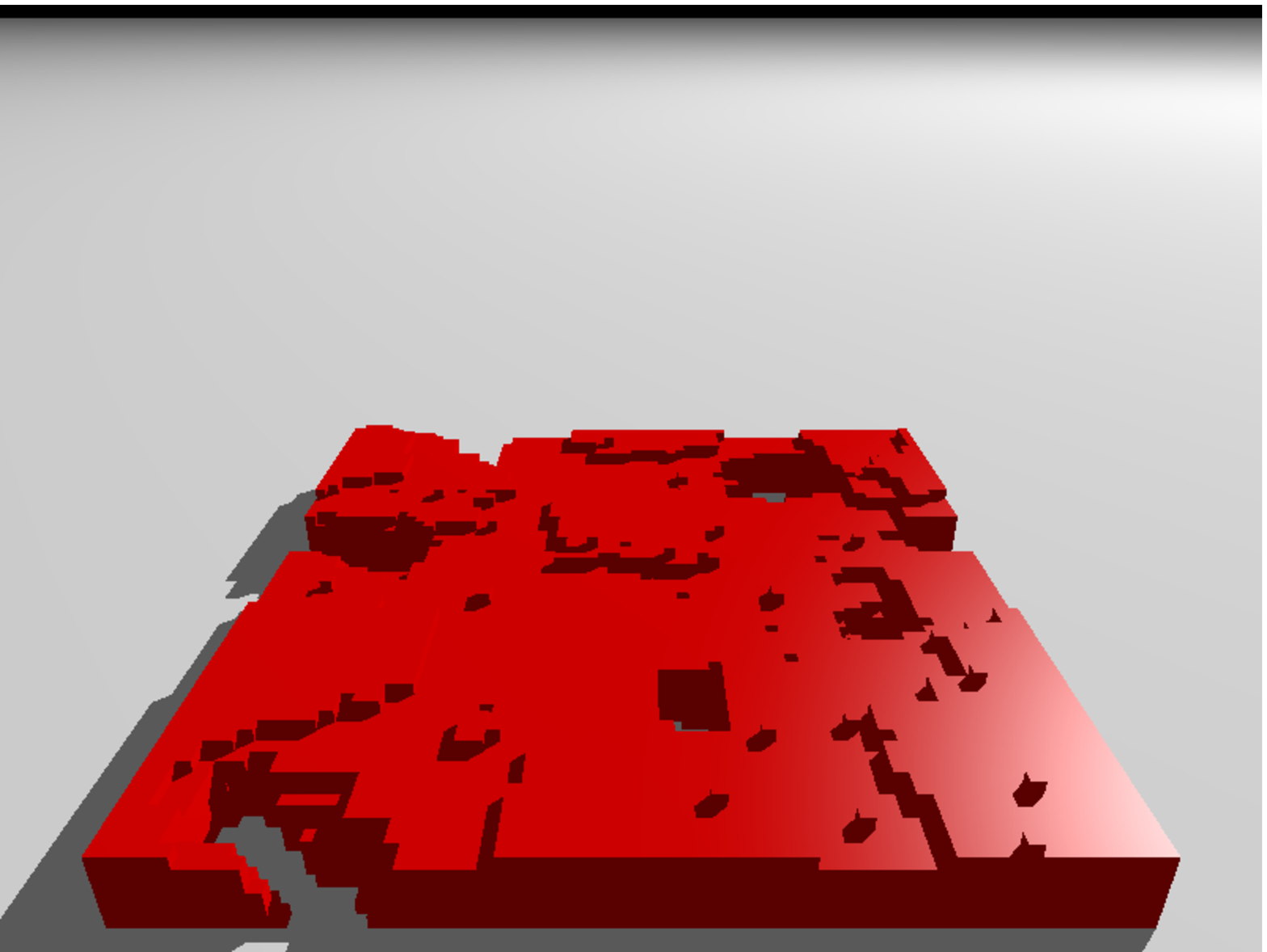}
	\caption{Snapshots of the system at $\epsilon=-3$, $\esub=-1.33$, $\Gamma=10^4$ after deposition of 5 MLs in the CGM (left) 
and the SOS model (right). Since particle moves are not restricted in the CGM, we can see in the CGM snapshot 
(i) the formation of a gas phase above the film and (ii) more pronounced partial wetting of the substrate}
\label{fig:cgm_sos}
\end{figure}
\subsection{General considerations}
KMC solid-on-solid (SOS) lattice models have a venerable history in the study of film growth.
\cite{evans_thiel_bartelt_2006} 
In the easiest realization on a cubic lattice,
each particle must be supported, i.e. has a particle or substrate
below it (no overhangs) and particles are not allowed to be located in the vapor phase 
(in some models, particles are
allowed to desorb from the film, but are then consequently removed from the simulation \cite{schinzer_sokolowski_biehl_kinzel_1999}). 
This leads to films that exclude vacancies and overhangs and imposes an imminent vacuum above the film. 
In many experimental scenarios, this is a reasonable approximation, since desorption is often negligible.

In the simplest realization of this model (stochastic growth), 
new particles are deposited at random sites on top of the growing film and stick there without 
diffusing any further.\cite{michely_krug_2004} In this model, the film grows in a stochastic manner driven by the deposition process only, 
and its roughness will behave as $\sigma \propto \Theta^\beta$ with $\beta=1/2$, where $\Theta$ is the number of deposited monolayers (MLs) and 
\begin{equation}
\sigma = \sqrt{1/N \sum_{i=1}^{N} (h_i - \bar{h})^2}
\label{eq:sigma}
\end{equation}
is the RMS of the deviation from the mean height, with $h_i$ being the film height (in lattice units) at lattice site
with label $i$ and $\bar{h}$ the mean film height.

Surface diffusion can be incorporated into the simulation in the following most intuitive way: during each time step either 
a new particle is inserted at a random site with probability $f$, or an existing particle is moved to a neighboring site 
with probability $1-f$ (Refs.\citenum{clark_vvedensky_1988,siegert_plischke_1994}). Here, it is crucial
how one handles inter-layer diffusion: If it is forbidden, particles will always remain in the layer into which they were deposited, and the 
roughness will show the same $\sigma \propto \Theta^{1/2}$ behavior seen in the stochastic growth models. 
If inter-layer transport is allowed, it can occur either at the same rate as diffusion within a layer, or one can assign an 
energy cost to this layer change (the so-called step edge or Ehrlich-Schw\"obel barrier).
Mound formation in epitaxial growth with an ensuing roughening exponent $\beta \neq 1/2$  
is strong evidence for the effects of such a barrier, see 
Refs.~\citenum{Krug_2002_PhysicaA,michely_krug_2004} for an overview.

This simple model has been modified in various ways over the years. One early modification of the stochastic growth model 
was the Wolf-Villain model.\cite{wolf_1990,sarma_tamborenea_1991} Here particles will diffuse immediately after 
deposition before becoming immobilized forever. More recently, the quantitative accuracy of the SOS model has been 
improved by implementing fluctuating inter-particle attractions.\cite{rosenfeld_2007} Other modifications 
include a first-passage time approach\cite{amar_2010} (which can lead to a significant acceleration of simulations), 
and the introduction of anisotropic interactions.\cite{martynec_klapp_2018}

In other lattices, e.g. face centered cubic, one finds additional effects: Since each particle now needs multiple occupied 
sites in the 
layer below to be supported, one has to consider what happens if not all supporting sites are occupied. 
A frequently invoked mechanism is downward funneling,\cite{evans_sanders_thiel_depristo_1990} where particles 
will ``fall'' into lower layers until they have reached a fully supported position. 
This leads to films growing in a much smoother fashion than in an SC lattice, but is also a strong simplification 
of the processes occurring during film growth.\cite{evans_thiel_bartelt_2006} A somewhat more sophisticated approach 
is to modify this downward-funneling behavior by trapping new particles at the sides of protrusions, leading to overhangs and 
consequently voids inside the film.\cite{evans_2001}

A number of studies addressed the evolution of 3D structure in homoepitaxial growth, see e.g.
\cite{Ferreira2011,Assis_2015} for recent examples with large substrates and rather thick films.  
Previous KMC works on genuine heteroepitaxial growth focused on the problem of SK growth by incorporating strain and stress 
release,\cite{Baskaran2010,schulze_smereka_2011,schulze_smereka_2012} 
or on simulating the behavior of  specific systems (e.g. C60 on Pentacene\cite{clancy_2016}) by
fitting the corresponding KMC parameters from atomistic calculations.

\subsection{Solid--on--solid model}
The simulation is divided into discrete but variable time steps, which are Poisson distributed. During each of these time steps exactly one event occurs.\cite{levi_kotrla_1997,kotrla_1996} Such events can be e.g. particle moves or
insertions. Each event occurs with an average rate $k_i$ , which is a parameter specified a priori. In
general, KMC being event-driven means that the algorithm is ``rejection-free'', since one tracks
all events that are possible in each step and then chooses one of these events.
Our simulations entail attractive interactions between neighboring particles $\epsilon$ and
particles on top of the substrate $\epsilon_\text{sub}$. All energies are given in units of the thermal
energy $k_\text{B} T$. These affect the
rates at which events occur depending on the local environments of the respective particles.
We realize this by introducing a Metropolis step, accepting each move with a probability 
$p = \min (1, \exp (- \Delta E)) $
depending on the change in internal energy this move would cause. Hence
our algorithm is ``rejection-free'' regarding steric repulsions only. While this of course leads to some
moves being rejected, this hybrid method\cite{adam_billard_lancon_1999} significantly reduces the overhead of keeping track
of state changes. After each move attempt, we then increase the time by a random $\Delta t = -\ln(r)/k_\text{tot}$  where $k_\text{tot} = \sum_i k_i$ is the total rate of possible events at the current state and $r \in (0,1]$ is a random, uniform number. The average length of these time steps is $1/k_\text{tot}$. This procedure leads to more accurate dynamics (incorporating more fluctuations) than using a fixed-length time step of length $1/k_\text{tot}$ (Ref. \citenum{ruiz_barlett_2009}).

\begin{figure*}
    \centering
    \includegraphics[width=0.7\linewidth]{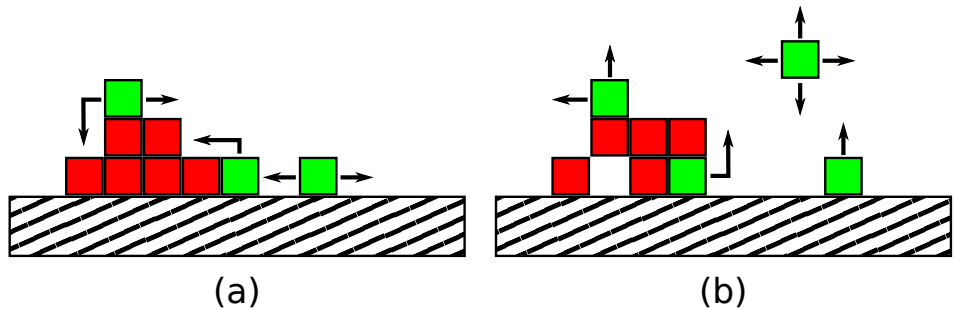}
    \caption{Schematic 2D drawings showing (a) the diffusion moves which are allowed in the SOS model (note that neither overhangs nor desorptions are possible), and (b) additional moves which are allowed in the CGM.}
    \label{fig:moves}
\end{figure*}

In the SOS version, we consider insertion moves at random lateral positions $(x,y)$ on top of the
growing film with rate $k_\text{ins}=F$,
diffusion moves to lateral next-neighbor positions with $\Delta z=0$, where $z$ is the vertical coordinate, with rate $k_\text{hop}$, and 
layer--changing moves to lateral next-neighbor positions with $\Delta z= \pm 1$ with rate 
$k_\text{ES} = k_\text{hop} \exp(-E_\text{ES})$ where $E_\text{ES}$ is the (dimensionless) Ehrlich--Schw\"obel barrier, which effectively leads to a rescaling of the acceptance probability $p$ to $p = \min (1, \exp (- \Delta E)) \cdot \exp(-E_\text{ES})$. We chose this implementation in order to ensure that detailed balance is obeyed.
All these moves have to respect the SOS condition of no vacancies/overhangs in the film; see Fig. \ref{fig:moves}(a) for a sketch of all possible diffusion moves.

The 2D diffusion constant (in lattice units) of a free film particle on top of the substrate or on top of a completely
filled layer is simply $D=k_\text{hop}$. Thus, the 4 parameters of the model are $\Gamma =D/F$
and the energies $\epsilon$, $\epsilon_\text{sub}$, $E_\text{ES}$.
To provide an experimental context,\cite{bommel_2014, clancy_2008} 
note that e.g. for room temperature growth of C60 on C60  one may estimate diffusion
coefficients $D = O(10^8)$ nm$^2$/s, and with fluxes $F=0.001...0.1$ monolayers/s
one finds $\Gamma = O(10^9) ... O(10^{11})$ (lattice units)$^{-4}$ where a real C60 lattice unit is approximately
1 nm.  Furthermore, using the Girifalco
potential,\cite{girifalco_1992} one can estimate the interaction 
between two neighboring C60 particles at room temperature as $\epsilon \approx -10$ in units of $k_BT$, where $k_B$ is the Boltzmann constant and $T$ is the (room) temperature. At this point we can remark that simulating
such high values of $\Gamma$ and $|\epsilon|$ is challenging, see Ref.~\citenum{Speck_2020} for a recent, state--of--the art example
examining few layer growth with C60. We will return to this problem below.

\subsection{Colloidal growth model}
\label{sec:cgm}

In the second version of the model, we relax the restriction on the next-neighboring site diffusion moves. Any such move
is now allowed if it is not blocked by another particle or the substrate. In addition to the moves allowed in the SOS model, this allows for desorption
from the film, diffusion in the vapor phase and re-adsorption at the film. Therefore, the growing film is covered by
a dynamically changing gas `cloud' and new particles are incorporated into the film through possibly direct deposition and
also adsorption from this gas layer. 
 A layer-changing particle move still requires
a support particle around which the particle moves up or jumps down one layer. See Fig. \ref{fig:moves}(b) for a sketch of all possible diffusion and evaporation moves in the CGM.

\begin{figure*}
    \centering
    \includegraphics[width=0.7\linewidth]{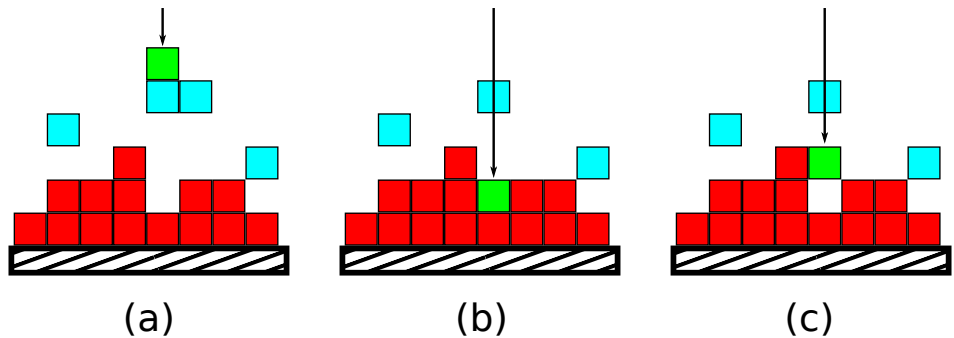}
    \caption{Different methods of inserting new particles into the simulation box in the CGM. Red particles are part of the film, blue particles are not, and the green particle is the newly inserted particle. (a) The newly inserted particle diffuses downwards until it touches any other particle (including particles in the gas). This leads to the formation of large clusters in the gas phase. (b) The new particle is inserted on the topmost film particle at the chosen lateral position. (c) The new particle is inserted at the topmost position with a film particle at a neighboring site (if that height is larger than the position of the topmost film particle). This is most similar to colloidal deposition.}
    \label{fig:insertions}
\end{figure*}

The insertion move now requires more discussion. In general, the film contains cavities and there is an adjacent gas phase
with freely floating particles.
Thus there are several sensible ways to insert new particles into the system, depicted in Fig. \ref{fig:insertions}. One natural implementation would be to insert 
particles at the top of the box and letting them sediment downwards until they meet any other particle (Fig. \ref{fig:insertions}(a)). However, this leads to the formation of large clusters 
inside the gas phase which do not dissolve during the simulation and consequently block most new particles from ever reaching the 
substrate or film. Since we have not implemented cluster moves, the clusters themselves are stationary and thus the blockage for newly deposited particles persists and leads to 
unphysical growth behavior.

More similar to insertion in the SOS model, we first define as ``film'' the set of particles which are connected
via next neighbors to the substrate.
We can choose a random position of the substrate and insert a new particle on 
top of the highest particle there which is part of the growing film (i.e. not in the gas phase (Fig. \ref{fig:insertions}(b)). 
Lastly, and closer in spirit to colloidal growth, we relax this SOS like condition insofar, as particles can also get ``caught'' 
at a certain height by neighboring particles which belong to
the growing film, leading to the formation of overhangs (Fig. \ref{fig:insertions}(c)). We choose this insertion method for all results shown below. 
All in all, this model is similar in spirit to colloidal deposition experiments in solution,\cite{Ganapathy_2010_Science} hence we call this
model a Colloidal Growth Model (CGM). 

The definition and calculation of the rates is unchanged compared to the SOS model. The particles can desorb into the gas phase with an attempt rate $D$. The dynamics in the gas are modeled by nearest-neighbor hops with rate $D$. The CGM obeys detailed balance (including with the gas) completely if the explicit insertion event is turned off.
We remark that the difference between the CGM and the SOS model lies in the catalog of allowed particle moves.
Sometimes, a difference between atomistic and colloidal growth is discussed with respect to the differing
interaction range in atomic/molecular systems and colloidal systems, see e.g. Ref.~\citenum{Kleppmann17}

We note that the CGM is a very generic extension of the SOS model without unphysical restrictions but
it leads to a substantial amount of simulation time being spent on simulating diffusion
moves inside the vapor, of particle desorption and re-adsorption, and necessitates more bookkeeping.
However, in contrast to SOS models, there is a well-defined equilibrium limit for the CG, since the whole phase space can be explored.

Most of the results in the CGM will be for a box size $L \times L \times L_z$ with $L=64$ and $L_z=200$. 
The upper boundary of the box is a hard wall, while
the lower boundary is the attractive substrate.
For the SOS model, only the lateral extension is important, here we use $L = 32 ... 300$.
The main computational observable is the roughness $\sigma$ (defined in Eq. \ref{eq:sigma}) as a function of time. The time is
proportional to the amount of deposited material (total coverage) $\Theta$ (in units of filled monolayers). We will also 
employ the
layer coverage $\Psi_i=N_i/L^2$, where $N_i$ is the number of particles in layer $i$. 
The rather moderate lateral system sizes are sufficient for studying the roughness behavior, as evidenced by the results below.

As an example for configurations occurring in growth, in Fig. \ref{fig:cgm_sos} we show a comparison of snapshots from the CGM and the SOS model for parameters corresponding
to initial island growth. The relaxed restrictions on diffusion moves in the CGM compared to the SOS model lead to
a more pronounced formation of islands.

\section{Results}

As we noted earlier, simulations using realistic values for $\Gamma$ and $\epsilon$ fitting actual growth experiments are 
numerically challenging. Nevertheless we conjecture that results from calculations at lower $\Gamma, |\epsilon|$ 
can be extrapolated to larger values using the following scaling arguments.

For homoepitaxial sub-monolayer growth, such a scaling relation can be derived.
In a one-component system, it has been shown\cite{michely_krug_2004} that the island density in the sub-monolayer at low 
densities scales with
\begin{equation}
n \propto \frac{\exp\left(-\frac{\epsilon_0}{k_BT(i^*+2)} \right)}{\Gamma^{i^*/(i^*+2)}}
\label{eq:is_dens}
\end{equation}
where $\epsilon_0$ (with proper energy units) is the interaction strength between nearest neighbors and $i^*+1$ is the size of the smallest 
stable cluster. If we now assume that $i^*=1$ (i.e. that dimers are stable, corresponding to large $|\epsilon_0|$), 
the island density is constant if:
\begin{align}
\frac{|\epsilon_0|}{k_BT} - \log\Gamma = \text{const}
\label{eq:scaling_submono}
\end{align}
i.e. a simulation with a $\Gamma < \Gamma'$ would yield the same results as one with $\Gamma'$, as long as one would use 
an appropriately scaled $|\epsilon_0| < |\epsilon_0'|$.
For the example above (C60 growth), island densities at the physical values $\epsilon_0' \approx -10$ $k_BT$ and 
$\Gamma' \approx 10^9$
should correspond to island densities at $\epsilon \approx -3$ $k_BT$ and $\Gamma \approx 10^6$.  


For the more general case considered here, we are treating multilayer growth, and the two additional energy 
parameters ($\epsilon_\text{sub}$, $E_\text{ES}$)
presumably enter a possible scaling relation. However, except for film-roughness scaling in the epitaxial case with no ES barrier 
(see below), scaling relations have not yet been identified for multilayer growth. 

In Sec. \ref{subs:dyn_layer} and \ref{subs:flat} below, we consider a vanishing ES barrier $E_\text{ES}=0$. 

\subsection{Dynamic layering transition}
\label{subs:dyn_layer}
\begin{figure*}[t]
    \centering
    \includegraphics[width=0.4\linewidth]{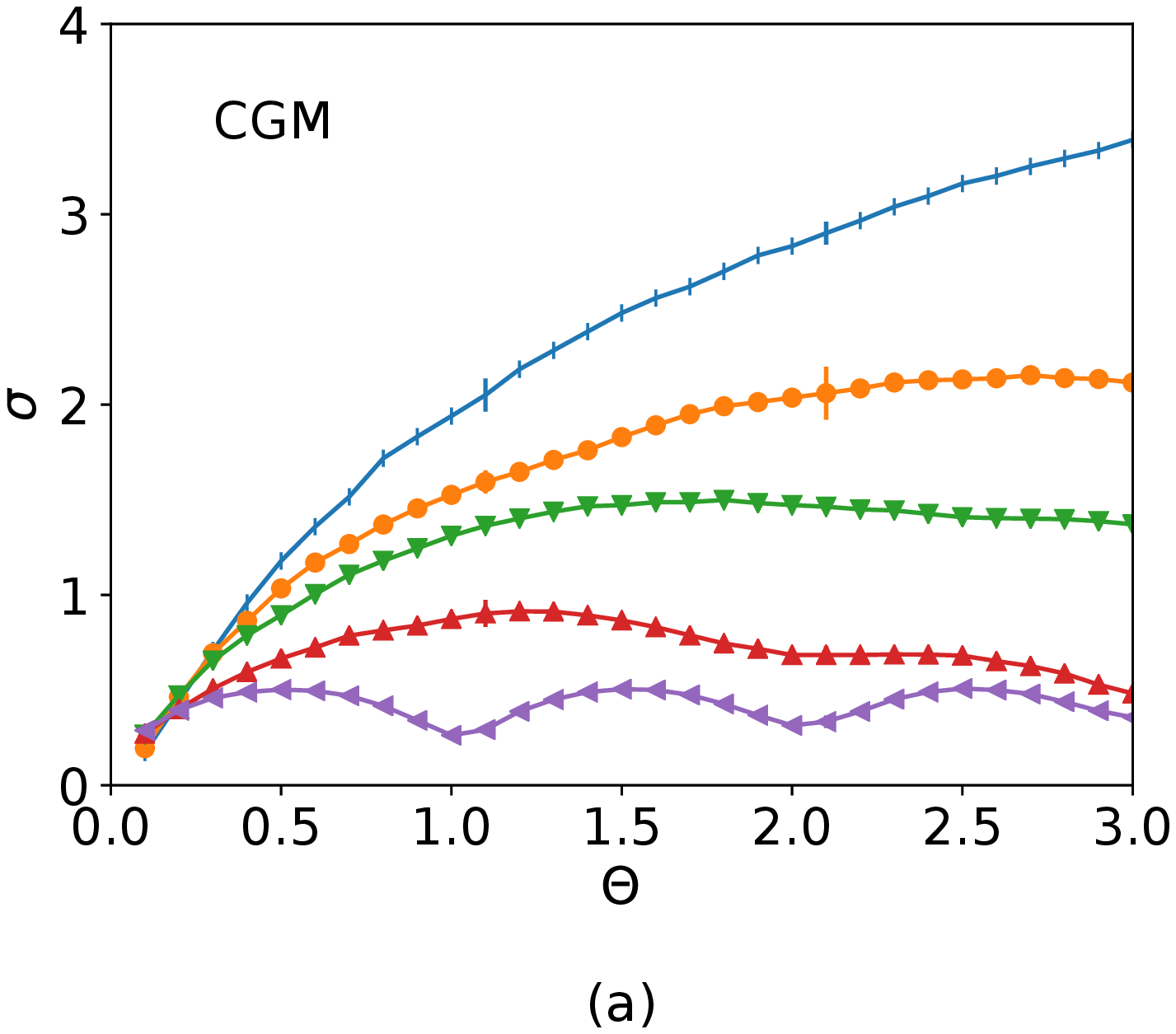}
    \includegraphics[width=0.4\linewidth]{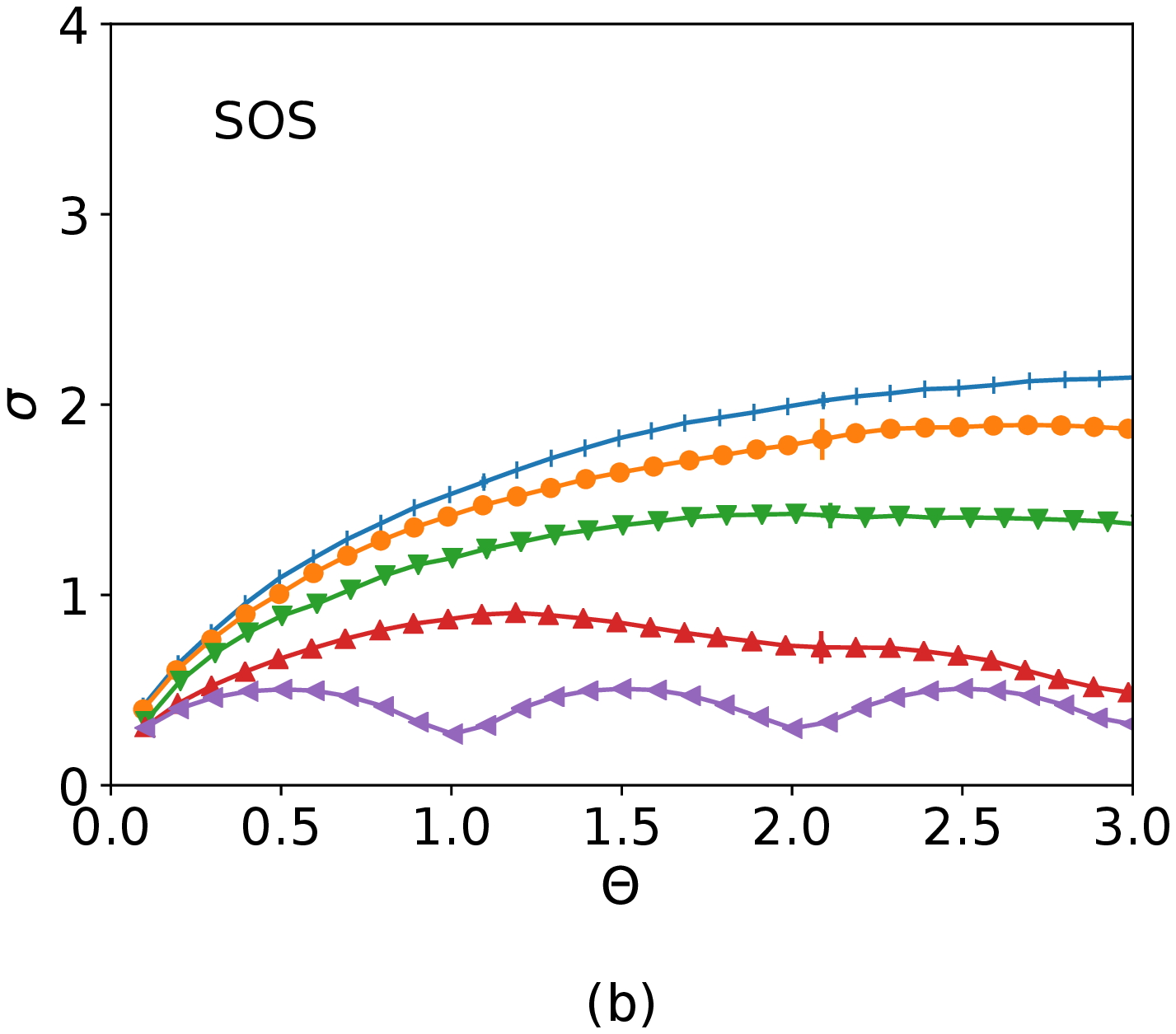}
    \includegraphics[width=0.4\linewidth]{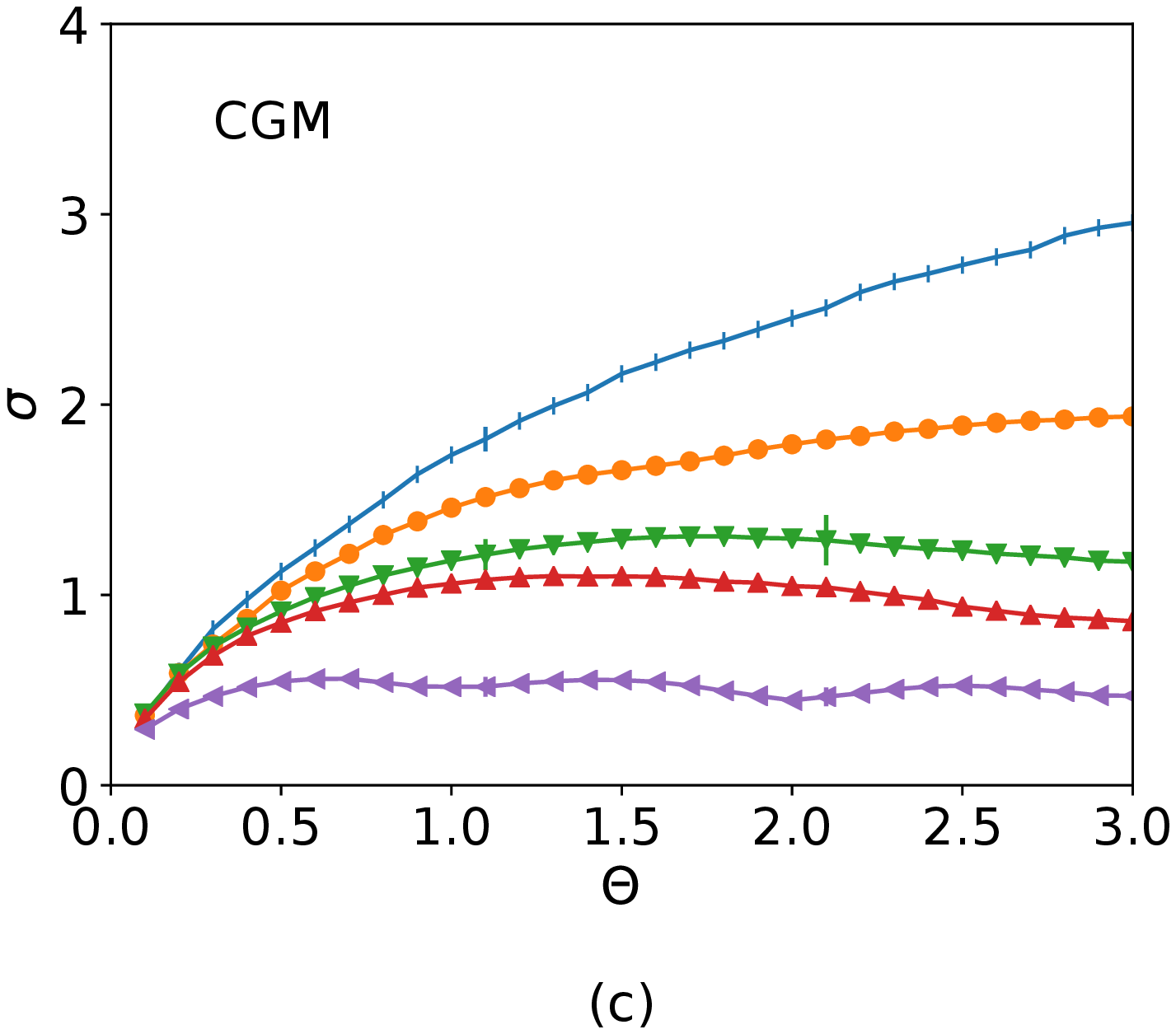}
    \includegraphics[width=0.4\linewidth]{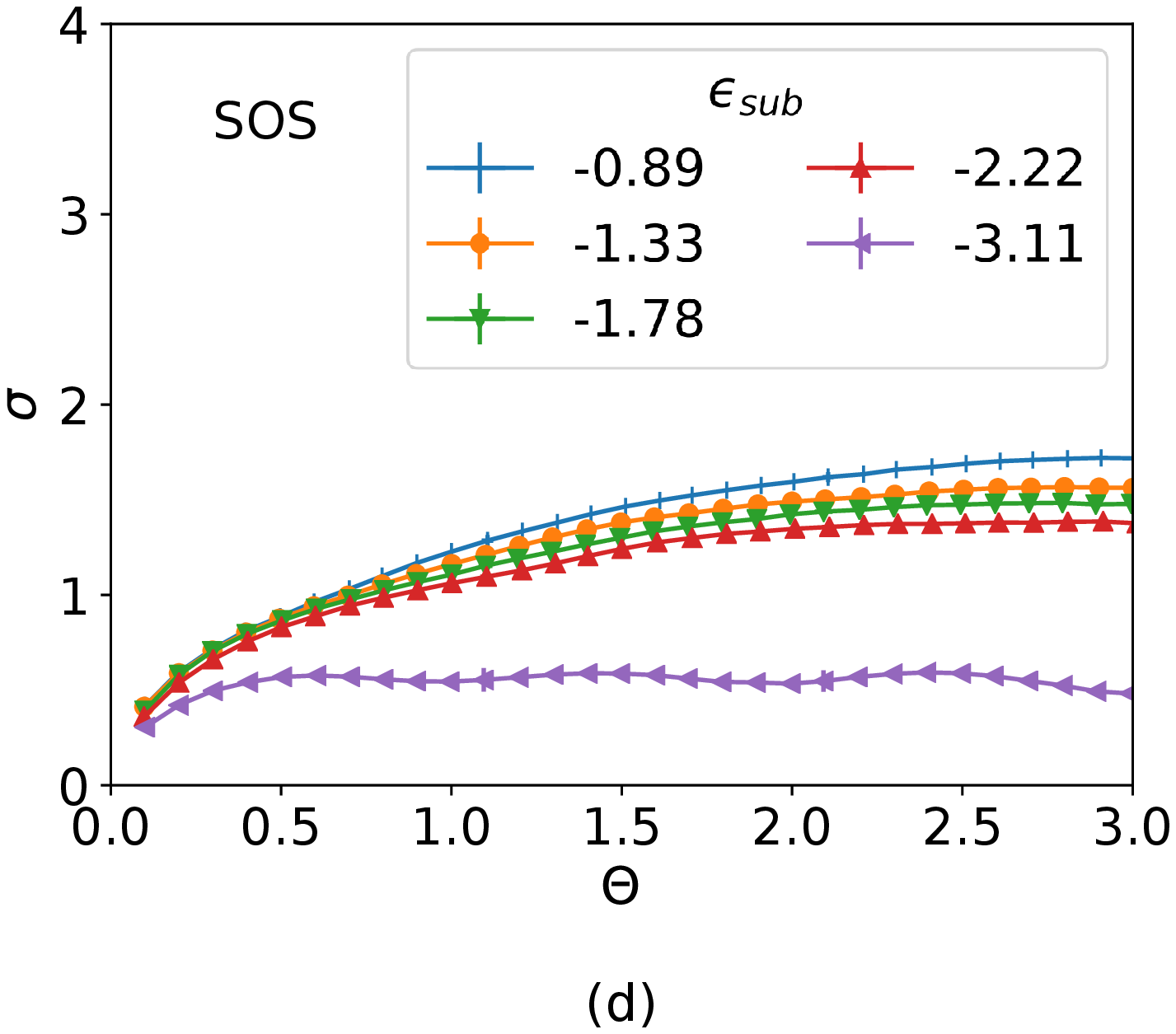}
    \caption{Evolution of roughness in the CGM and the SOS model for $\Gamma=10^4$ and different substrate strengths $\esub$, averaged over 5 runs.
        (a) $\epsilon=-3$, CGM, (b) $\epsilon=-3$, SOS, (c) $\epsilon=-5$, CGM, (d) $\epsilon=-5$, SOS. The legend in (d) also applies to (a)--(c). Vertical lines indicate statistical errors, deduced from 5 independent runs for each parameter set. 
    }
    \label{fig:rough_comp}
\end{figure*}

In Fig. \ref{fig:rough_comp} we show the roughness evolution up to a total coverage $\theta=3$ both for the SOS model and the CGM, 
for $\epsilon=-3$ and $-5$, respectively, each for a range of substrate attractions $\epsilon_\text{sub}$.
We found that in the CGM and for very weak particle-substrate interactions $|\esub| \lesssim 0.8$, the initially deposited particles will form floating clusters which then coalesce into a film which might or might not be connected to the substrate, which yields strongly varying 
results for $\sigma$. Hence we only consider values of $|\esub| \gtrsim 0.9$.
For both models it is seen that upon  increasing the magnitude of $\epsilon_\text{sub}$, the system will go from 
an evolution with increasing 
roughness to an evolution with oscillating roughness, which
indicates a transition from islands forming on top of the substrate to LBL growth. 
This transition (abbreviated as \textbf{ISL $\leftrightarrow$ LBL)} can be considered as a dynamic layering transition.

\begin{figure*}[t]
    \centering
    \includegraphics[width=0.4\linewidth]{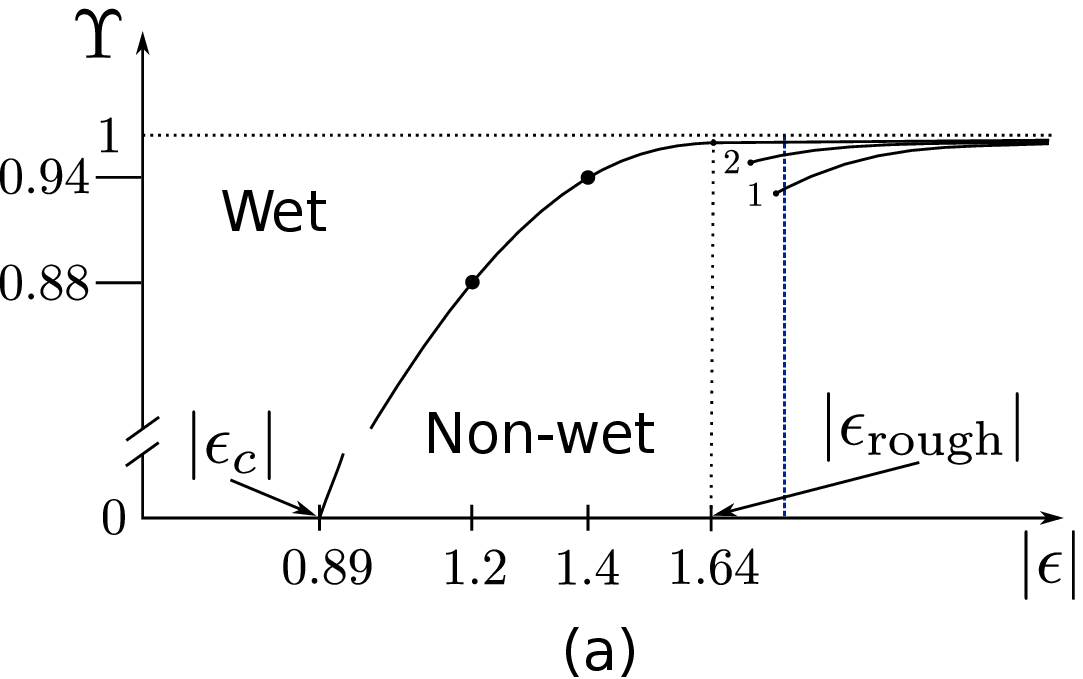}
    \includegraphics[width=0.2\linewidth]{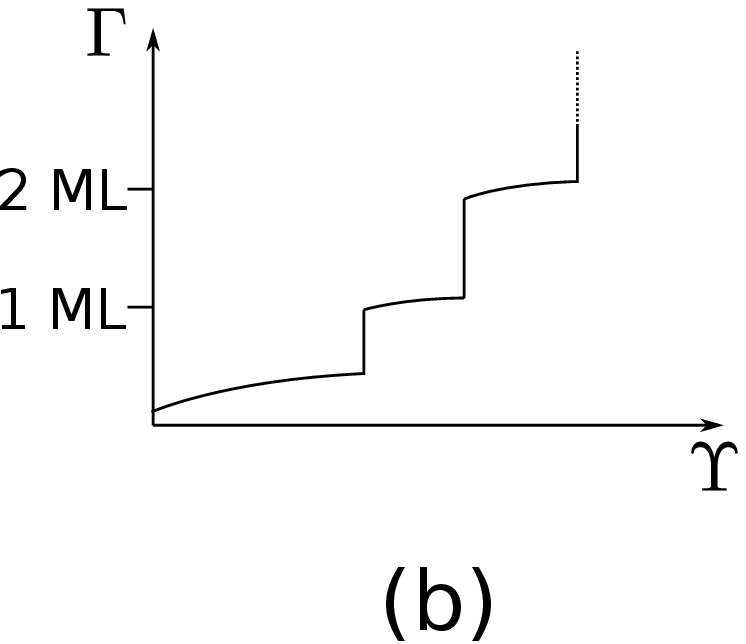}
    \caption{(a)
        Schematic equilibrium wetting/layering phase diagram for the lattice model in 3D for $\Upsilon \equiv \esub / \epsilon$ vs. $\epsilon$.
        The critical point for liquid-vapor phase separation is at $\epsilon=\epsilon_\text{c}$. The curve
        originating at $|\epsilon_\text{c}|$ is the line of the wetting transition (thin films of adsorbed liquid
        at substrates with attraction strength $\esub$ below that line, thick films for $\esub$ above that line).
        This is second order until $\epsilon = \epsilon_\text{rough} \approx -1.64$, which denotes the
        critical attraction strength for roughening.
        For $|\epsilon| > |\epsilon_\text{rough}|$ layering transitions occur (indicated with 1 and 2) where the effective thickness
        of the adsorbed film jumps to values close to 1 or 2 lattice units. The explicit data points are from simulations,\cite{binder_landau_1988} for more details and discussion see Ref.~\citenum{binder_landau_1992}. 
        (b) Schematic equilibrium adsorption $\Gamma$ as function of $\Upsilon$ for the path marked by the blue dashed line in
        (a).
    }
    \label{fig:schem_phase}
\end{figure*}

Layering transitions can also be found in {\em equilibrium} systems as a particular form of a wetting transition.
It is useful to recall the equilibrium wetting and layering behavior of the lattice gas model before
discussing further the dynamic layering transition. 
Wetting transitions are conveniently discussed in a diagram with temperature and the reduced substrate attraction strength 
$\Upsilon=\epsilon_\text{sub}/\epsilon$ as axes. For the lattice gas, here we use a $\epsilon$--$\Upsilon$ diagram
which corresponds to a $1/T$--$\Upsilon$ representation.
Wetting transitions can only occur for particle--particle interaction strengths $|\epsilon| > |\epsilon_\text{c}|$
where $\epsilon_\text{c} \approx -0.89$ is the bulk critical attraction strength for the gas--liquid separation.
Here, we interpret the high--density liquid phase as the condensed (solid) phase in the growing film.
One studies the system at the substrate at coexistence conditions such that far away 
the system is in the gas phase.
 The equilibrium net adsorption at the substrate 
displays a characteristic behavior near a critical reduced strength $\Upsilon_\text{c}(\epsilon)$: it may diverge continuously
to infinity upon $\Upsilon \to \Upsilon_\text{c}$ (critical wetting), it may jump discontinuously to infinity
(first order wetting), or it may jump discontinuously to a value corresponding to a net coverage of $n$ layers 
($n^\text{th}$ layering transition).
The lattice gas model as used here is equivalent to the Ising model
upon a few redefinitions (summarized in Appendix \ref{app:ising}), its wetting behavior has been studied in 
Refs.~\citenum{binder_landau_1988,binder_landau_1992}.
The schematic wetting/layering phase diagram (i.e. the curve $\Upsilon_\text{c}(\epsilon)$ for wetting/layering transitions)
is sketched in
Fig. \ref{fig:schem_phase}(a).
A second interaction strength relevant for the wetting transition is the  roughening
point $\epsilon_\text{rough} \approx -1.64$. For $|\epsilon| < |\epsilon_\text{rough}|$, steps on the film surface
can be created with no free energy cost whereas for $|\epsilon| > |\epsilon_\text{rough}|$ the step free energy
is finite. The line $\Upsilon_\text{c}(|\epsilon|)$ for $|\epsilon_\text{c}|< |\epsilon| < |\epsilon_\text{rough}|$
describes critical wetting. For $|\epsilon| > |\epsilon_\text{rough}|$ layering transitions occur
(the lines labeled with 1 and 2 for the first and second layering transition in Fig. \ref{fig:schem_phase}(a)).
In Fig. \ref{fig:schem_phase}(b), we show the characteristic behavior of film adsorption upon varying
the control parameter $\Upsilon$ when it crosses two layering transitions (blue dashed line in Fig. \ref{fig:schem_phase}(a)).
It is important for the subsequent discussion of the dynamic layering transition that for the
high attraction strengths considered there ($|\epsilon| \ge 3$), all equilibrium layering transitions occur very close to 
$\Upsilon_\text{c}=1$
(which is the intuitive zero-temperature, or $\epsilon \to -\infty$, limit).

\begin{figure*}[t]
    \centering
    \includegraphics[width=0.4\linewidth]{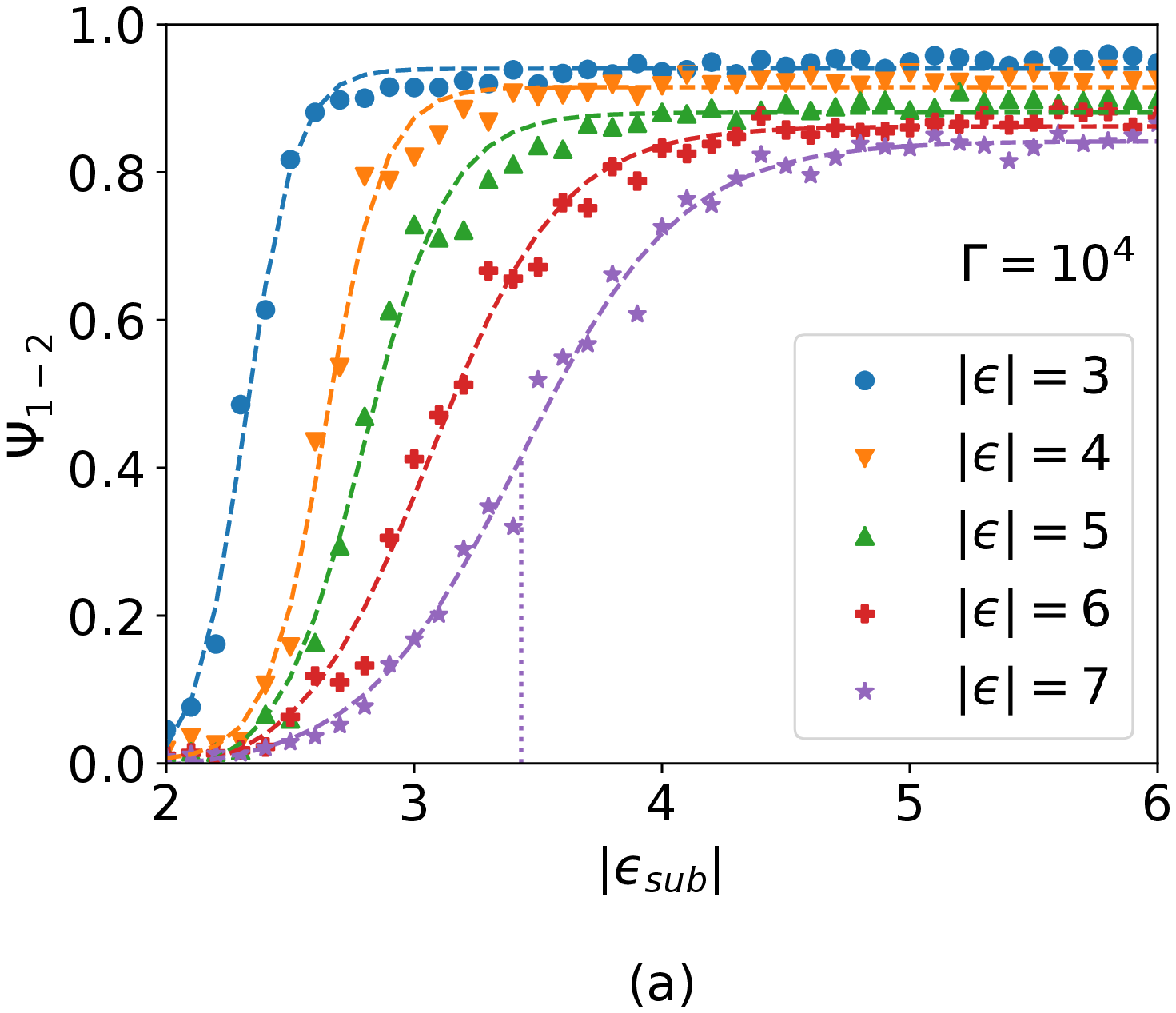}
    \includegraphics[width=0.4\linewidth]{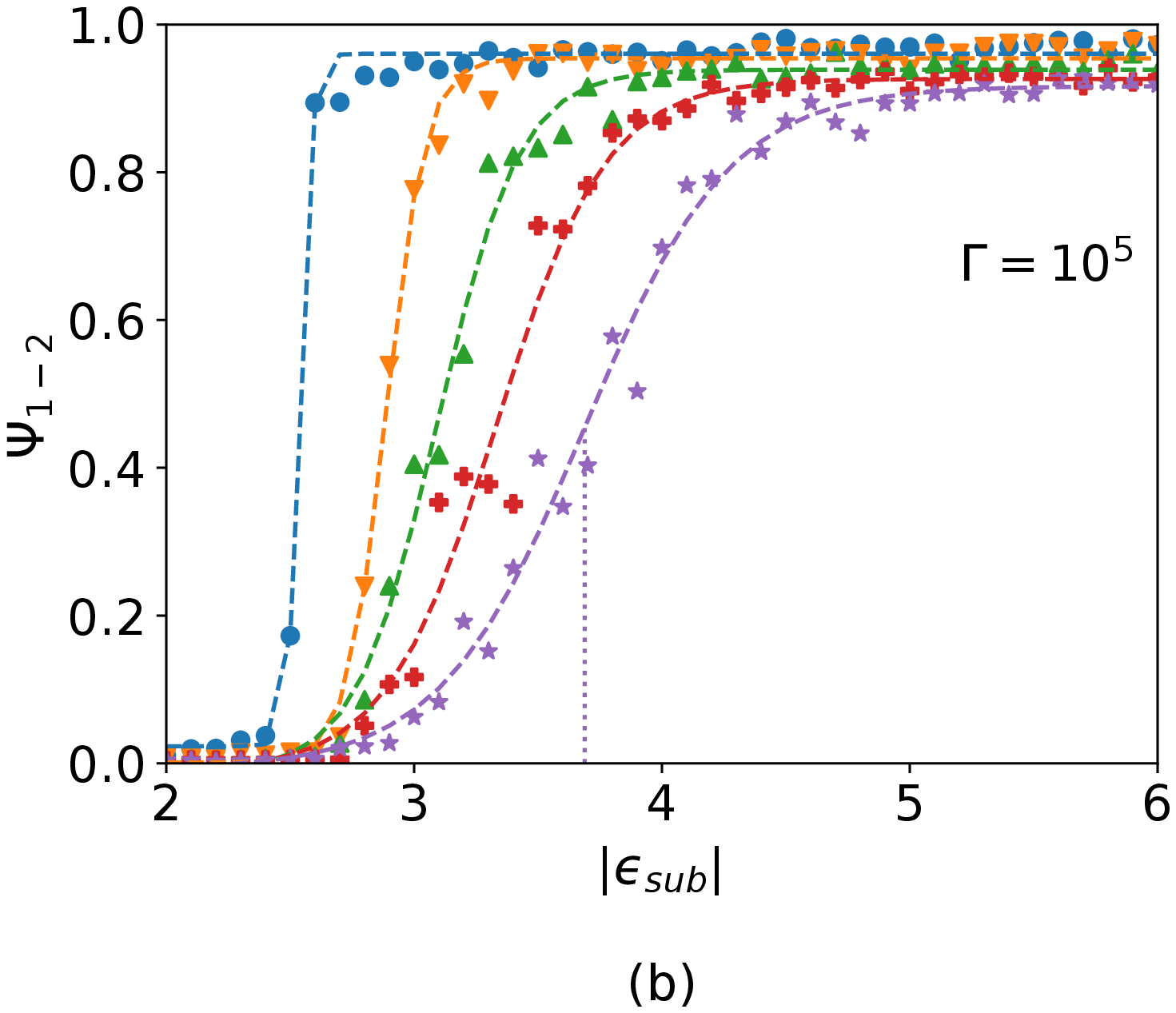}
    \caption{$\psid$ after deposition of 1 ML in the CGM with corresponding fits for (a) $\Gamma = 10^4$ and (b) $\Gamma = 10^5$, averaged over 5 runs.
        The critical substrate attraction strength for the dynamic layering transition is determined
        by the $x$--coordinate of the inflection point (dotted lines for $|\epsilon|=7$ as an example). The legend in (a) also applies to (b).  We only show data for the CGM since those of the SOS model are very similar.
    }
    \label{fig:psidiff_fits}
\end{figure*}

\begin{figure}[b]
    \centering
    \includegraphics[width=\linewidth]{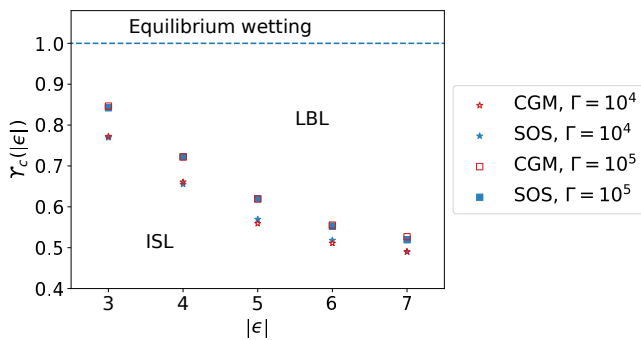}
    \caption{
        The critical ratio $\Upsilon_\text{c}(|\epsilon|)$ for the dynamic layering transition at $\Gamma=10^4$ 
        and $10^5$ in both the CGM and the SOS model. The dashed line is the approximate value 1 for the equilibrium layering transition. 
    }
    \label{fig:crit_esubs}
\end{figure}

In the case of film growth, it is not clear a priori
how to determine the substrate attraction strength $\epsilon_\text{sub,crit}$ for the dynamic layering transition. 
From the roughness behavior in Fig. \ref{fig:rough_comp} it can be deduced that at rather early stages of growth (i.e.\ at or below monolayer deposition) the transition from ISL to LBL is decided. We have investigated various observables
(roughness, coverages of the first and second layer, anti--phase Bragg intensity and growth number) and found that in
particular for $\Theta=1$ (i.e. after deposition of one monolayer), all these quantities show a pronounced qualitative change in behavior when
plotted as functions of $\epsilon_\text{sub}$ (see Appendix \ref{app:obs})). Furthermore, we find a
very suitable order parameter, namely the observable $\psid = \Psi_1 - \Psi_2$, with $\Psi_i,\; i \in \lbrace 1,2\rbrace$ quantifying the net occupancy or filling
of the first or second layer. 
In an LBL scenario $\psid$ will be $1$, whereas in an island forming scenario its value will be close to $0$. 
If dynamic layering is connected to the
sharp equilibrium layering transition one would expect its value to jump from $0$ to $1$ at 
$\epsilon_\text{sub,crit} \approx \epsilon$, and that the transition is rounded by finite--size effects. 
In our simulations we have found that the $\psid(\esub)$ can be fitted quite well 
with a tanh curve (this is similar to the behavior of an order parameter for an equilibrium transition):
We determined $\epsilon_\text{sub,crit}$ as the inflection point of the fitting curve
(see Fig. \ref{fig:psidiff_fits}: There we only show data for the CGM since the results for the SOS model are very similar).

The dependence of the critical ratio $\Upsilon_\text{c}=\epsilon_\text{sub,crit}/\epsilon$ on $|\epsilon|$ (shown in Fig. \ref{fig:crit_esubs}) demonstrates (i) that
there is only a small difference between the CGM and the SOS model, and more importantly (ii)
that the difference to the equilibrium value $\Upsilon=1$ increases with increasing $\epsilon$
at fixed $\Gamma$. This corresponds to an increasing ``dynamic gap'' in the onset of layering with increasing
attraction strength.

\begin{figure*}[t]
    \centering
    \includegraphics[width=0.4\linewidth]{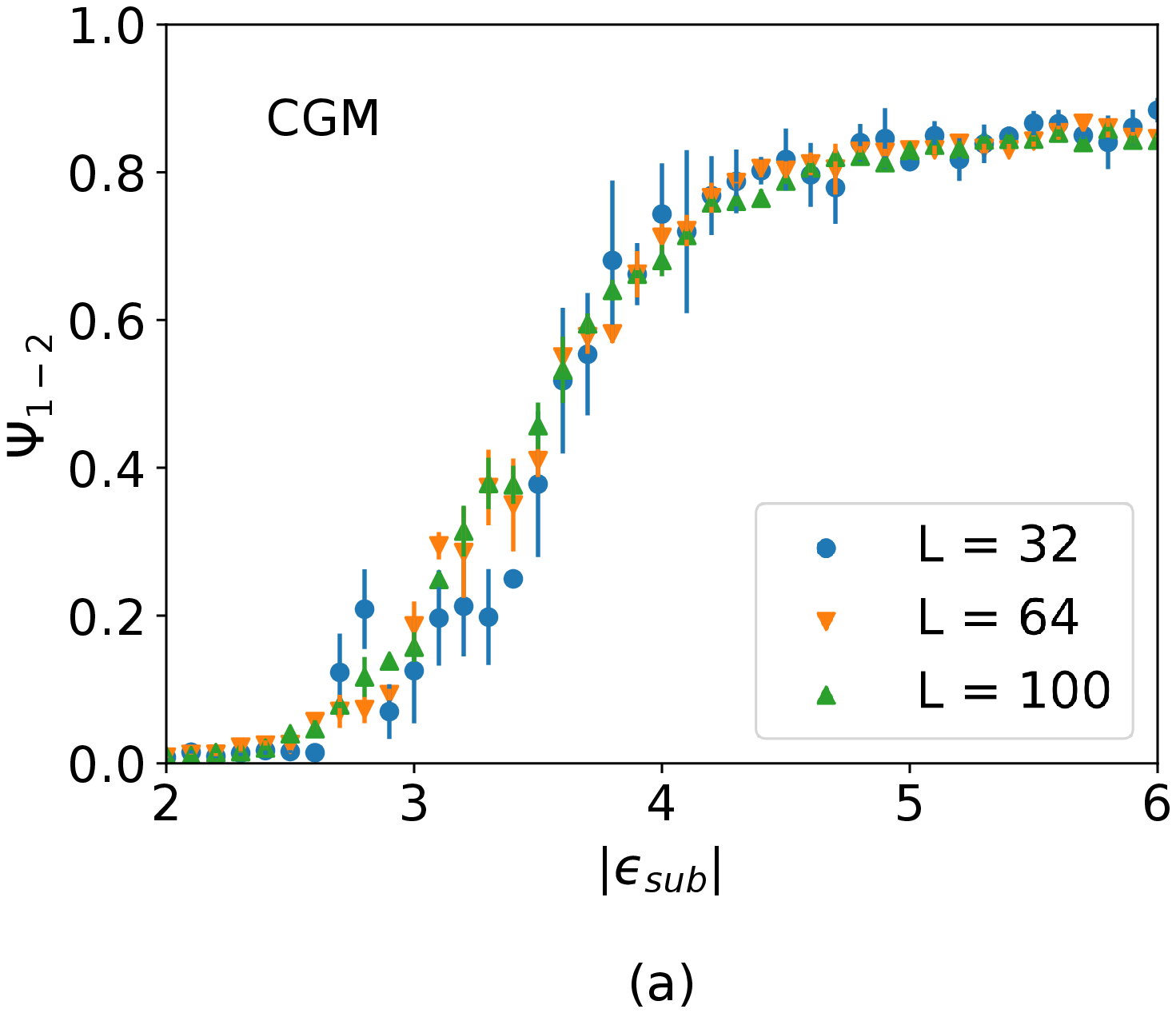}
    \includegraphics[width=0.4\linewidth]{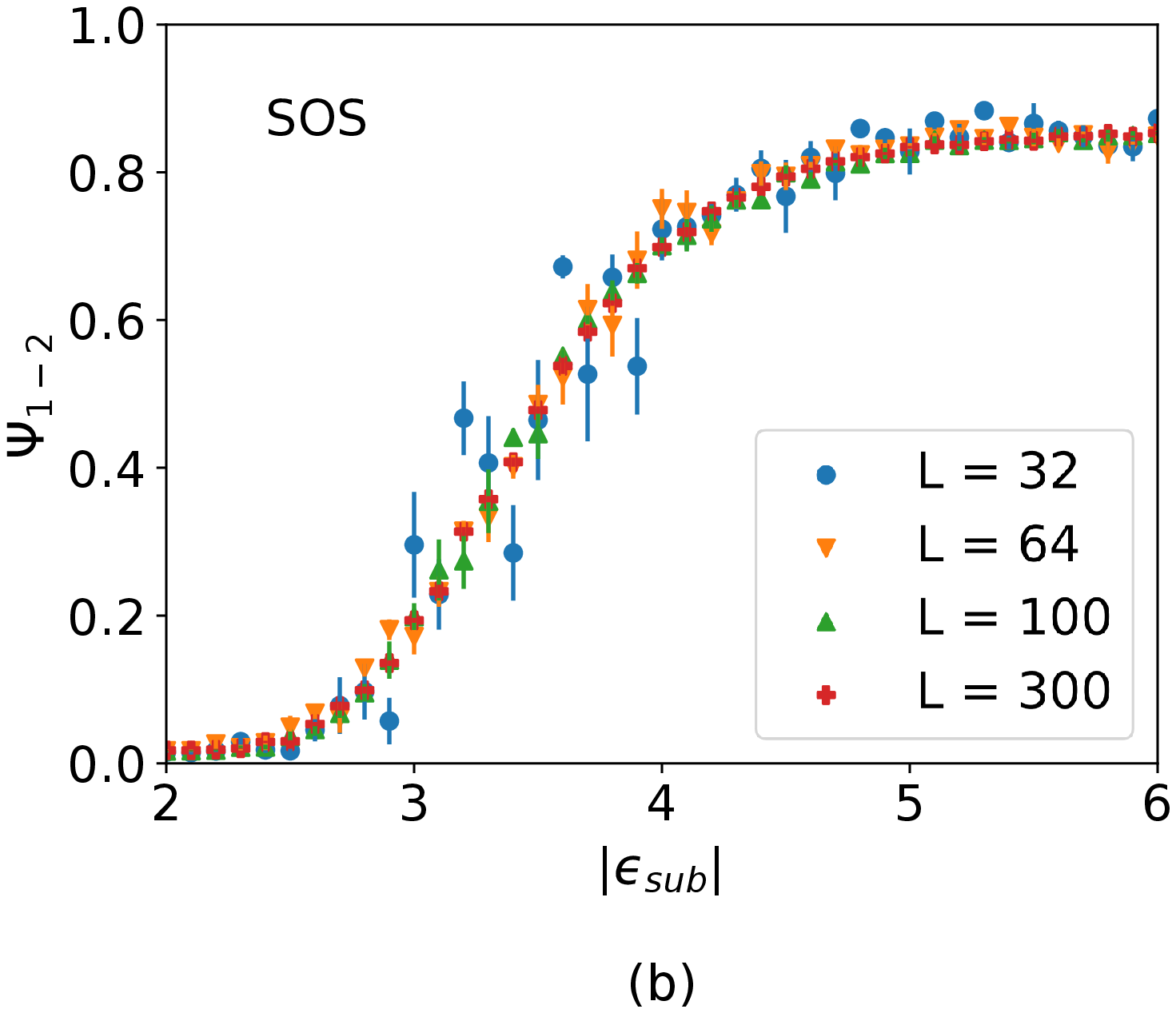}
    \caption{
        The behavior of $\psid(|\esub|)$ at $\Gamma = 10^4$ and $\epsilon = -7$ for different system sizes.
        (a) CGM, (b) SOS. Data are averaged over 5 runs.
    }
    \label{fig:scaling}
\end{figure*}

The tanh--fits for $\psid(|\esub|)$ (see Fig. \ref{fig:psidiff_fits})) show that the width is increasing for increasing
$|\epsilon|$ (i.e. for increasing ``distance to equilibrium'', since a stronger $\epsilon$ leads to a slower exploration of the phase space). This finite width is not a finite--size effect as
in equilibrium transitions, it is  
largely independent of the size of the lattice. This is illustrated in Fig. \ref{fig:scaling} which demonstrates 
that both in the SOS and the CG model for $L \ge 64$  there is no significant change in $\psid(|\esub|)$ when increasing $L$,
 which might be surprising considering that $L=64$ is comparatively small. For $L \lesssim 32 $
the data are very noisy and cannot be fitted very well. 

\begin{figure}[b]
    \centering
    \includegraphics[width=\linewidth]{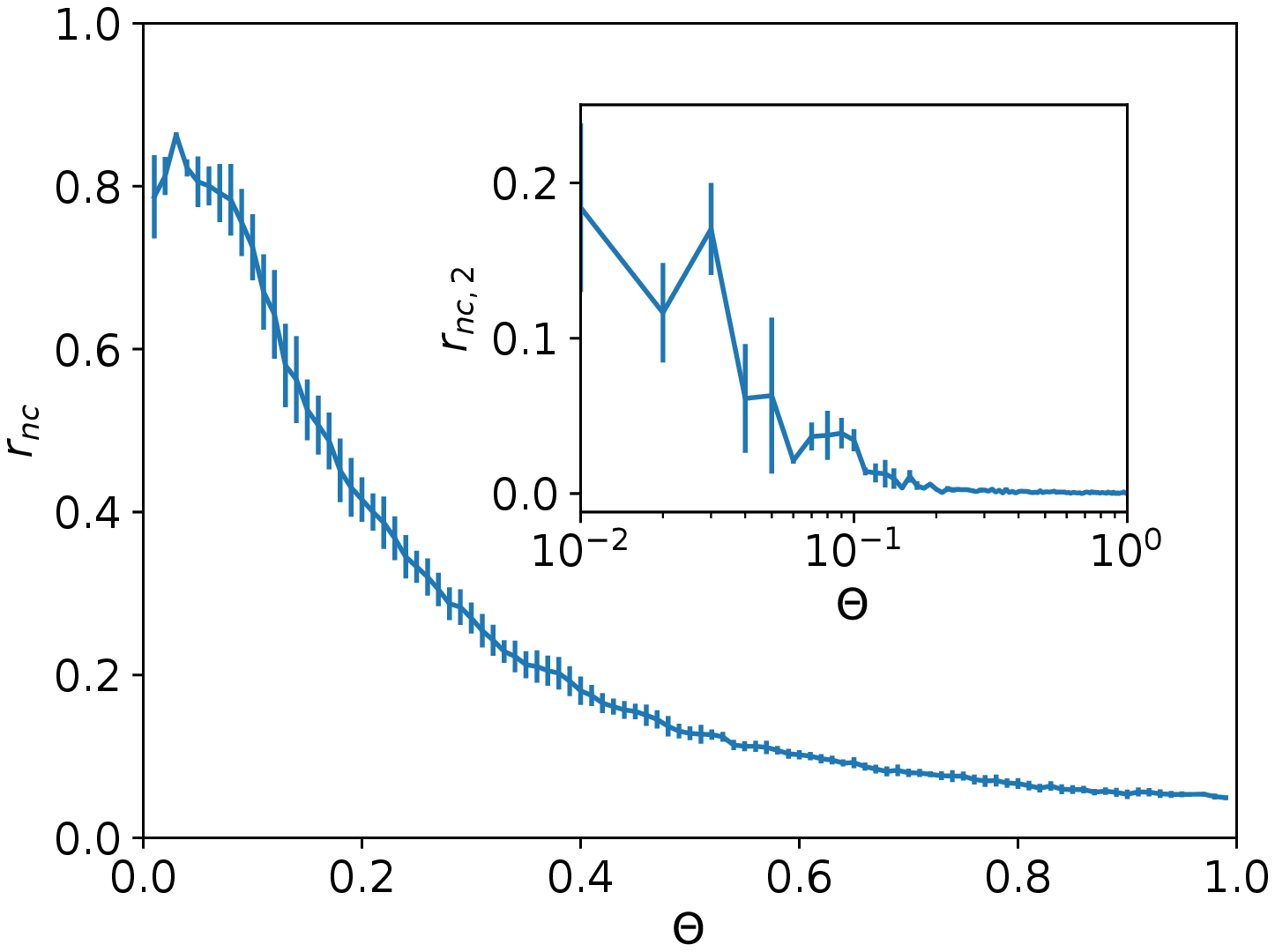}
    \caption{The fraction $r_\text{nc}$ of particles not connected to the substrate vs. $\Theta$ 
        (total amount of deposited MLs) at $\Gamma=10^5$, $\epsilon=-3$, $\esub=-2$. The inset shows the corresponding fraction,
        if only disconnected particles in the first two layers are considered. We see that at $\Theta=1$ all particles 
        within the first two layers are in some way connected to the substrate. }
    \label{fig:flrat}
\end{figure}

The proximity of the dynamic layering transition for the CGM and the SOS models appears surprising at first glance,
considering that in the CGM particles can desorb from the substrate which should be a non-negligible process at 
lower $\epsilon$. From the roughness evolution in Fig. \ref{fig:rough_comp} we see, however, that
the behavior of $\sigma(\Theta)$ is very similar for both models for substrate attraction strengths around the
transition value. We rationalize this by studying in the CGM the fraction $r_\text{nc}$, the number of particles not connected 
to the substrate divided by the
total number of deposited particles (in the SOS model, $r_\text{nc}=0$ by definition).
Figure \ref{fig:flrat} shows this for exemplary parameters: While during the very early stages of growth $r_\text{nc}$ is substantial, it
quickly drops to zero when enough material is deposited for one monolayer, i.e. when $\Theta \to 1$. Considering only disconnected particles in the first two layers (these
contribute to the order parameter $\psid$), we see that the corresponding ratio is small from the start of growth and essentially
zero at $\Theta=1$ (inset of Fig. \ref{fig:flrat}). Therefore, we can expect that the occupation in the first two layers
of the growing film is very similar for the CGM and the SOS models, resulting in similar roughness and order parameters.   

We have analyzed the dynamic layering transition using the control parameter $\Upsilon=\esub/\epsilon$, in particular to facilitate
a comparison to the equilibrium wetting/layering transition of the lattice gas model.
In the dynamic model, the critical ratio most generally depends on 
three parameters, $\Upsilon_c = \Upsilon_c (\epsilon, \Gamma, E_\text{ES})$.
Experimentally, the substrate attraction strength $\esub$ appears to be difficult to tune for locating the transition.
Within certain limits, it is easier to tune $\Gamma$ by changing the deposition rate, but seeing the transition would require
that the chosen substrate is not too deep in the LBL or the ISL regime. It is quite easy to control the substrate temperature $T$;
however, changing $T$ influences all three variables: $\epsilon,\ E_\text{ES} \propto 1/T$ and 
$\Gamma=D/F \propto T \exp(-E_\text{D}/(k_BT))$ where $E_\text{D}$ is an energetic barrier for surface diffusion.\footnote{For colloidal diffusion, one typically assumes $D \propto T \exp(-E_D/(k_BT))$. In contrast, for metal-on-metal diffusion, one assumes $D \propto \exp(-E_D/(k_BT))$}
Therefore, temperature variations are not very suitable for locating the transition.

\subsection{Flattening transition}
\label{subs:flat}

\begin{figure*}[t]
    \centering
    \includegraphics[width=0.4\linewidth]{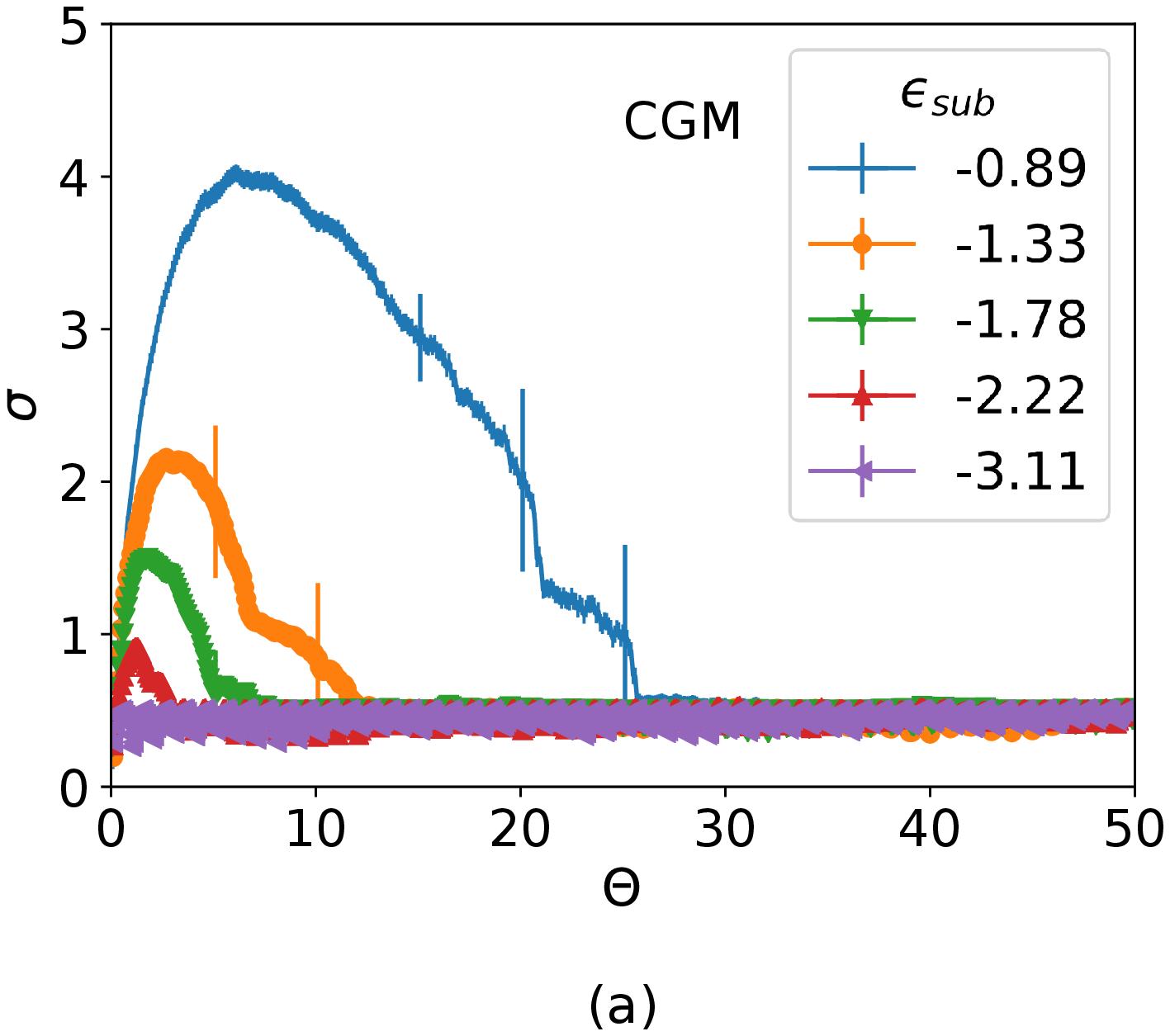}
    \includegraphics[width=0.4\linewidth]{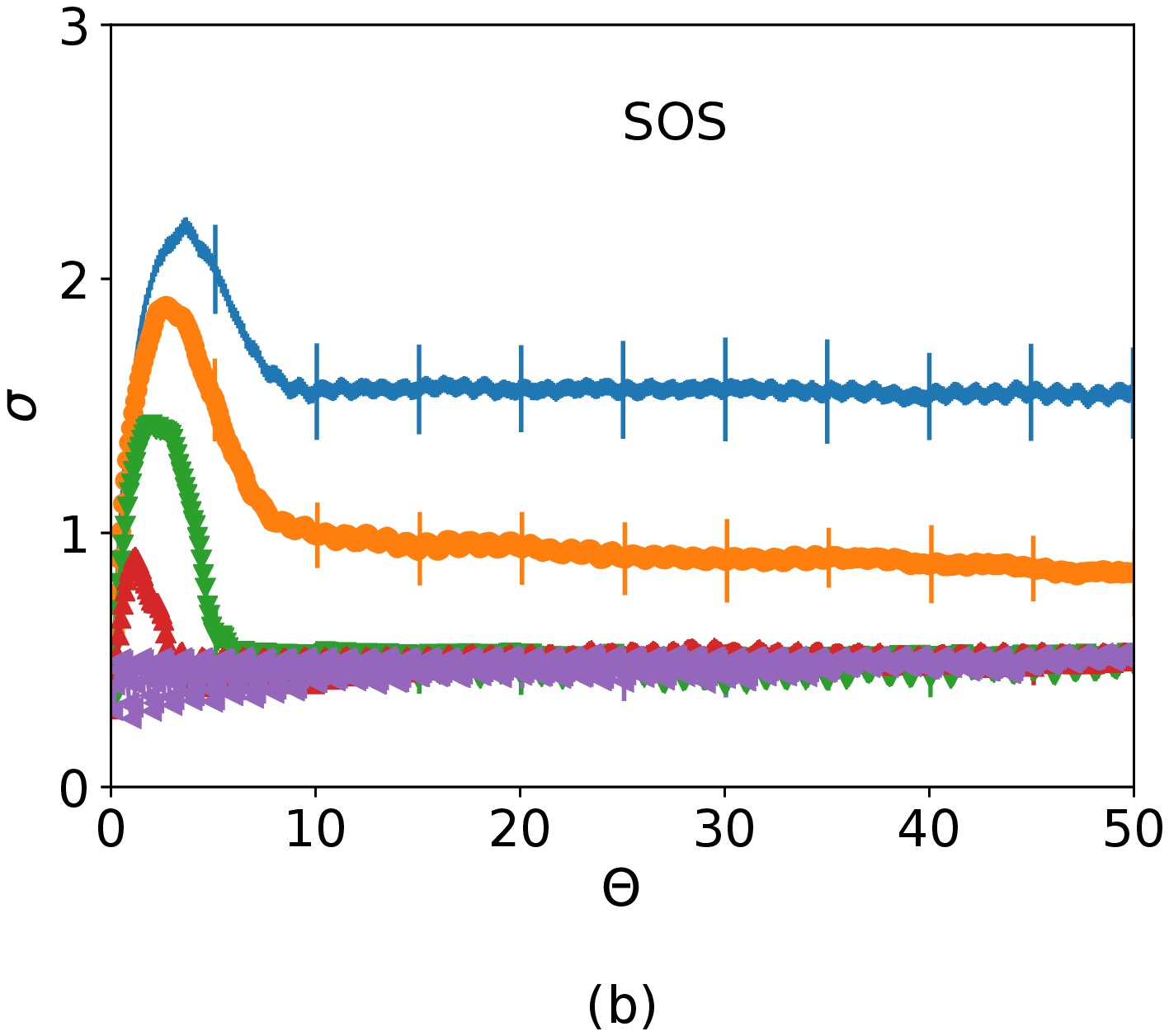}
    \includegraphics[width=0.4\linewidth]{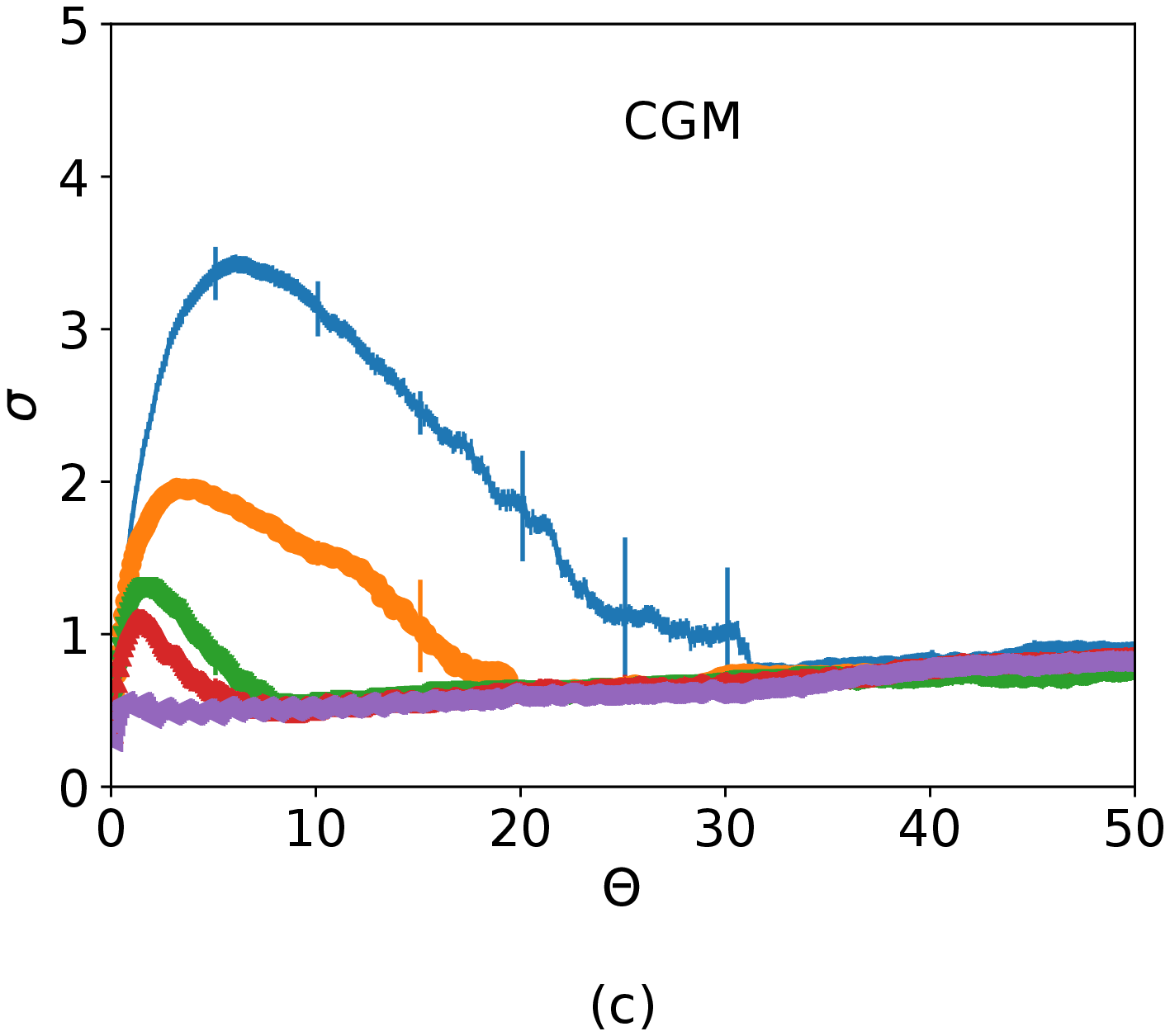}
    \includegraphics[width=0.4\linewidth]{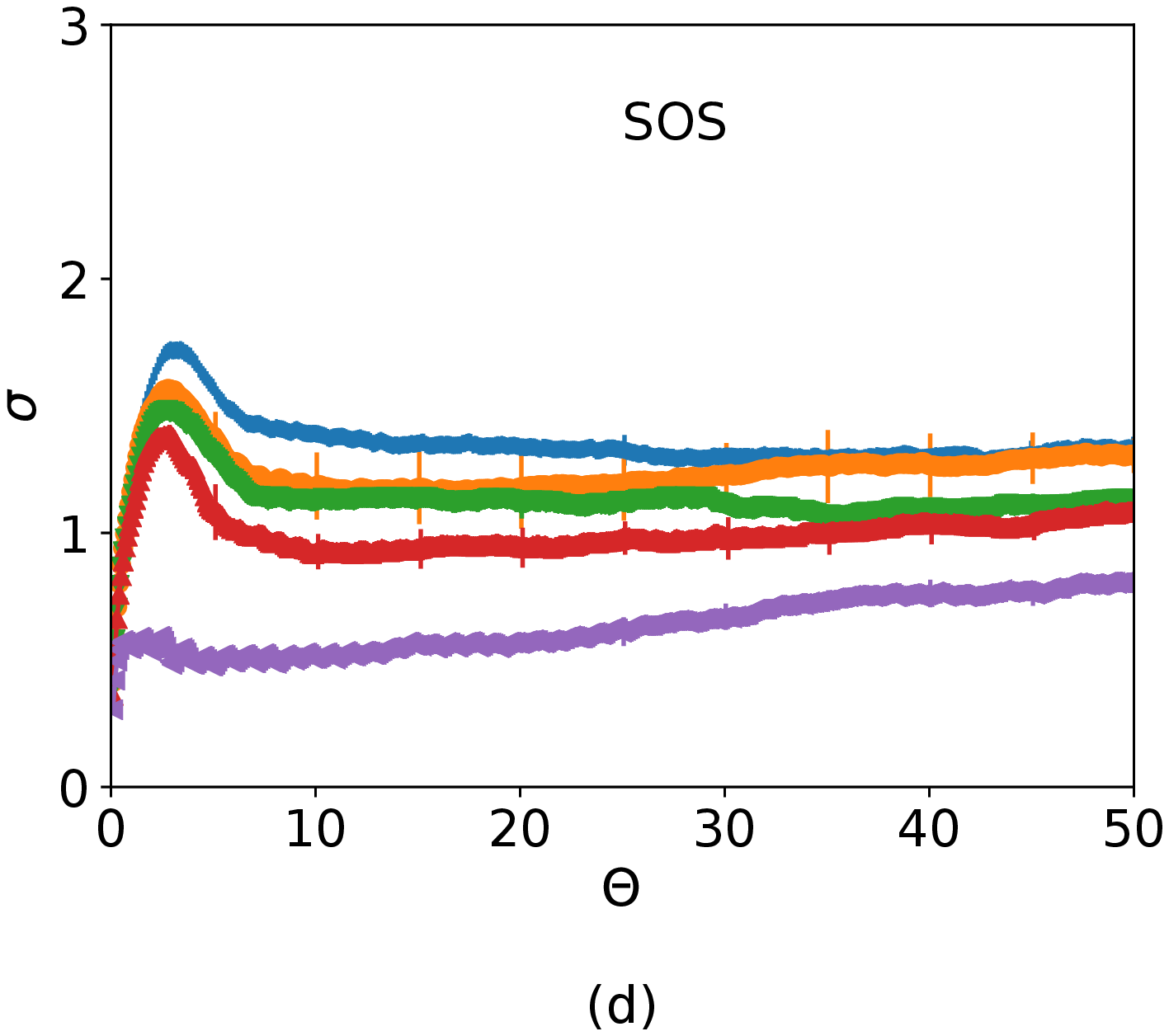}
    \caption{ 
        Roughness evolution for deposition up to 50 ML for $\Gamma = 10^4$ and different substrate strengths $\esub$.
        (a) $\epsilon=-3$, CGM, (b) $\epsilon=-3$, SOS, (c) $\epsilon=-5$, CGM, (d) $\epsilon=-5$, SOS. The legend in (a) also applies to (b)--(d).
        Note that the maximum roughness on the $y$-axis differs between the CGM and the SOS model. The CGM data are averaged over 3 independent runs, the SOS data over 5.
    }
    \label{fig:manylayers}
\end{figure*}

In Fig. \ref{fig:manylayers} we show the roughness evolution up to a total coverage $\Theta=50$ both for the SOS model and the CGM,
again for $\epsilon=-3$ and $-5$, respectively, and each for a range of substrate attractions $\epsilon_\text{sub}$. 
For weaker $\esub$ (initial growth in the ISL mode), 
after an initial increase of the roughness (reflecting the island formation), it then decreases for intermediate times and reaches $\sigma \lesssim 1$, indicating a change to growth in LBL fashion. 
This behavior can be seen in both the CGM and the SOS model, however we will see that in the SOS model 
it can only occur at sufficiently weak $\epsilon$ and strong $\esub$.\\
Reduction of the roughness occurs due to a ``flattening'' of the film and can be explained in simple terms by the following picture: Initial island growth at a certain deposition rate results in a finite coverage of the substrate with islands. Newly arriving particles ``see'' an effective substrate which is a mixture of the original one and the islands with a correspondingly increased effective $\esub$. This triggers dynamic layering, eventually leading to the substrate being completely covered in particles. For any particles arriving afterwards, growth continues as it does in a homoepitaxial system. However, this can only occur if the effects of 3D growth at short times are weak.

We call this transition \textbf{ISL $\to$ LBL}. In the CGM it can be seen for all combinations of $\epsilon$ and $\esub$ 
for which the system initially shows island formation.
In the SOS model, however, we see that at these parameters (Fig. \ref{fig:manylayers}(b)) the roughness will 
(after the initial increase) decrease towards a constant value larger than 1.
We call this transition \textbf{ISL $\to$ CONST}.  
The appearance of this transition is a consequence of the restrictions in the layer-changing move 
in the SOS model which can only 
proceed one layer up or down. At weak $\esub$, the particles will initially form islands on the substrate. At later
times these will start to merge, but do so only incompletely. Trenches of depth $>1$ remain between them, which can only be filled
with deposition moves. This is different in the CGM where height differences (such as in the trenches) 
can be compensated more easily due to particles 
desorbing into the gas phase and statistically re-adsorbing at the film at positions with higher binding number.
For weak $\esub$ this has the dual effect of initially forming larger and fewer islands compared to the SOS model
(with ensuing higher roughness) and later on smoothening the film of merged islands, resulting in LBL growth. 

\begin{figure*}[t]
    \centering
    \includegraphics[width=0.4\linewidth]{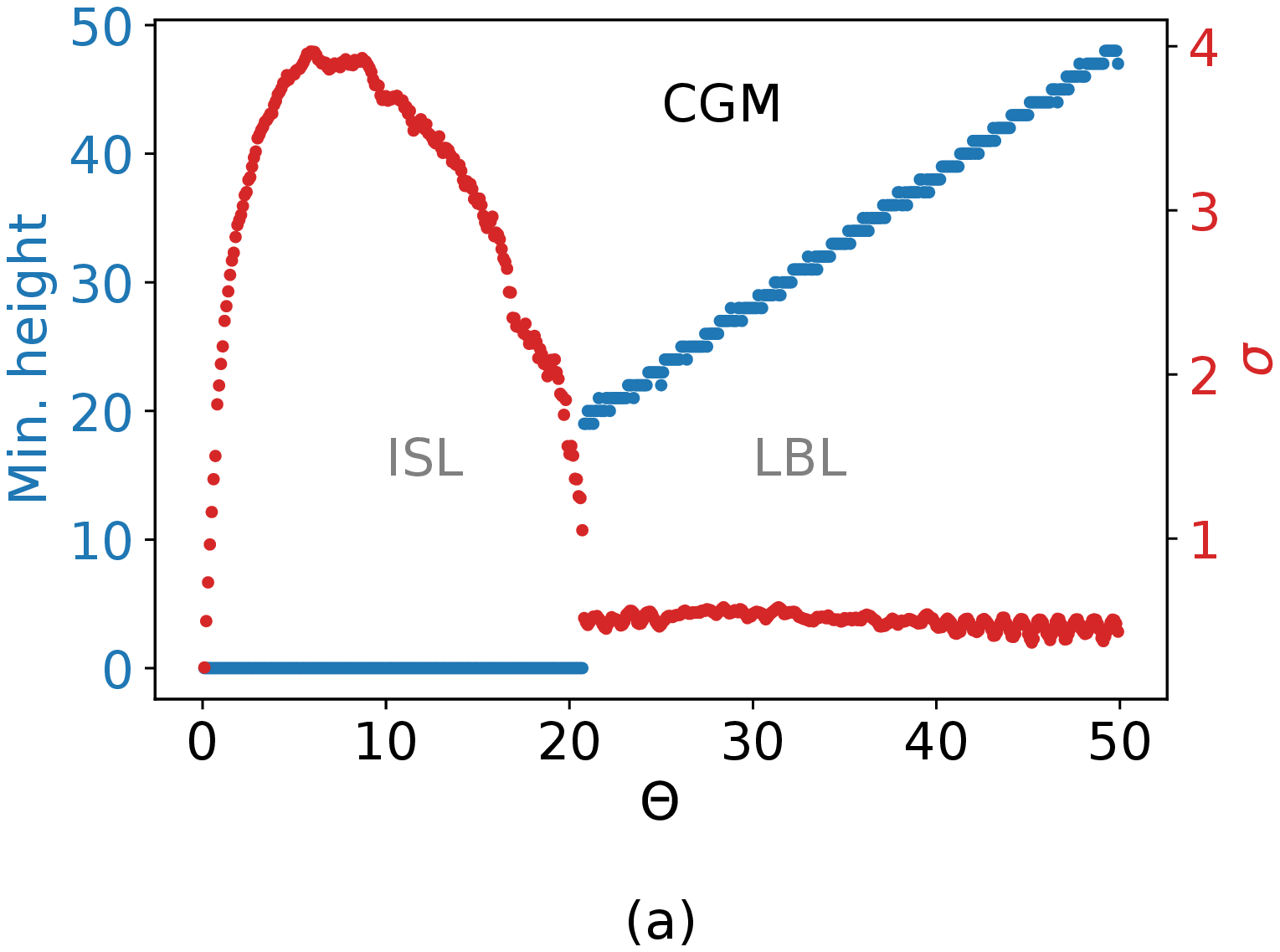}
    \includegraphics[width=0.4\linewidth]{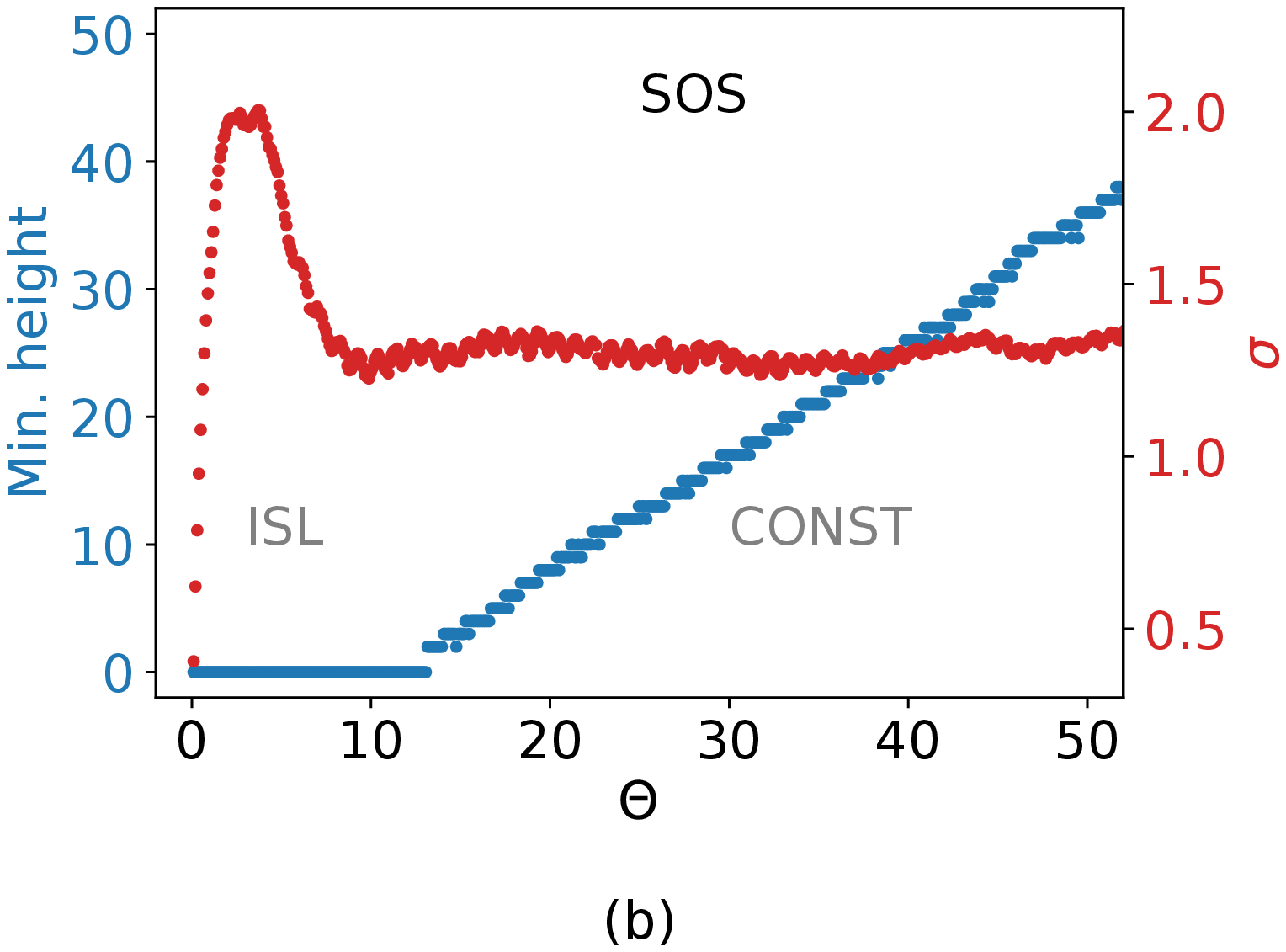}
    \includegraphics[width=0.4\linewidth]{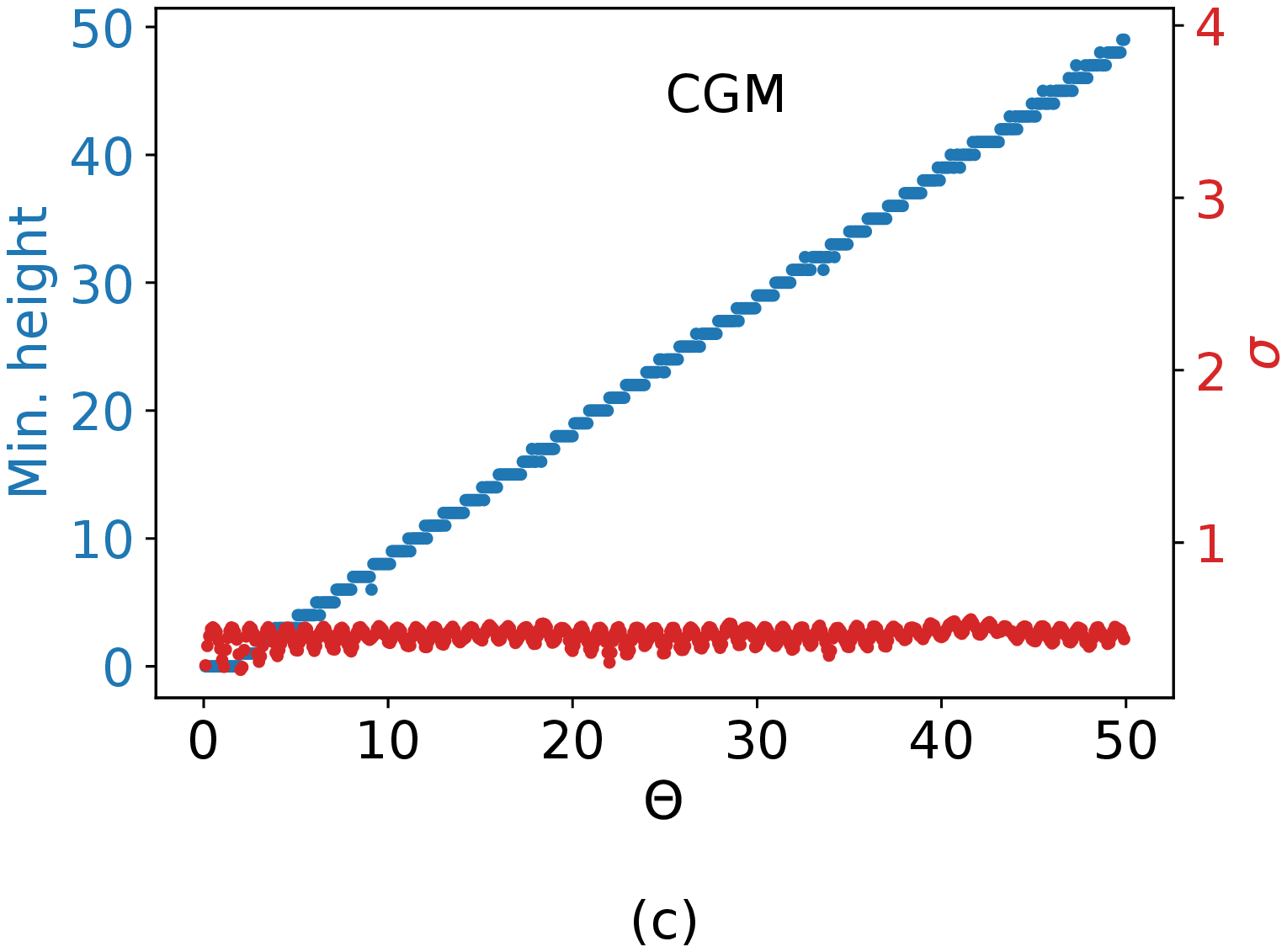}
    \includegraphics[width=0.4\linewidth]{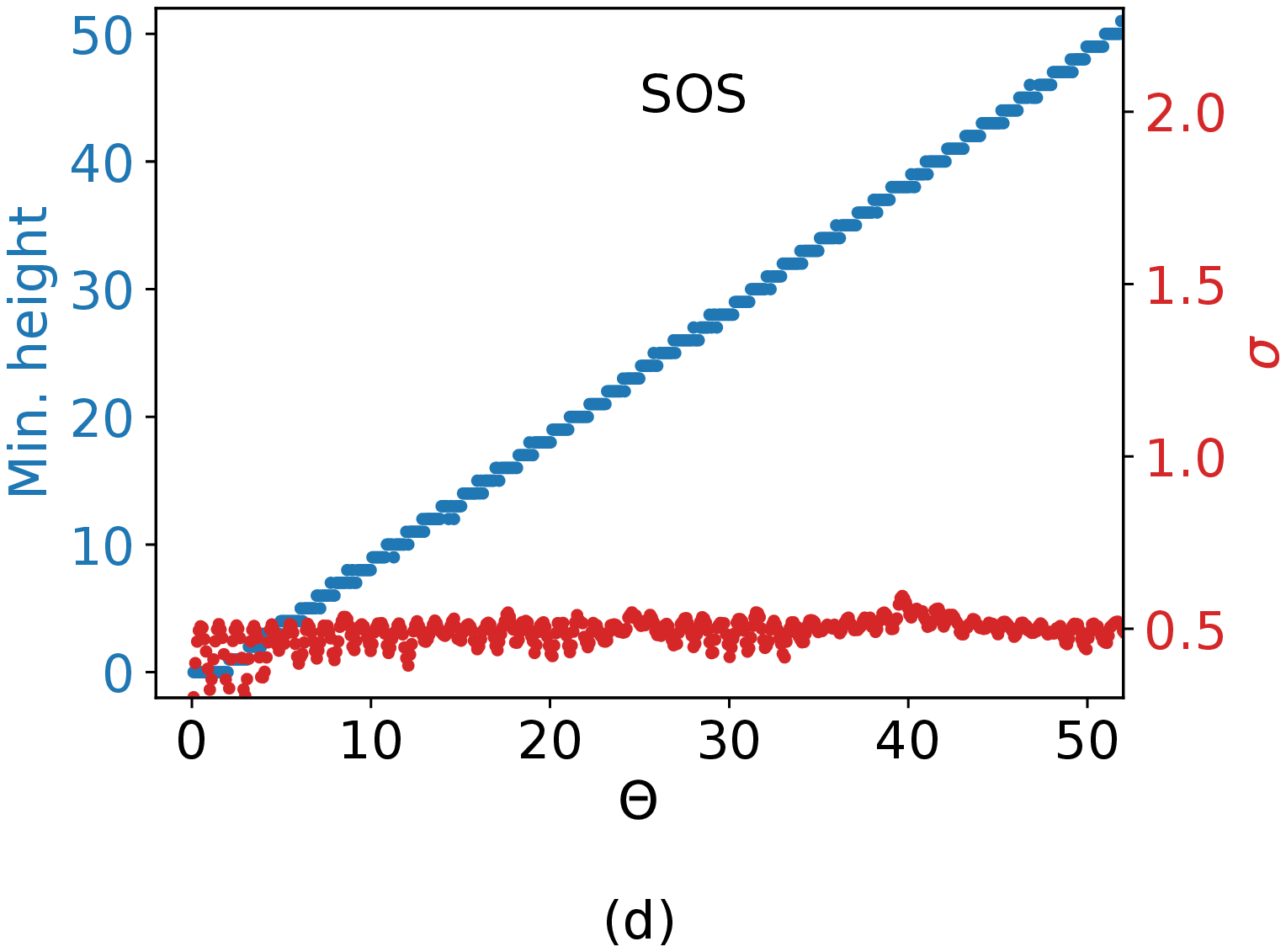}
    \caption{ Minimum film height and roughness vs $\Theta$ in the CGM and the SOS model for (a),(b) $\esub=-0.89$ and (c),(d) $\esub=-2.67$. For $\esub = -0.89$ we see that in the CGM the min. height will be $0$ and then show a large jump around $\Theta = 20$, while in the SOS model there is no jump, and the min. height increases continuously after a certain point. For $\esub=-2.67$ we can see such a continuous increase in both models from an early time on.}
    \label{fig:minh_comp}
\end{figure*}

\begin{figure*}[t]
    \centering

    \includegraphics[width=0.4\linewidth]{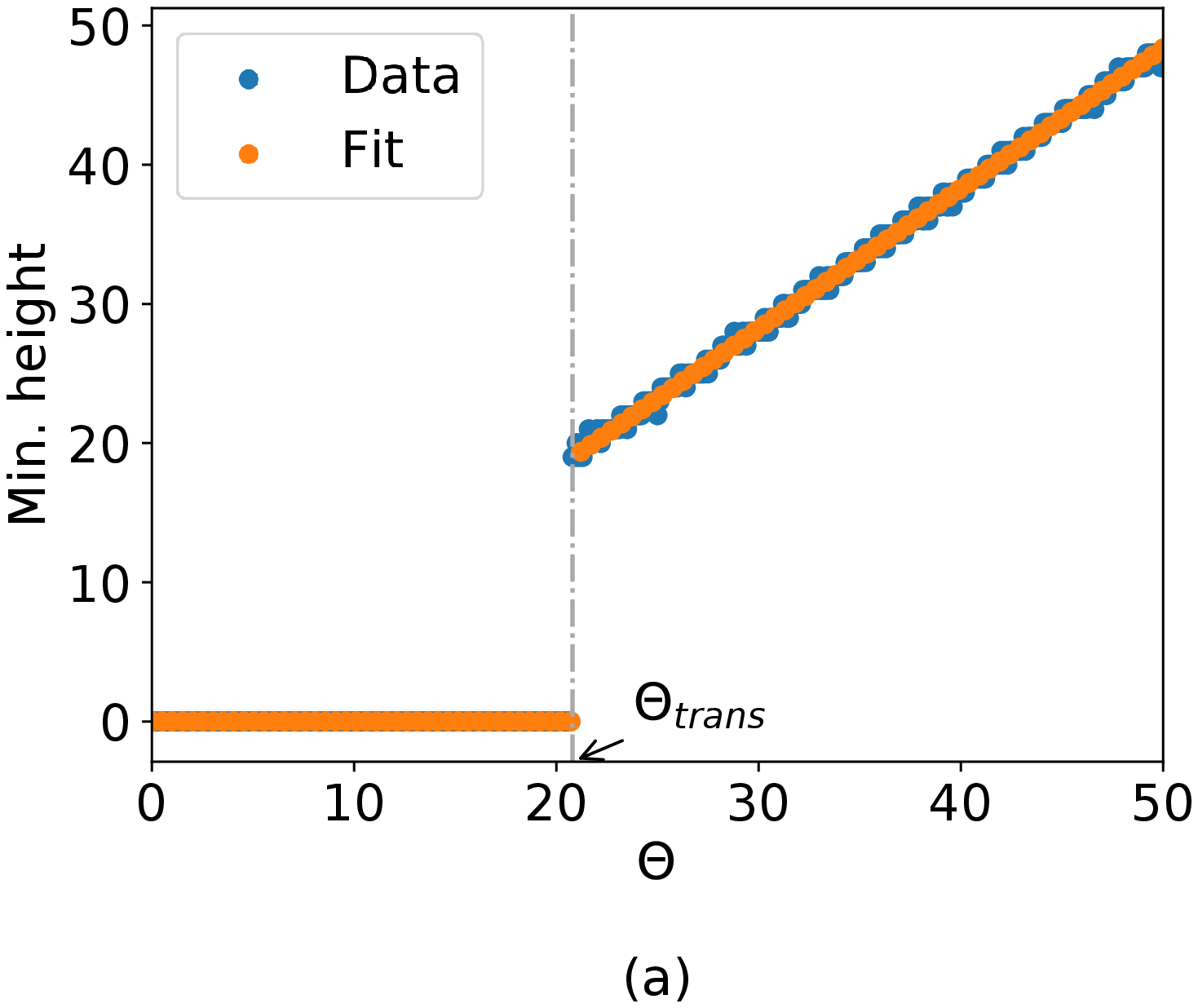}
    \includegraphics[width=0.4\linewidth]{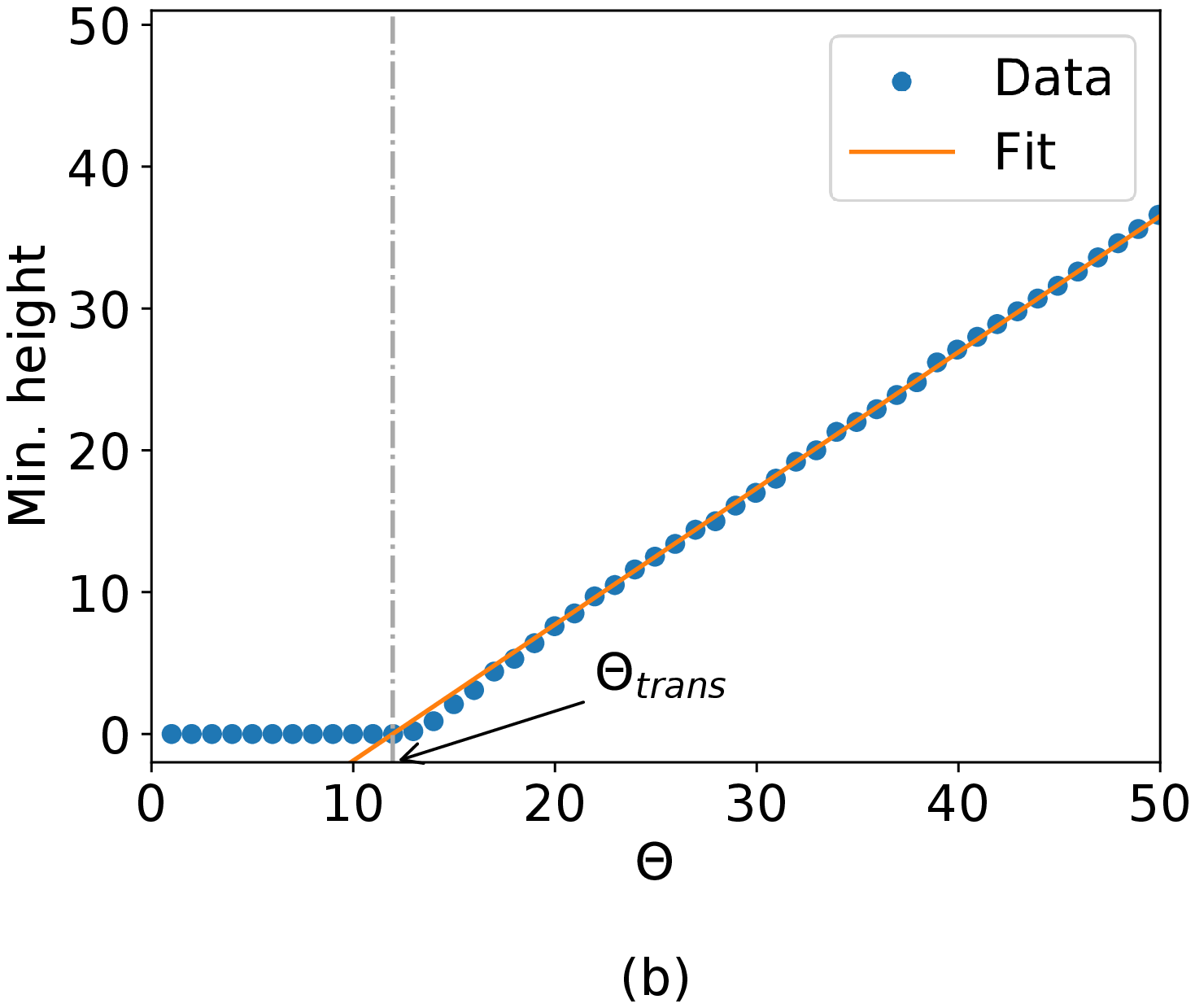}
    \caption{Examples of fitted $\Theta_\text{trans}$ for $\epsilon=-3$, $\esub=-0.89$, $\Gamma = 10^4$ in (a) the CGM and (b) the SOS model. The plots show the data and the fit used to determine the respective  $\Theta_\text{trans}$.}
    \label{fig:minh_fits}
\end{figure*}

To quantify the times at which these transitions will occur, we found the behavior of the minimum film height or of the kurtosis 
of the height distribution (fourth normalized moment), depending on $\esub$, to be effective. For quantitative analysis, we use the minimum height, 
since this yields clearer results than the kurtosis (see Appendix \ref{app:kurt}).
In Fig. \ref{fig:minh_comp} we show examples for the evolution of the minimum film height overlaid with that of the roughness. 
In the CGM, the minimum film height jumps at the point where the roughness drops to values $<1$ (indicating LBL behavior) and then
increases linearly. In the SOS model, the minimum film height changes from being flat zero to a linear increase near the corresponding
roughness drop.
In order to obtain numerical values for the transition time (coverage $\Theta_\text{trans}$), we used different methods for the two models. 
For the SOS model, we average $\text{min.\ \{ h}(\Theta) \}$ of several runs, fit a line to the region where it increases and extract the intersection of this line with the x-axis. 
For the CGM results,
we define a fit function of the form
\begin{equation}
f(\Theta) = \left\{
\begin{array}{ll}
0, & \text{if}\ \Theta < \Theta_\text{trans}\\
a \cdot \Theta + b, & \text{else}
\end{array}
\right.
\end{equation}
where $\Theta_\text{trans}$ is the transition point ISL $\to$ LBL. Here $a$,$b$, and $\Theta_\text{trans}$ are free parameters which we
fitted for each run. The final result for the transition time is the average over the different runs.  
Examples for these fits are shown in Fig. \ref{fig:minh_fits}.

\begin{figure*}[t]
    \centering

    \includegraphics[width=0.4\linewidth]{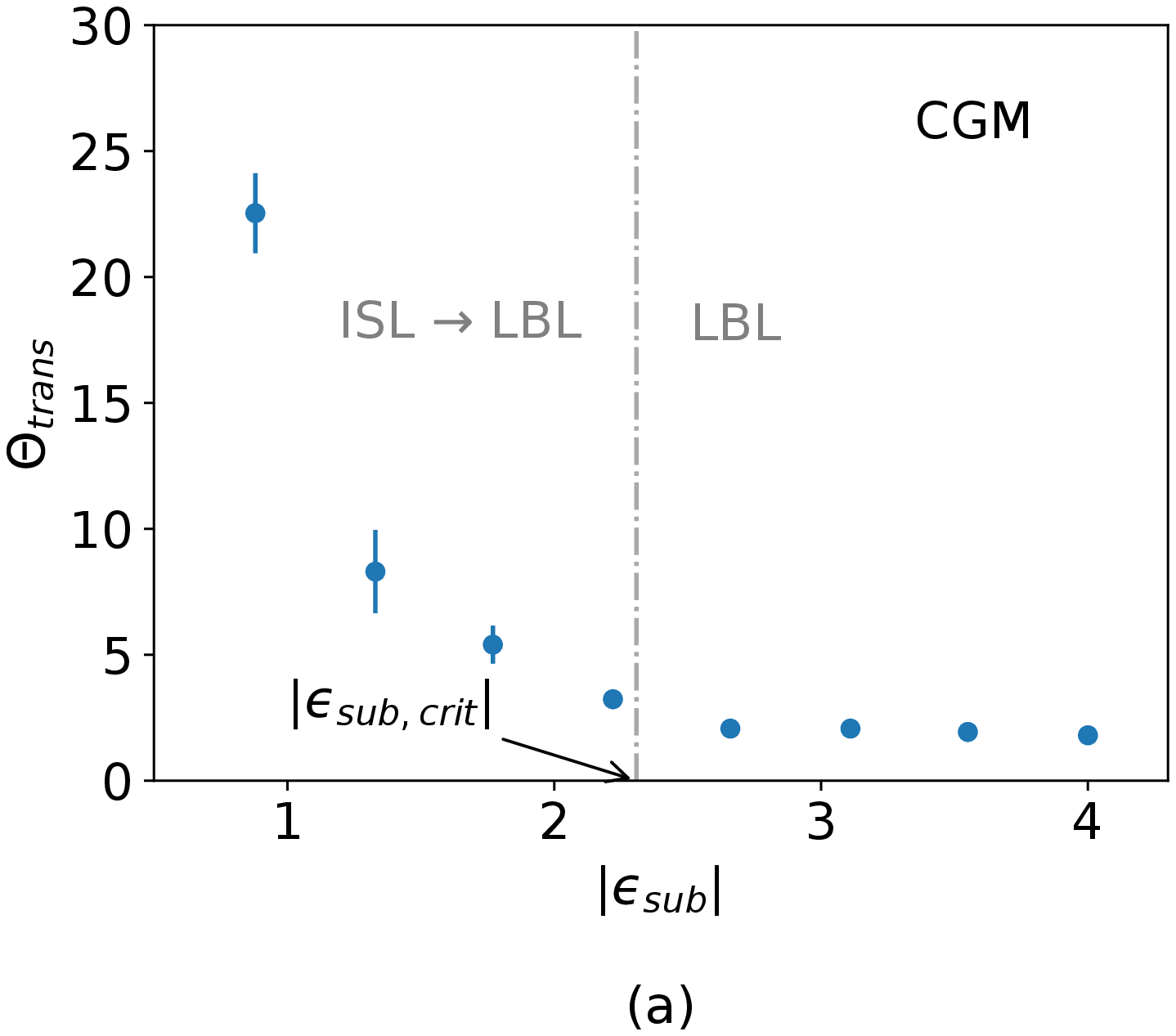}
    \includegraphics[width=0.4\linewidth]{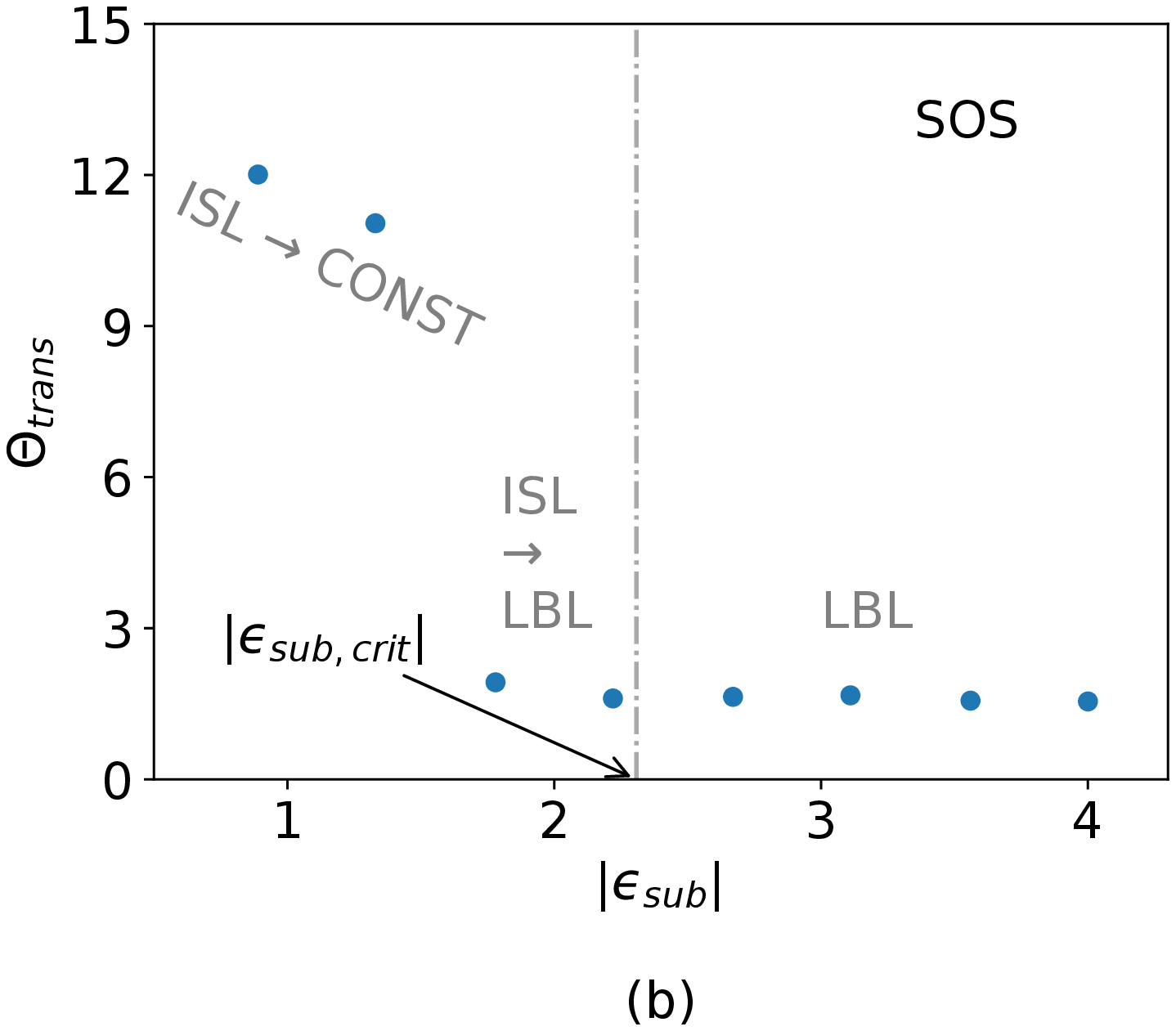}
    \includegraphics[width=0.4\linewidth]{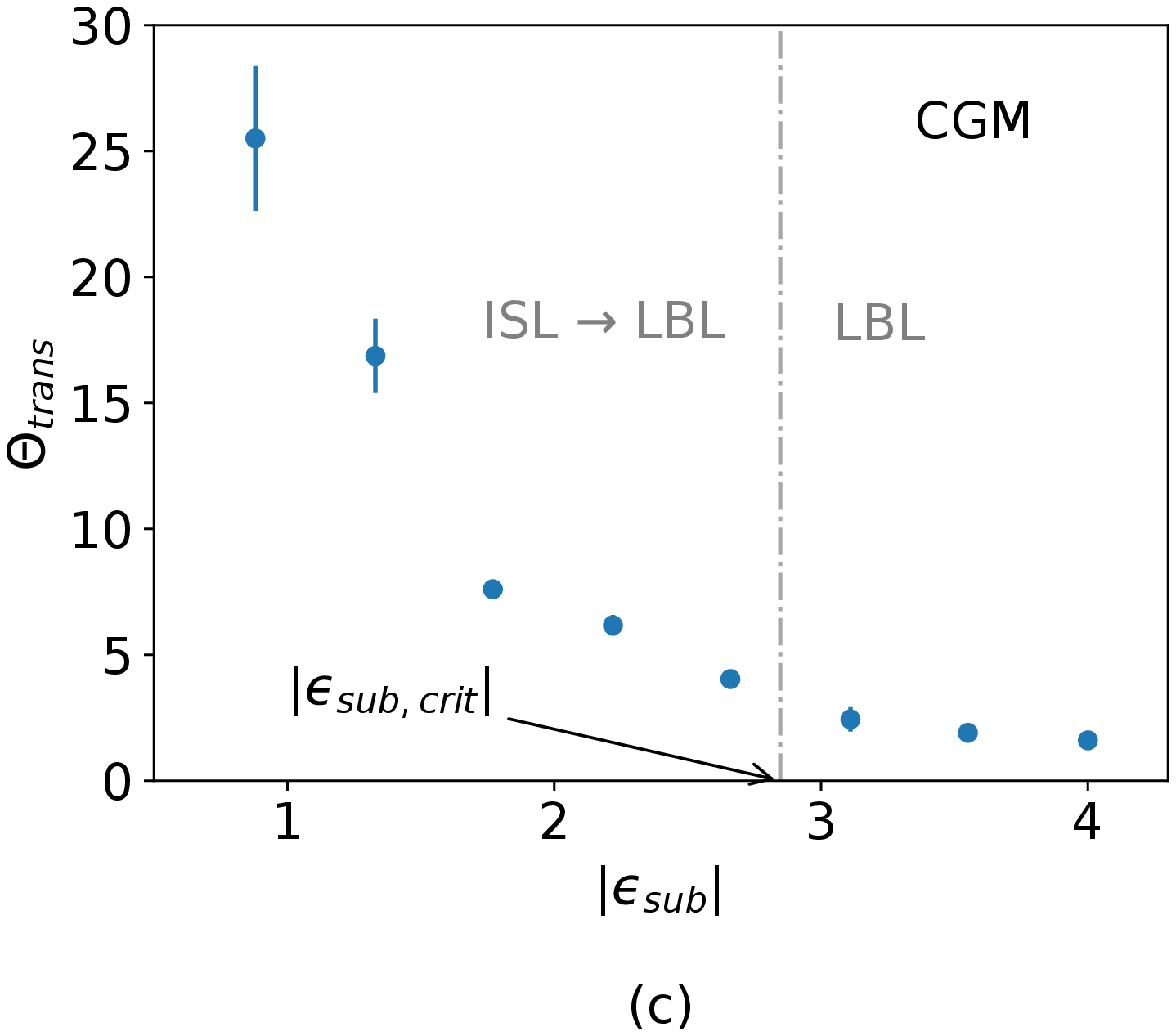}
    \includegraphics[width=0.4\linewidth]{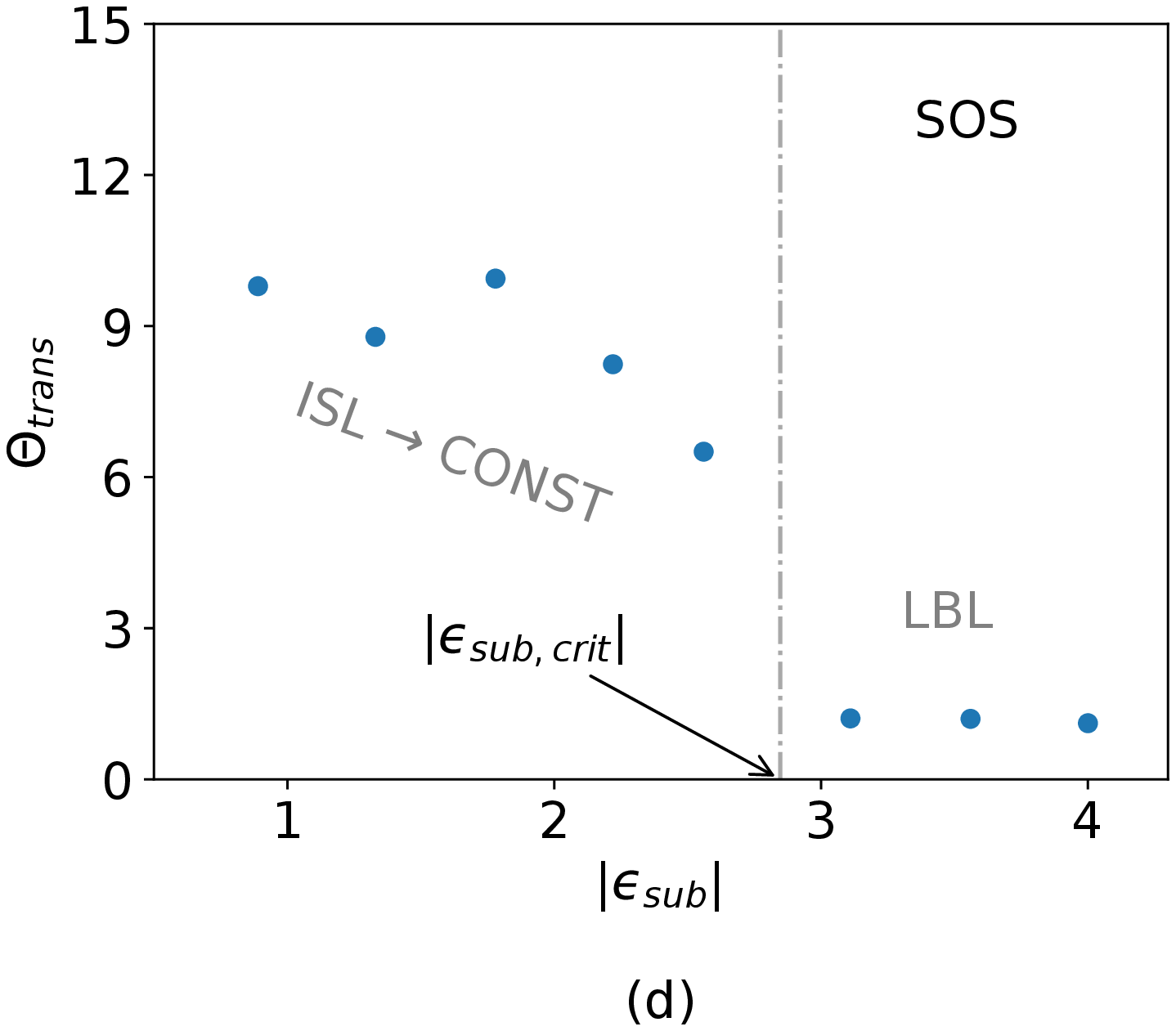}
    \caption{$\Theta_\text{trans}$ plotted vs $\esub$ in the CGM and SOS model respectively for (a),(b) $\epsilon = -3$ and (c),(d) $\epsilon = -5$. For $\epsilon = -3$ we see that in the SOS model there is a jump in $\Theta_\text{trans}$ at an $\esub$ which is lower than that of the dynamic wetting transition, while for $\epsilon = -5$ this jump is at a $|\esub| \approx |\epsilon_\text{sub,crit}|$, causing the ISL $\to$ LBL transition to disappear. In the CGM we see a smooth decrease of $\Theta_\text{trans}$ with an increasing $\esub$. The CGM data are averaged over 3 runs, while the SOS data are averaged over 10 runs.}
    \label{fig:sos_multi_trans}
\end{figure*}

The variation of the transition time $\Theta_\text{trans}$ with $\esub$ reveals interesting behavior
As can be seen in Fig. \ref{fig:sos_multi_trans}, $\Theta_\text{trans}$ in the SOS model changes rather abruptly from a value around 10
to a value below 2. $\Theta_\text{trans} \sim 10$ is connected to the occurrence of the \textbf{ISL $\to$ CONST} transition, 
whereas $\Theta_\text{trans} < 2$ is connected to the \textbf{ISL $\to$ LBL} transition. The abrupt change thus means that there
is a ``transition (upon substrate change) between transition scenarios (in temporal roughness evolution)''. 
For $\epsilon=-3$, the change
in $\Theta_\text{trans}$ occurs at a $|\esub| < |\epsilon_\text{sub,crit}|$, i.e. at a substrate attraction strength
weaker than the critical attraction strength for the dynamic layering transition. Consequently one observes
the sequence in growth modes \textbf{ISL $\to$ CONST} to  \textbf{ISL $\to$ LBL} to \textbf{LBL} 
upon increasing $|\esub|$ (see
Fig. \ref{fig:manylayers}(b), going from the top curve to the bottom curve). For $\epsilon=-5$, the change
in $\Theta_\text{trans}$ occurs at a $|\esub| \sim |\epsilon_\text{sub,crit}|$. This leads to a 
disappearance of the \textbf{ISL $\to$ LBL} transition (which is ''swallowed'' by LBL growth from the start), 
i.e. one observes  only the sequence \textbf{ISL $\to$ CONST} to \textbf{LBL} upon increasing $|\esub|$ (see
Fig. \ref{fig:manylayers}(d)). 
In the CGM, there is only one transition scenario (\textbf{ISL $\to$ LBL}), and the variation of $\Theta_\text{trans}$ with $\esub$
is smoother (although a drop with increasing substrate attraction strength is seen as well).

\subsection{Asymptotic growth behavior}
\label{subs:asymptotic}

We have analyzed quantitatively the dynamic layering transition and the flattening transition
for vanishing Ehrlich--Schwöbel barrier. If $\Gamma$ is not too small, film growth in the CGM
will always return to LBL growth at intermediate times, and in the SOS model there will be a finite intermediate
time when the first layer is fully covered. This means that we return to homoepitaxial growth where
substrate and film material are the same. In this case, we expect that the film will always roughen for very 
long deposition times, although for $E_\text{ES}=0$ the effect is quite weak.  
This has been recently studied for the SOS model,\cite{Assis_2015} and the following approximate
scaling relation has been found:
\begin{equation}  
\sigma \propto \Theta^\beta/(\Gamma^{3/2}(\exp(-|\epsilon|) +a)\;,
\end{equation} 
with $\beta \approx 0.2$ and $a = 0.025$. For $|\epsilon|<-\ln a \approx 3.7$ this implies equivalent roughness
evolution if
\begin{equation}
|\epsilon| - \frac{3}{2} \log\Gamma = \text{const}
\label{eq:scaling_epi} \;,
\end{equation}
which should be compared to Eq. \ref{eq:scaling_submono} for the scaling of the island density in the
sub-monolayer regime where the factor 3/2 is absent.   

The limit $\Gamma \to 0$ in the SOS model corresponds to stochastic growth ($\beta=1/2$). Thus, for small $\Gamma$,
there is roughening or 3D growth from the start which upon increasing $\Gamma$ crosses over to the scenario
described above.

A non--vanishing Ehrlich--Schwöbel barrier modifies these scenarios quantitatively but not qualitatively. 
For a given $E_\text{ES}>0$ and for very small $\Gamma$ (fast deposition), we start with stochastic growth. The
gradual crossover to either ISL or LBL growth upon increasing $\Gamma$ is shifted to larger $\Gamma$ compared to the case $E_\text{ES}=0$. 
In the homoepitaxial case with $\epsilon=\esub$, the gradual crossover to LBL growth has
been studied in the analytic rate equation model by Trofimov et al. \cite{trofimov_2000} and shows the continuous
shift to higher $\Gamma$. This corresponds very well to the SOS simulation phenomenology.  
It is also clear that the \textbf{ISL $\to$ LBL} transition at intermediate times still exists. It possibly occurs at larger
$\Gamma$, since the necessary inter-layer transport for this transition is slowed down by a nonzero
Ehrlich--Schwöbel barrier, which can be compensated by larger a diffusion constant.
In general, for  $E_\text{ES}>0$ the CGM and SOS model should also show the same qualitative phenomenology but there
is an interesting difference for large ES barriers ($E_\text{ES} \to \infty$). In the CGM, there is always a net
inter-layer transport even if $E_\text{ES} = \infty$ due to desorption/re-adsorption processes and this leads
to the phenomenon of another transition, namely between LBL growth and stochastic growth. This is discussed in Appendix
\ref{app:ESinfinite}.

\begin{figure*}[t]
    \centering  \includegraphics[width=0.8\linewidth]{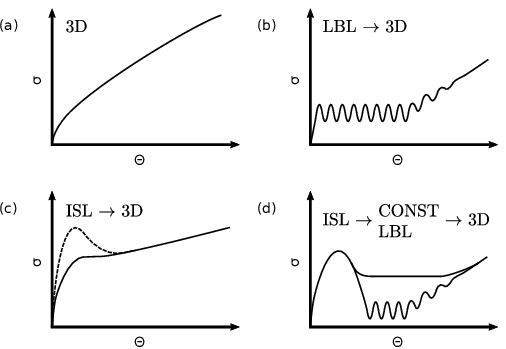}
    \caption{Schematic representations of possible transitions. 
        (a) \textbf{3D} growth from the start for small $\Gamma$ (very high deposition rates). 
        (b) initial \textbf{LBL} growth, followed by \textbf{3D} growth for larger $\Gamma$ and $|\esub| > |\epsilon_\text{sub,crit}|$.
        (c) initial \textbf{ISL} growth, followed by \textbf{3D} growth for intermediate $\Gamma$ and $|\esub| < |\epsilon_\text{sub,crit}|$.
        (d) initial \textbf{ISL} growth, intermediate \textbf{LBL} or \textbf{CONST} growth, followed by \textbf{3D} growth for
        high $\Gamma$ and $|\esub| < |\epsilon_\text{sub,crit}|$. The intermediate growth modes can be summarily termed as \textbf{2D} growth.}
    \label{fig:rough_schemes}
\end{figure*}

Putting all these findings together, we have the following global scenarios for the roughness evolution in the
CGM and SOS model studied here (see Fig. \ref{fig:rough_schemes}). In all scenarios, there is 3D growth
for asymptotic times.
For small $\Gamma$ (very high deposition rates) one finds \textbf{3D} growth from the start, see
Fig. \ref{fig:rough_schemes}(a). For larger $\Gamma$, one must distinguish between 
$|\esub| > |\epsilon_\text{sub,crit}(\epsilon,\Gamma,E_\text{ES})|$ and 
$|\esub| < |\epsilon_\text{sub,crit}(\epsilon,\Gamma,E_\text{ES})|$. In the first case, one finds
\textbf{LBL $\to$ 3D} (initial LBL growth, followed by 3D growth, see Fig. \ref{fig:rough_schemes}(b)).
In the second case, island formation occurs from the start. For intermediate $\Gamma$, the transition
from ISL back to LBL growth does not happen and the growth sequence is \textbf{ISL $\to$ 3D} 
(see Fig. \ref{fig:rough_schemes}(c)). For high $\Gamma$, this transition does happen and the
growth sequence is \textbf{ISL $\to$ LBL/CONST $\to$ 3D} (see Fig. \ref{fig:rough_schemes}(d)). In Sec. \ref{subs:flat}
we have discussed (for $E_\text{ES}=0$) under which condition the intermediate LBL or CONST regime occurs.

\begin{figure*}[t]
    \centering
    \includegraphics[width=0.4\linewidth]{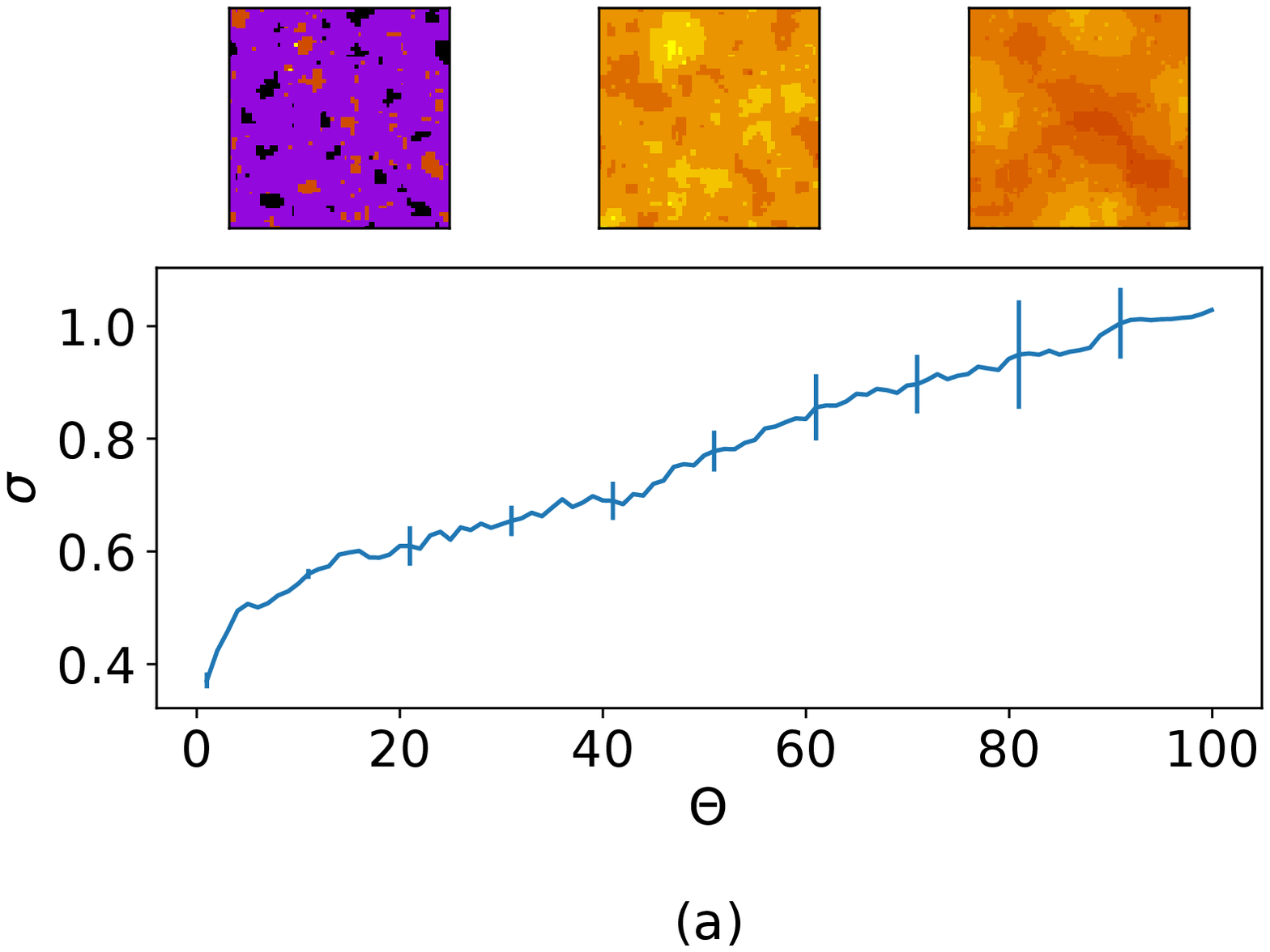}
    \includegraphics[width=0.4\linewidth]{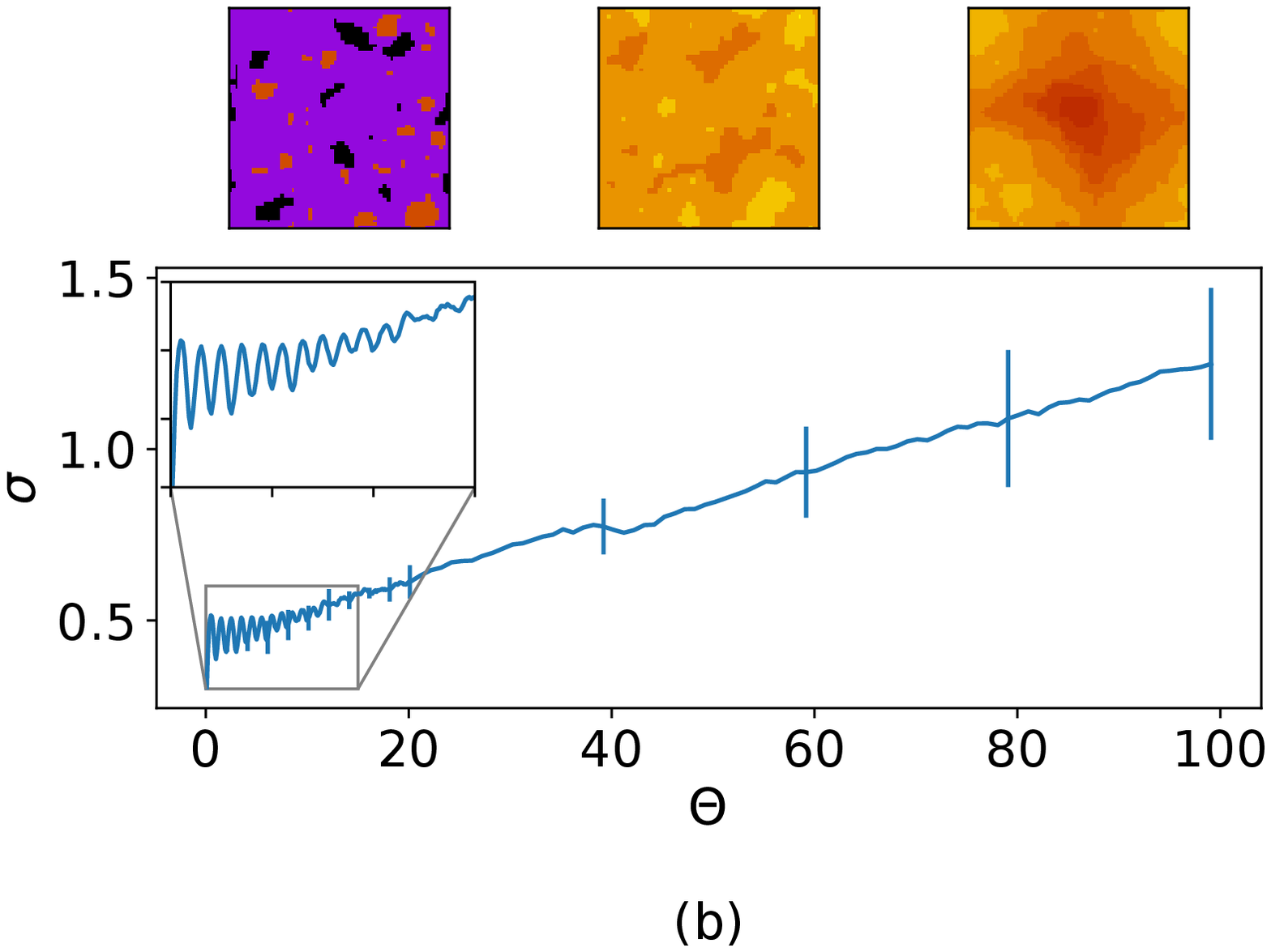}
    \includegraphics[width=0.4\linewidth]{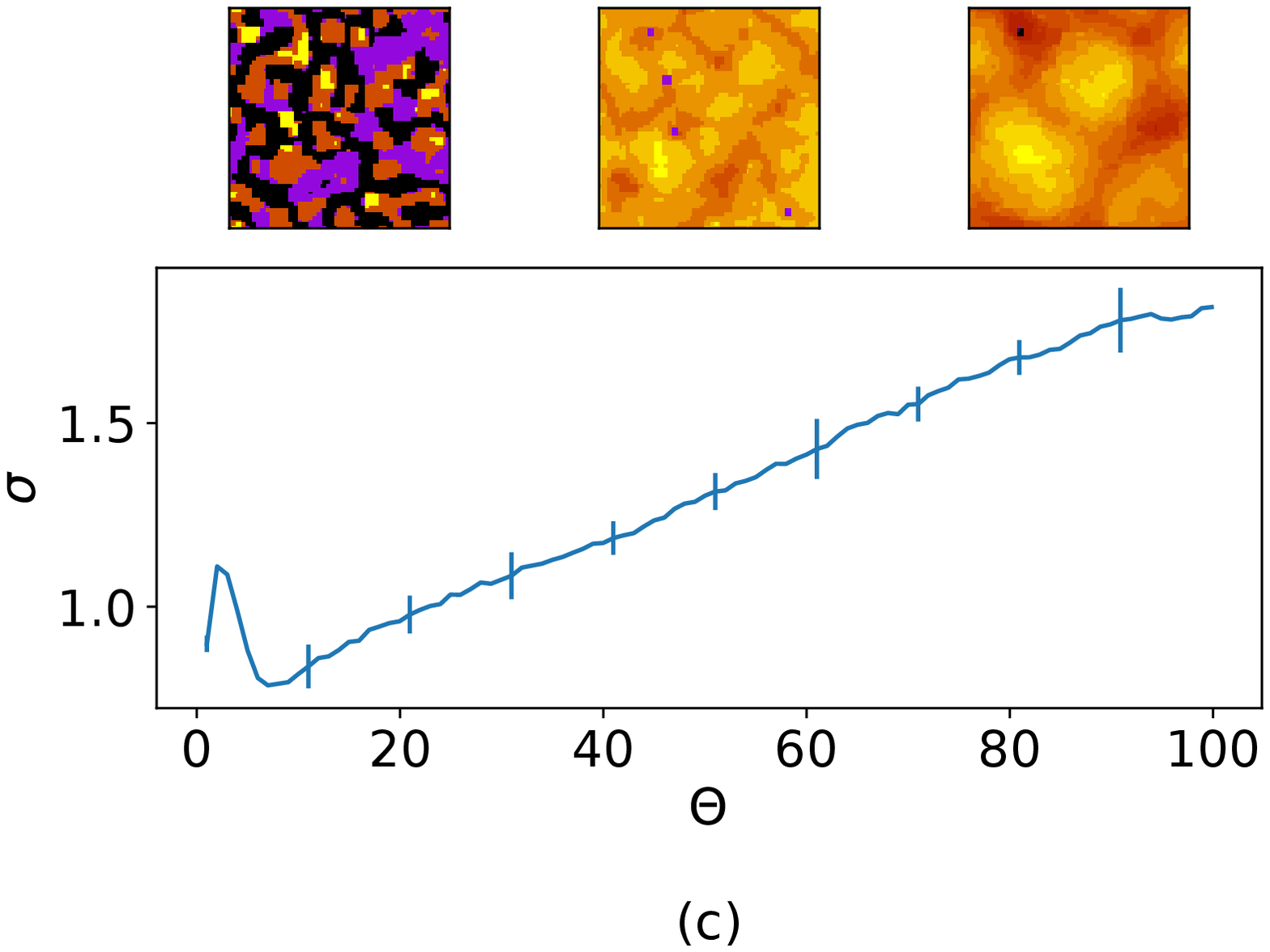}
    \includegraphics[width=0.4\linewidth]{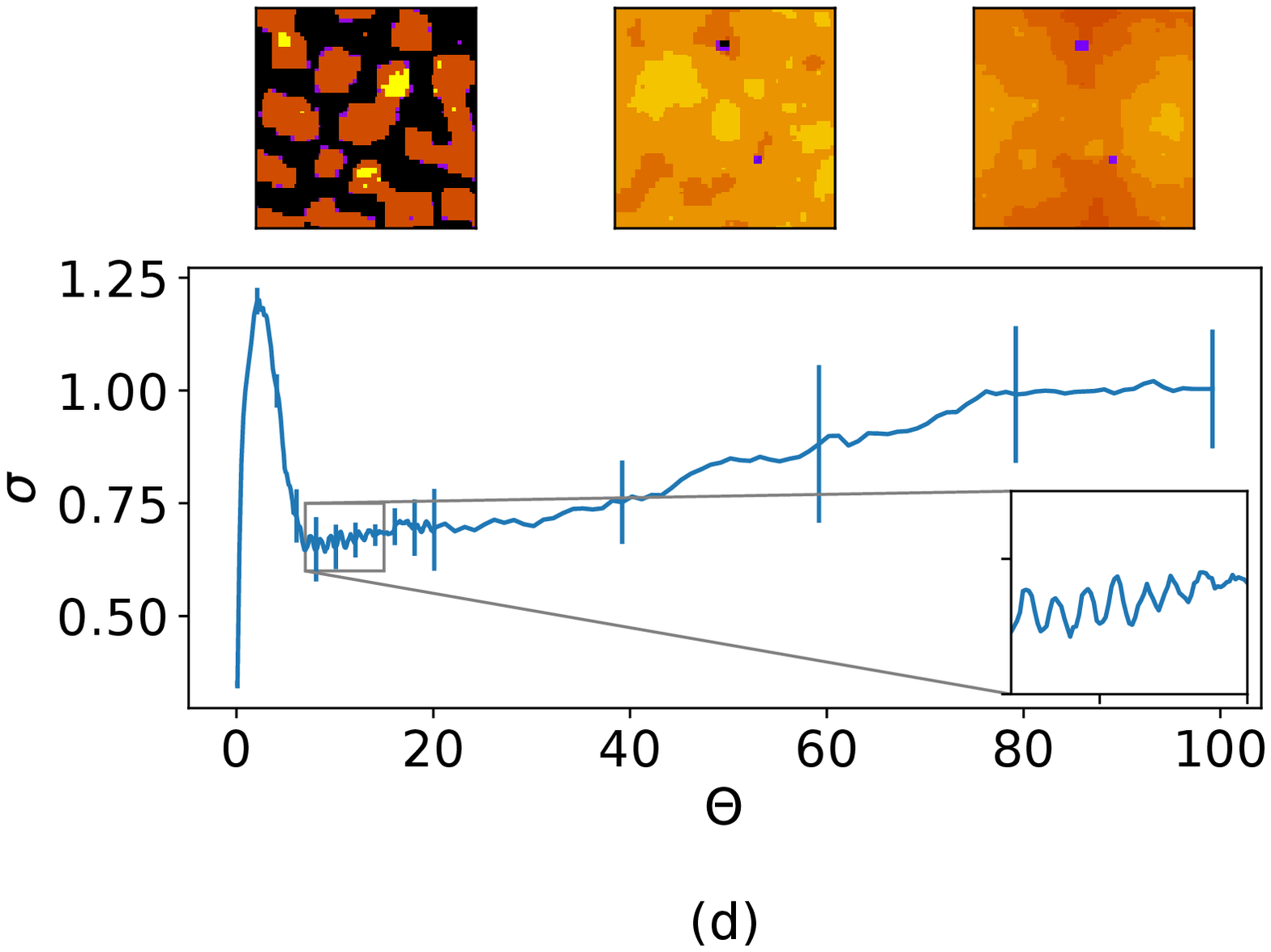}
    \includegraphics[width=0.3\linewidth]{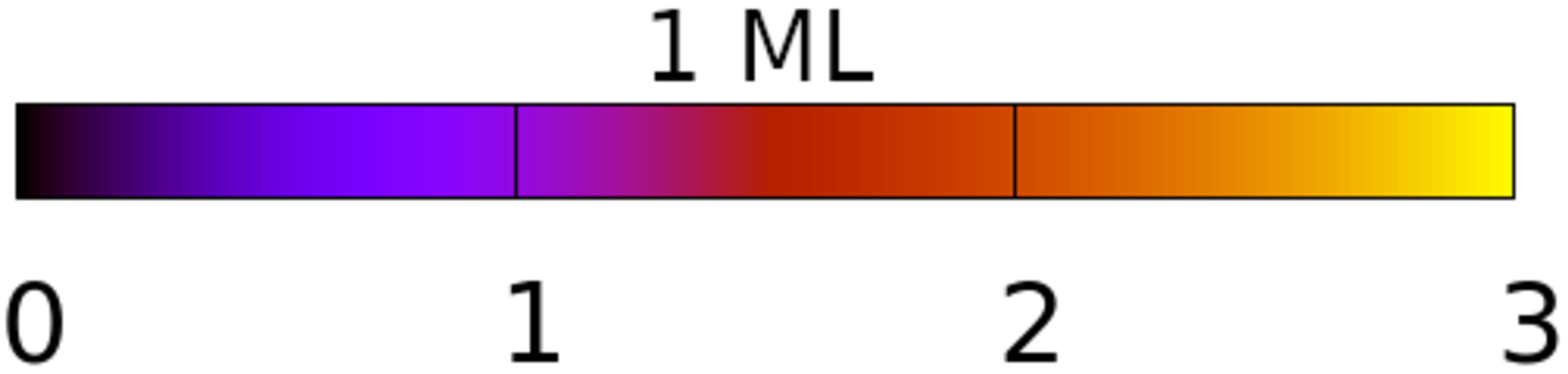}
    \includegraphics[width=0.3\linewidth]{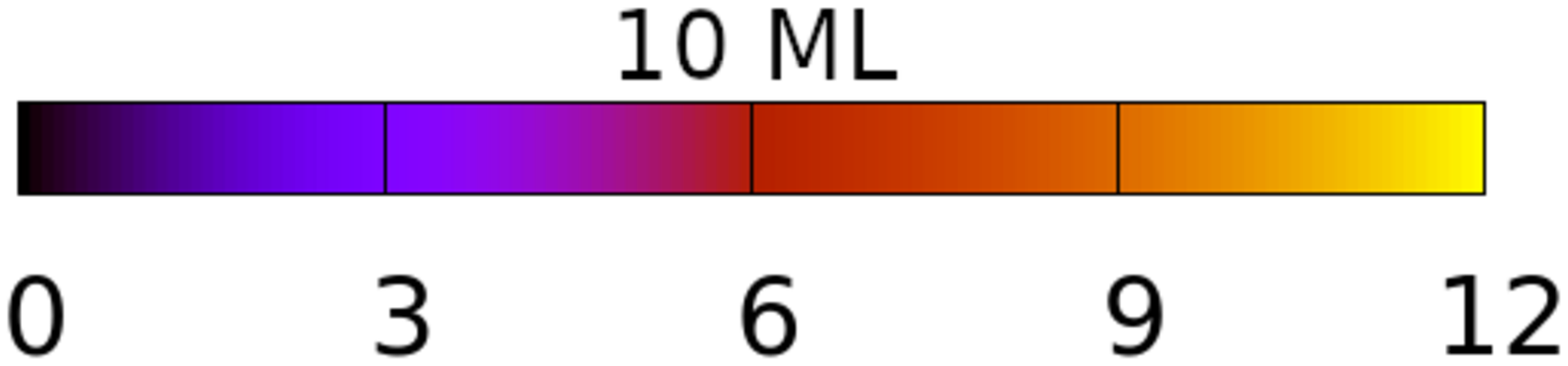}
    \includegraphics[width=0.3\linewidth]{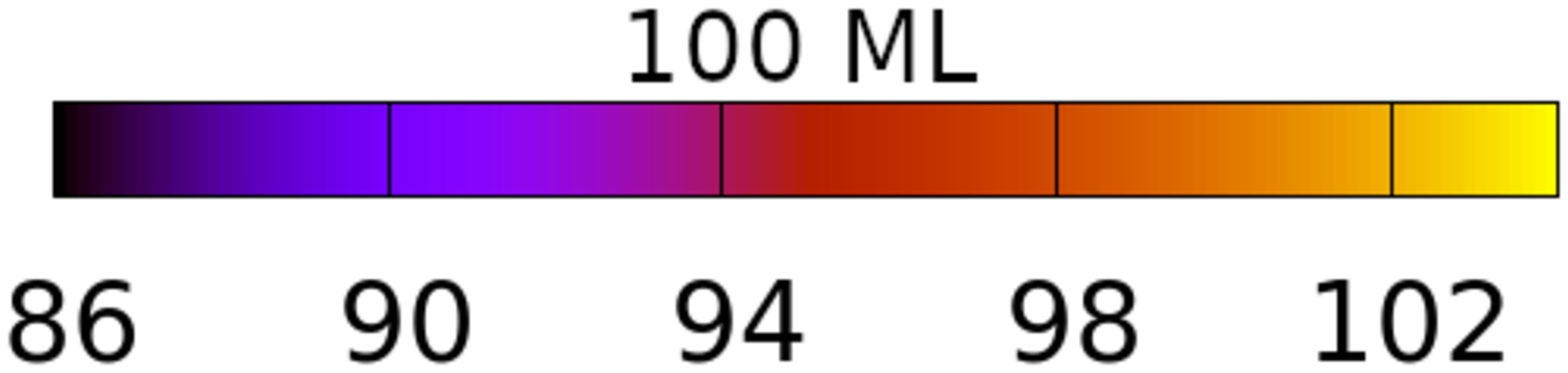}
    \caption{Examples for the growth modes presented in Fig. \ref{fig:rough_schemes}, all for the SOS model. Vertical lines indicate statistical errors, deduced from 5 independent runs for each parameter set. (a) 3D growth at $\Gamma=10^3, \epsilon=-3, \esub=-3.56$, (b) LBL $\to$ 3D at $\Gamma=10^4, \epsilon=-5,\esub = -3.56$, (c) ISL $\to$ 3D at $\Gamma=10^3,\epsilon=-5,\esub=-2.22$, (d) ISL $\to$ LBL $\to$ 3D at $\Gamma=10^4,\epsilon=-4,\esub=-2.22$. Above each roughness plot are three height maps of representative runs at $\Theta=1,10,100$}
    \label{fig:sim_transitions}
\end{figure*}

The schematic roughness evolution of Fig. \ref{fig:rough_schemes} is illustrated by simulation examples using the SOS model, see Fig. \ref{fig:sim_transitions}. Here we have chosen $E_\text{ES} = 0$, but, as discussed before, the same scenarios can also be found for a finite ES barrier.


\section{Comparison to experimental results}
\label{sec:comparison}

\begin{figure*}[t]
    \centering
    \includegraphics[width=0.3\linewidth]{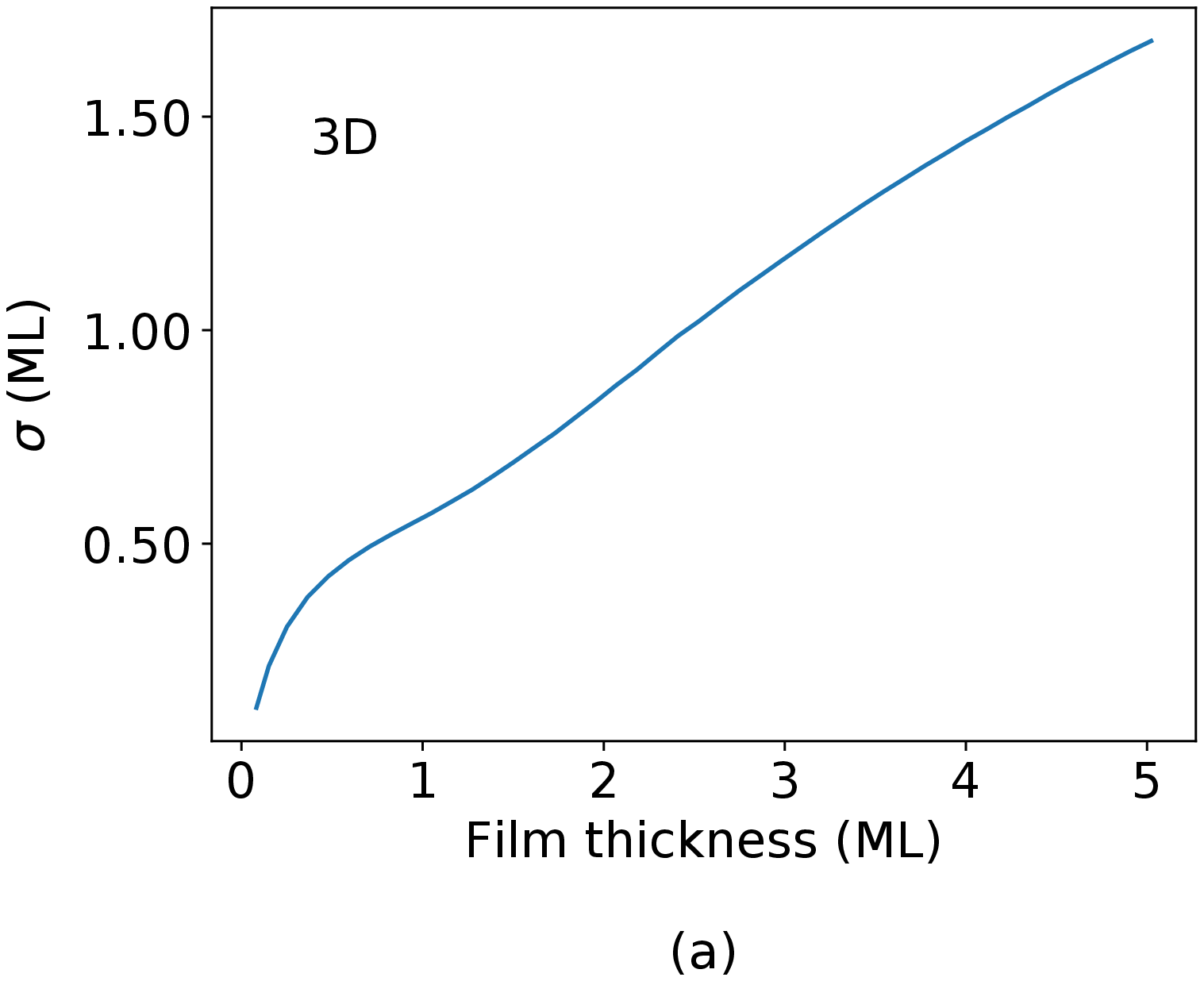}
    \includegraphics[width=0.3\linewidth]{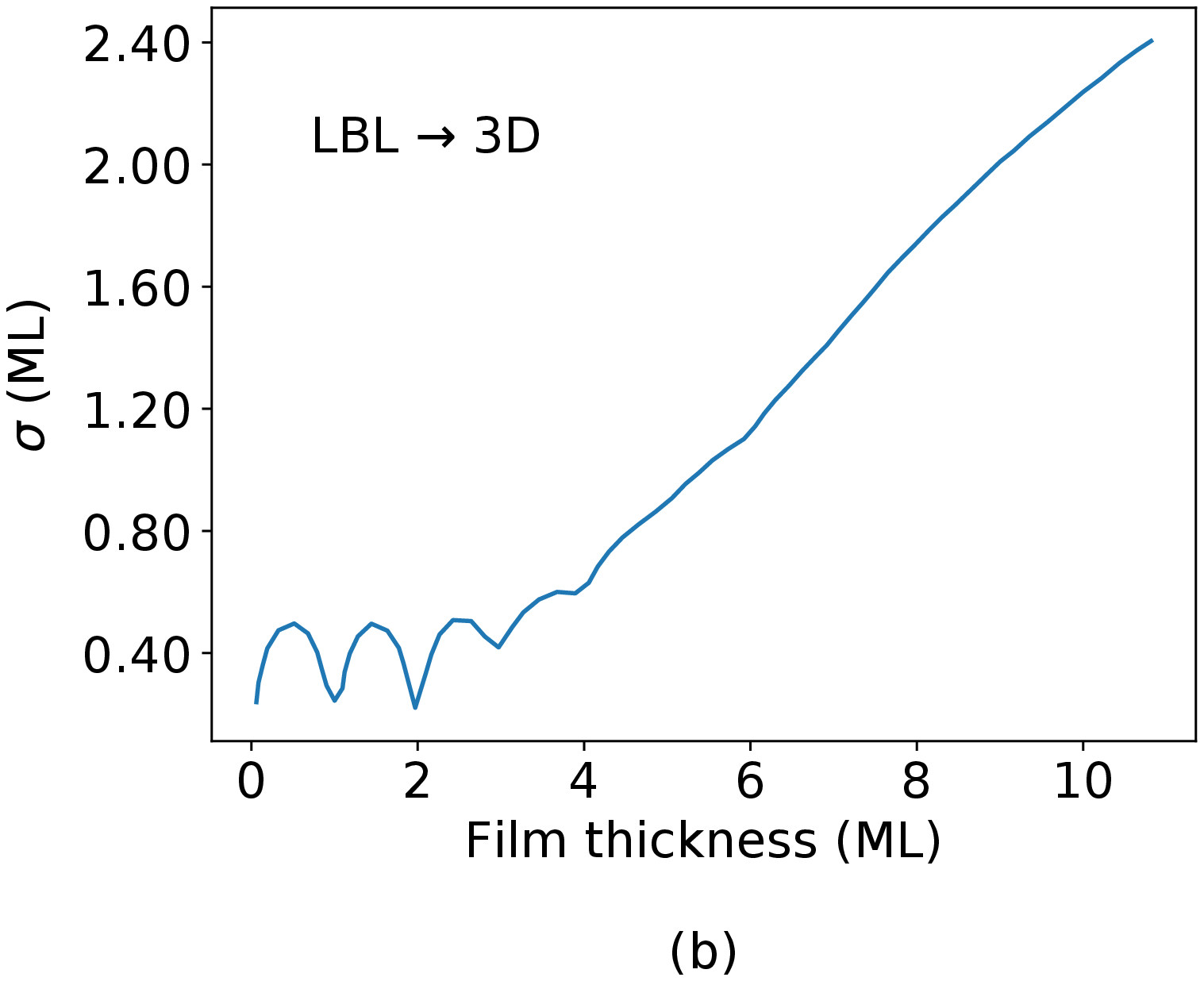}
    \includegraphics[width=0.3\linewidth]{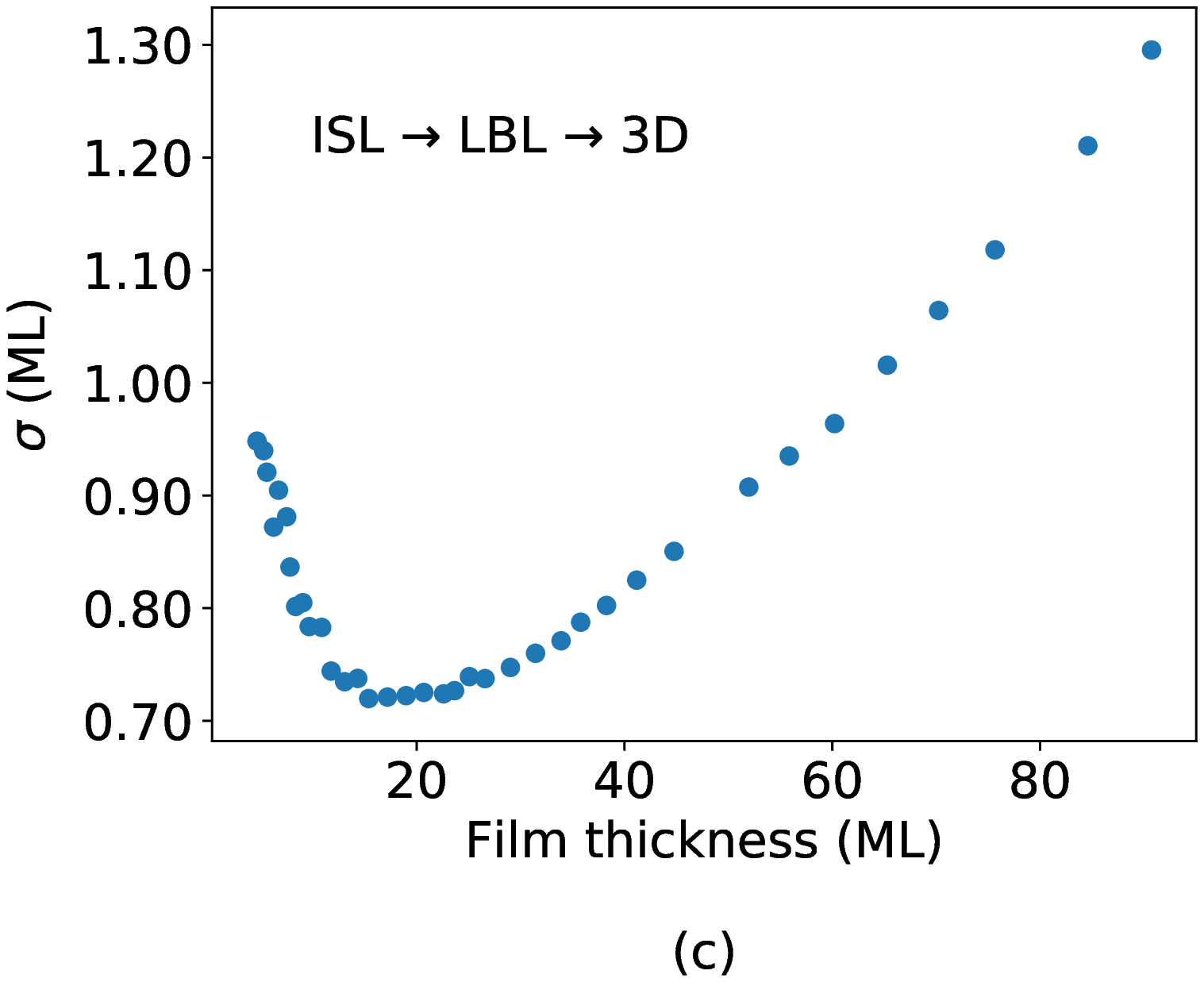}
    \caption{Experimental in-situ measurements of roughness vs. coverage showing different growth modes 
(a) tetracene on SiO$_2$ showing 3D growth\cite{Nahm_2017_JPhysChemC} (1 ML $\equiv\ 13.4\ \AA$\cite{PhysRevB.73.121303}), 
(b) pentacene on SiO$_2$ showing LBL $\to$ 3D behavior\cite{Kowarik_2007_ThinSolidFilms} (1 ML $\equiv\ 15.4\ \AA$), 
(c) rubrene on SiO$_2$ showing ISL $\to$ LBL $\to$ 3D behavior\cite{Kowarik_2006_PhysChemChemPhys} 
(1 ML $\equiv\ 13.4\ \AA$ in an orthorhombic polymorph crystal.\cite{doi:10.1021/acs.nanolett.7b00380} 
Note, however, that in these experiments the rubrene films were amorphous)}
    \label{fig:expgmodes}
\end{figure*}
In order to put the theory in a broader perspective, we shall discuss experimental results from different areas,
namely
``molecular thin films'' and ``atomic thin films''. 

Growth of thin films in both categories differ strongly, since, in comparison, molecules are generally anisotropic with 
comparably weak interaction strengths (dominantly of van-der--Waals type), whereas atoms are isotropic and typically exhibit stronger, covalent or ionic, interaction. 
For the comparison with experimental systems, an obvious choice would also be heteroepitaxial growth of metallic thin films.
There are indeed many studies (see, e.g., Refs.\citenum{chason_2016, thompson_2000} for an overview and general considerations), 
but of course in these the evolution of elastic strain usually plays an important role.

Since the lattice parameter is an intrinsic property of the materials involved, which will generally differ, the role of strain cannot easily be 'switched off' in the experiment. This applies in a similar and indeed more serious manner to classical semiconductor systems,
for which lattice strain and the elastic response is typically much stronger than
for molecular systems (see below).
Our theoretical considerations might be employed for systems
which incidentally have the same lattice parameter for substrate and film,
such as, e.g., Au and Ag.
Unfortunately, studies of this type are rare,
and of course they would also not allow any tuning of the substrate energies.

Nevertheless, some of the general scenarios of Fig. \ref{fig:rough_schemes} are also found in systems in which
strain is part of the picture. For example, Al grown on sapphire shows a transition from 3D to 2D growth (Fig. \ref{fig:rough_schemes}(d)) after several hundred monolayers.\cite{zhu_2016}
In contrast, Al grown on Si(111) at high deposition rates shows LBL growth followed by a slow roughening transition (Fig. \ref{fig:rough_schemes}(b)).\cite{levine_2012}%
These growth modes can also be found in systems with metals evaporated onto soft and disordered substrates,
where strain is expected to be less important.\cite{lita_2000,lita_sanchez_1999,ruffino_2011} 

Molecular thin films frequently are more tolerant against mechanical strain,
which makes a comparison to theory without considerations of epitaxy potentially very suitable. In fact, all of the growth 
scenarios depicted in Fig. \ref{fig:rough_schemes} can be realized.
Pure 3D growth or strong islanding without coalescing (Fig. \ref{fig:rough_schemes}(a)) is regularly observed for growth on 2D materials, e.g.\ diindenoperylene on MoS$_2$ \cite{Mrkyvkova_2019_ApplPhysLett} and Pentacene or Oligothiophenes on Graphene.\cite{Huss-Hansen_2020_Langmuir, Hodas_2018_ACSApplNanoMater}
In addition, it is found that substrates strongly interacting with the deposited molecules often exhibit a strongly bound wetting layer which saturates the reactive surface. In the subsequent multilayer regime, again, pure 3D growth is found. 

Since the first strongly bound monolayer has usually a different molecular orientation and features a completely different interaction, we have to distinguish this from classical Stranski-Krastanov growth, with a transition from LBL to 3D growth (Fig. \ref{fig:rough_schemes}(b)). Instead, we regard the first bound monolayer here as a surface modification that induces 3D growth of the same material in the multilayer regime. Typical examples for this type of growth are pentacene on Au\cite{Kafer_2007_PhysRevB} or diindenoperylene on Au.\cite{Durr_2003_PhysRevB}

A transition from LBL growth to 3D growth is typically observed for growth on weakly interacting substrates like SiO$_2$. 
Well-studied examples are the growth of pentacene,\cite{Ruiz_2004_ChemMater,Kowarik_2007_ThinSolidFilms} perfluoropentacene\cite{Kowarik_2008_PhysStatusSolidiRRL} and PTCDI-C$_x$\cite{Desai_2011_JPhysChemC,Zykov_2017_JChemPhys} or alpha-sexithiophene\cite{Chiodini_2020_JPhysChemC} which all feature the transition from LBL to 3D within a few closed layers. The dependence of the LBL to 3D transition on growth rate and temperature was studied for diindenoperylene\cite{Durr_2003_PhysRevLett,Woll_2011_PhysRevB,Kowarik_2006_PhysRevLett} and tetracene.\cite{Nahm_2017_JPhysChemC} The latter example also demonstrates that by tuning growth rate and temperature the transition of 
LBL to 3D can be below one monolayer changing the growth behavior effectively to pure 3D (Fig. \ref{fig:rough_schemes}(a)). Also surface modification by self-assembled monolayers (SAM) may modify the transition thickness from LBL to 3D.\cite{Desai_2011_JChemPhys,Yavuz_2019_JPhysChemC}

Rod-like compounds which feature an LBL to 3D transition are mostly those growing in an upright-standing mode on weakly interacting substrates. 
In that case the molecule--molecule interaction in the plane ($\pi$-$\pi$ overlap) is stronger than the substrate--molecule 
interaction resulting in LBL growth. In contrast, growth of such a material on a reactive metal surface usually results in 
3D growth as explained above. Exceptions are either compounds with a strong tendency to lie flat on the substrate such as PTCDA, 
which features an LBL to 3D transition on Ag\cite{Krause_2004} or the spherical compound C60 which has an isotropic potential\cite{bommel_2014}.


Coalescing of islands at later stages with a clearly observed subsequent LBL mode (Fig. \ref{fig:rough_schemes}(d)) is observed rarely. 
One example are nearly amorphous thin films of rubrene on SiO$_2$ were the coalescing starts after approximately 5-10 monolayers.\cite{Kowarik_2006_PhysChemChemPhys} In addition, for crystalline films of picene the island coalescing was observed at approximately 20-40 layers depending on growth conditions.\cite{Hosokai_2015_JPhysChemC,Kurihara_2013_MolCrystLiqCryst,Gottardi_2012_JPhysChemC}

For the sake of a final quantitative illustration summarizing this overview,
in Fig. \ref{fig:expgmodes} we show experimental roughness data from the growth of molecular thin films on amorphous substrates.
These correspond to three of
the four growth scenarios of Fig. \ref{fig:rough_schemes}. For the ISL$\to$3D
growth scenario, we were unable to find experimental data
covering the range of film heights needed to detect
the transition from island to 3D growth.
However, we think that the growth of C$_{60}$ on SiO$_2$ and the growth of picene at ambient temperatures are good candidates for this scenario. It is known that C$_{60}$ is a strong island former. For picene the coalescing of islands is demonstrated in Refs \citenum{Hosokai_2015_JPhysChemC, Kurihara_2013_MolCrystLiqCryst, Gottardi_2012_JPhysChemC}. However, no continuous data set for all thicknesses is available. Due to limitations of the XRR methods quantitative results for the roughness are available at intermediate time (where
it is larger than expected from Poisson growth hence supporting the island picture). It is not possible to extract the roughness of such films at very
short times nor at long times where the film is very rough.
At long times, AFM pictures show significant roughening but post growth effects would need to be considered. We remark in passing that quantitative roughness data are only available for a fraction of the plethora of thin film studies. 

\section{Summary and conclusion}
\label{sec:conclusion}

In this work we have investigated thin film growth in simple lattice gas models where the substrate is energetically different from the
film, and substrate and film phase are defined on the same simple--cubic lattice 
(no genuine heteroepitaxy, i.e. no strain effects). The investigated
models are a solid--on--solid (SOS) model (with no vacancies/overhangs in the growing film) and a colloidal growth model (CGM)
where particles can desorb from the film, diffuse in the gas phase above the film, and re-adsorb again on the film.
The latter is suitable for describing colloidal film growth in solutions. 

For small to intermediate deposition times (up to an equivalent of about 10--20 monolayers) and not too fast deposition rates,
the growth modes are island (ISL) and layer--by--layer (LBL) growth.
In both the CGM and the SOS model, we have identified two dynamical transitions in this regime
of small to intermediate deposition times. The first (``dynamic layering transition'') describes a transition from ISL
to LBL growth as a function of the reduced substrate strength $\Upsilon=\esub/\epsilon$ and can be viewed as the dynamical
counterpart of the equilibrium wetting/layering transition. The latter can only depend on
the particle interaction strength $\epsilon$, and is nearly independent of it for larger $|\epsilon|$ (located at $\Upsilon(\epsilon) \approx 1$). The dynamic transition, however, depends
in general on the three parameters $\epsilon$, $\Gamma$ (ratio of diffusion to deposition rates) and $E_\text{ES}$
(Ehrlich--Schwöbel barrier for inter-layer diffusion). It is found at lower values of $\Upsilon$ compared to the equilibrium transition,
and the difference increases with increasing $\epsilon$ and decreasing $\Gamma$.   
The second transition (``flattening transition'') describes the transition from initial ISL growth back to LBL growth at an
intermediate transition time. Physically, the transition is connected with the coalescence of islands and manifests itself in a
drop in film roughness at the transition time. In the SOS model and depending on the specific parameters, 
the roughness occasionally only drops to a constant
value, reflecting island coalescence with residual trenches which can be filled only by deposition.  

For very long deposition times, film growth will always show roughening (3D growth) which has been already studied earlier 
(see e.g. Ref.~\citenum{evans_thiel_bartelt_2006} for an overview).  
Combined with the results for small to intermediate deposition times, we have identified four global scenarios for the evolution of roughness,
which are depicted in Fig. \ref{fig:rough_schemes}. These are (a) 3D growth for all times, (b) initial LBL growth followed by (weak)
3D growth, (c) initial ISL growth followed by 3D growth and (d) initial ISL growth with a transition to intermediate LBL growth 
(or growth with trenches) and followed by 3D growth. Interestingly, scenario (b) is akin to Stranski-Krastanov (SK) growth. SK growth is commonly related to genuine heteroepitaxy: The incommensurability of the substrate and film lattices leads to the build-up of mechanical stress in LBL growth, which is released after deposition of a few layers, causing the smooth film to break-up into islands(Refs.~\citenum{Baskaran2010, schulze_smereka_2011, schulze_smereka_2012}). Our results suggest that SK growth can also occur in growth on amorphous substrates where strain would be absent.

We have discussed these results with respect to existing experimental findings.
In thin film molecular growth (with weaker inter-particle interactions), the four growth scenarios can all be identified, and these
also depend on the substrate interaction energy and growth kinetics. This points to a relative unimportance of molecular anisotropy
with regard to the global roughness evolution.  
The epitaxial growth of strongly interacting compounds like metals and inorganic semiconductors depends critically on the 
lattice matching of substrate and thin film and strain related issues, which are not incorporated in our simple approach.
Nevertheless, there are several examples in the literature, which exhibit 3D to 2D or 2D to 3D growth mode transitions similar 
to our description. 

While the dynamic transitions and the global growth scenarios are very similar for the CGM and the SOS model, differences
can be found in cases where desorption and re-adsorption are important. As an example, we discussed the case of infinite Ehrlich--Schwöbel
barrier (no direct inter-layer changing moves). In the SOS model, this leads to stochastic growth, while in the CGM inter--layer 
diffusion is still possible as a multi--step process via the gas phase. In the CGM this leads
to the appearance of another transition for strongly attractive substrates: 
For low interaction strengths $|\epsilon|$ there is initial island growth which rather abruptly changes to stochastic growth
upon increasing $|\epsilon|$.

The ``dynamic gap'' between the dynamic and equilibrium layering transition also implies that monolayer films can be prepared
by deposition, but would be subject to dewetting if deposition was stopped. This process has been studied e.g. in 
Refs.~\citenum{pierrelouis_2007,pierrelouis_2010}. Likewise, smooth multilayer films may be subject to strong post-growth roughening. This is known for the growth of rubrene films: As discussed, these show the ISL $\to$ LBL $\to$ 3D growth mode (see Fig. \ref{fig:expgmodes}(c)) when measured in real-time, but strong dewetting of the smooth film into a variety of patterns is observed on the timescale of a few days to a month.\cite{AngererDiss} Leaving the subject of one-component films, one can expect that the phenomenology of dynamic transitions and dewetting
behavior becomes much richer when going to growth in binary systems.\cite{reis_2020_binary,reisz_2020_c60_cupc}
With regard to growth with organic molecules, an extension of the lattice models to anisotropic interactions 
would also be desirable which can be accomplished by using anisotropic, energetic interaction parameters or using lattice rods
to capture steric effects.\cite{Choudhary06,oettel_2016,Klopotek:2017}

\begin{acknowledgments}
\noindent
We gratefully acknowledge the financial support of the German Research Foundation (Deutsche Forschungsgemeinschaft, DFG).
\end{acknowledgments}



\bibliography{Simulation_refs_new}
\clearpage
\appendix

\section{Ising model and lattice gas}
\label{app:ising}
The Ising model in 3D on a semi--infinite SC lattice bounded by a planar substrate is defined by the Hamiltonian
\begin{equation}
 {\cal H}_\text{Is} = J \sum_{\langle ij\rangle} \sigma_i \sigma_j + H \sum_i \sigma_i + H_1 \sum_{\text{surf},i} \sigma_i\;, 
\end{equation}
where $\sigma_i=\{1,-1\}$ is a spin variable,
$J$ is the nearest neighbor coupling strength ($\langle ij\rangle$ label nearest neighbor sites), 
$H$ is a bulk (magnetic) field and $H_1$ is a surface field (the sum over spins in the corresponding term only extends 
over spins adjacent to the substrate).

On the other hand, the Hamiltonian of the lattice gas (as used here, with a bulk external field $V^\text{ext}/(k_BT) \equiv -\mu$) is defined by
\begin{equation}
 \frac{{\cal H}_\text{lg}}{k_BT} = \epsilon \sum_{\langle ij\rangle} n_i n_j - \mu \sum_i \sigma_i + \esub \sum_{\text{surf},i} \sigma_i\;, 
\end{equation}
where $n_i=\{1,0\}$ is a lattice site occupation variable.
Upon defining $\sigma_i = 2n_i-1$, both Hamiltonians are equivalent (up to an unimportant constant) if the
following identifications are made:
\begin{equation}
  \epsilon = \frac{4J}{k_BT} \;, \quad \esub = \frac{2J+2H_1}{k_BT} \;, \quad \mu = \frac{-2H+12J}{k_BT} \;.
\end{equation}

Using these identifications, the wetting/layering diagram of Ref.~\citenum{binder_landau_1992} (Fig. 1(c)) corresponds
to Fig. \ref{fig:schem_phase} in the present work.

\begin{figure}[t]
	\centering
	\includegraphics[width=\linewidth]{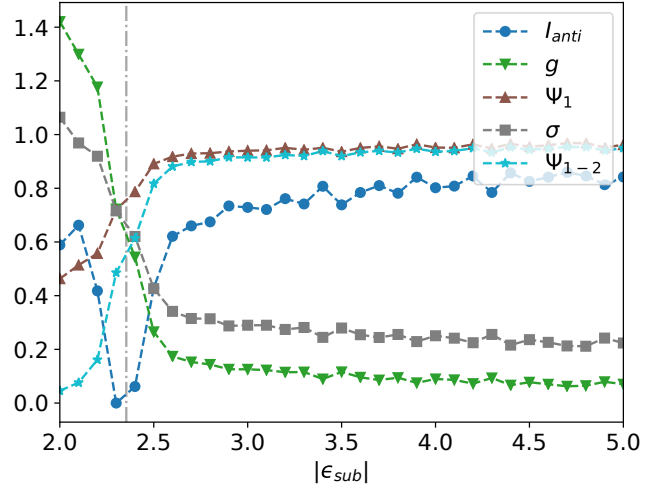}
	\caption{
Values of several eligible observables after deposition of 1 ML in the CGM. Plotted are the anti-phase Bragg intensity $I_\text{anti}$, the filling of the first layer $\Psi_1$, the film roughness $\sigma$, and the difference of the filling of the second and first monolayer $\psid$. All these observables show a change in behavior roughly at the same $\esub$, however $\psid$ varies smoothly around this point and allows us to fit a tanh to the values and extract the inflection point.}
	\label{fig:all_obs}
\end{figure}

\section{Comparison of possible order parameters for locating the dynamic layering transition}
\label{app:obs}
In order to quantify the critical $\esub$ for the dynamic layering transition, we compared the values of several possible observables after deposition of 1 ML. These observables included among others the roughness $\sigma$, the layer filling $\Psi_i$ of the first and second layer, the growth number\cite{michely_krug_2004}
\begin{equation}
g = \frac{\sum_{n=1}^{\infty}\left|\Theta_{n} - \Theta_{n,LBL}\right|}{\sum_{n=1}^\infty \left| \Theta_{n,stat} - \Theta_{n,LBL} \right|}
\end{equation}
(where $\Theta_{n}$ is the measured coverage in layer $n$, $\Theta_{n,LBL}$ is the coverage at this height assuming perfect LBL growth, and $\Theta_{n,stat}$ is the coverage at this height assuming completely statistical growth), 
and the anti-phase Bragg intensity, which is defined\cite{michely_krug_2004} as
\begin{equation}
I_\text{anti}(\Theta) = \left| \sum_{i=0}^\infty (-1)^i (\Psi_i(\Theta) - \Psi_{i+1}(\Theta)) \right|^2
\end{equation}
(where $i=0$ denotes the substrate layer, i.e. $\Psi_{0}$ is always $1$). The growth number $g$ is a measure of whether a film grows in an LBL fashion ($g=0$), in a Poisson manner ($g=1$) or in an intermediate manner.
The anti-phase Bragg intensity is the intensity of e.g. reflected X-rays at the anti-Bragg point, where reflections from neighboring layers interfere destructively. This leads e.g. to $I_\text{anti}(\Theta)$ showing oscillations when observing films growing in an LBL fashion.\cite{Kowarik_2008}

All of these show a change in behavior around the same $\esub$, however, in Fig. \ref{fig:all_obs} we can see that the behavior of $\psid$
is most intuitive for signaling a transition: we may comfortably fit a tanh to the data points. Hence we chose to use this observable to quantify the dynamic transition point.


\section{Comparison of multilayer order parameters}
\label{app:kurt}
To quantify where the ISL $\to$ LBL and the ISL $\to$ const transition occur, we considered the minimum height of the film and the kurtosis which is the normalized fourth moment of the height distribution.

 \begin{figure*}[t]
    \centering
    \includegraphics[width=0.49\linewidth]{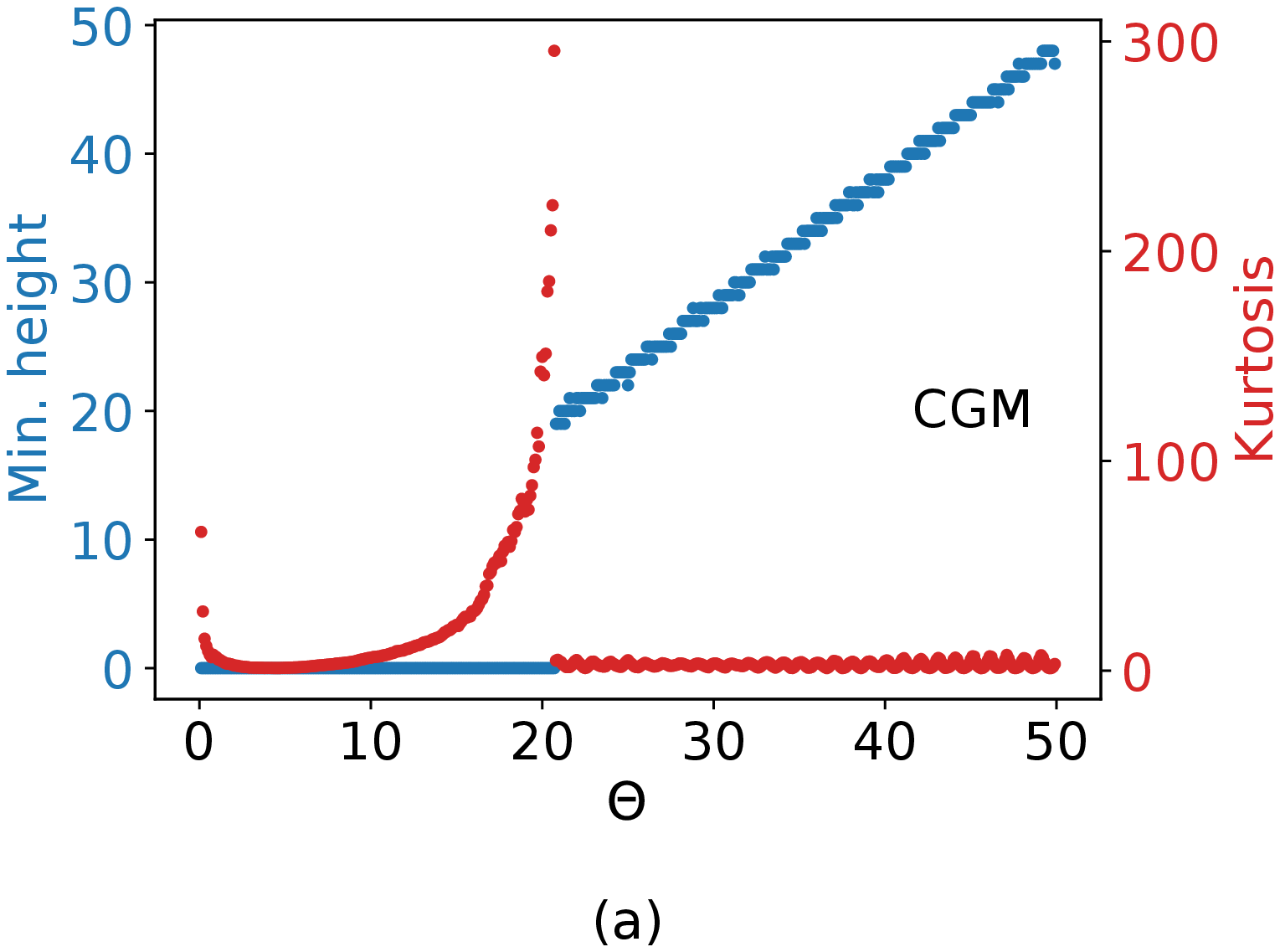}
    \includegraphics[width=0.49\linewidth]{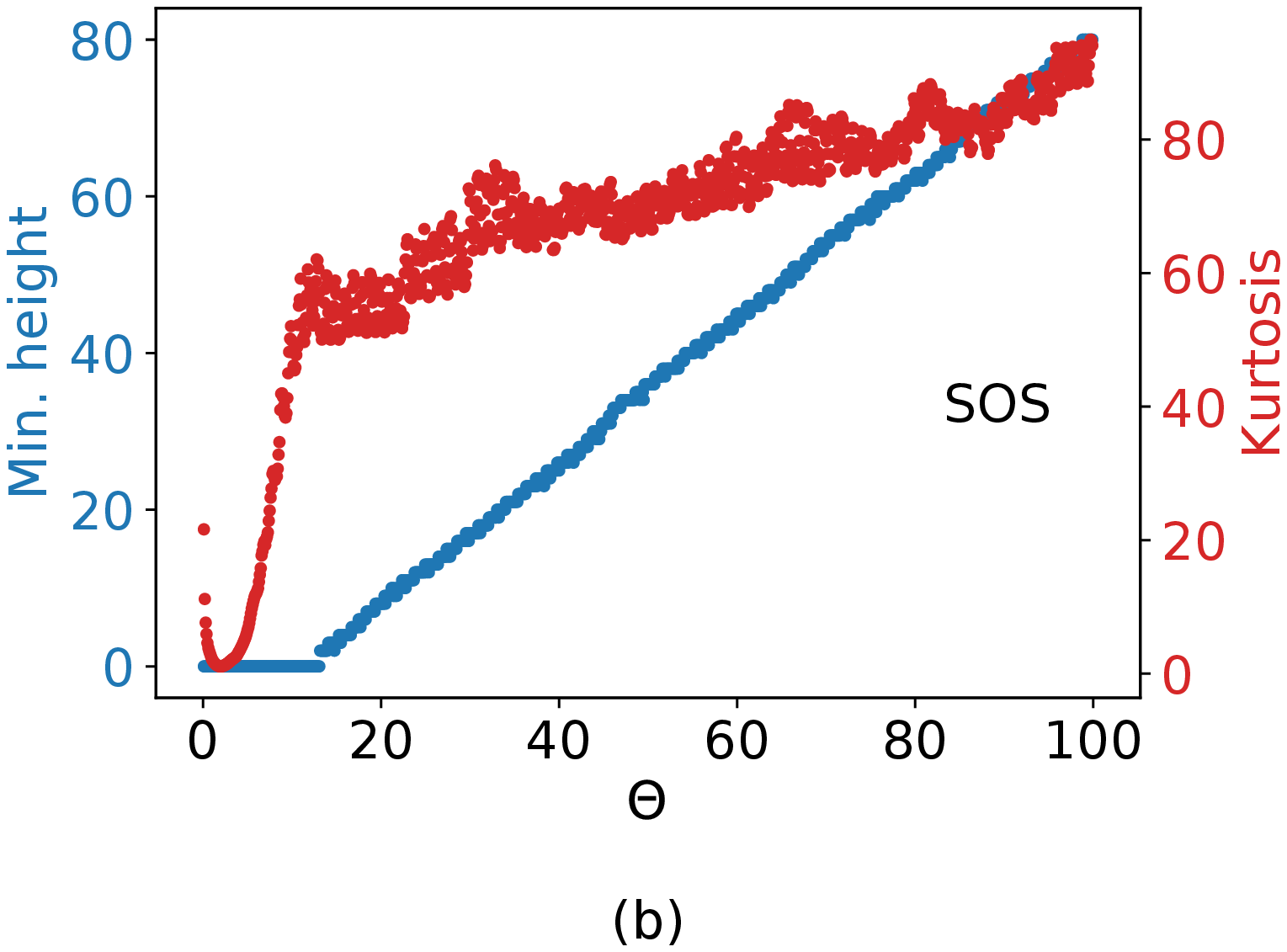}
    \caption{Comparison of Min. height and kurtosis of the height distribution, characterizing the flattening transition, at $\Gamma = 10^4, \epsilon=-3, \esub=-0.89$ in (a) the CGM and (b) the SOS model.}
    \label{fig:kurt_minh}
\end{figure*}

In Fig. \ref{fig:kurt_minh} we can see that both observables show a change in behavior at the same coverage $\Theta$. However, only the minimum height allows us to clearly pinpoint the exact transition time (the earliest time at which min h > 0), while the change in behavior of the kurtosis is e.g. not easily quantifiable in the SOS model.

\section{Infinite Ehrlich-Schwöbel barrier}
\label{app:ESinfinite}

As discussed in Sec. \ref{subs:asymptotic} the generic growth modes for both the CGM and the SOS model
at not too small $\Gamma$ are as follows: At short times, both systems will show either island growth or 
LBL growth, while at long times, they will both show 3D growth. 
We can, however, already see a deviation at short times in the special case of an infinite ES barrier. 
In the SOS model, this means that inter-layer diffusion is prohibited, which leads to the well known roughening 
behavior of $\sigma \propto \Theta^{1/2}$. 

\begin{figure*}[t]
	\centering
		\includegraphics[width=0.49\linewidth]{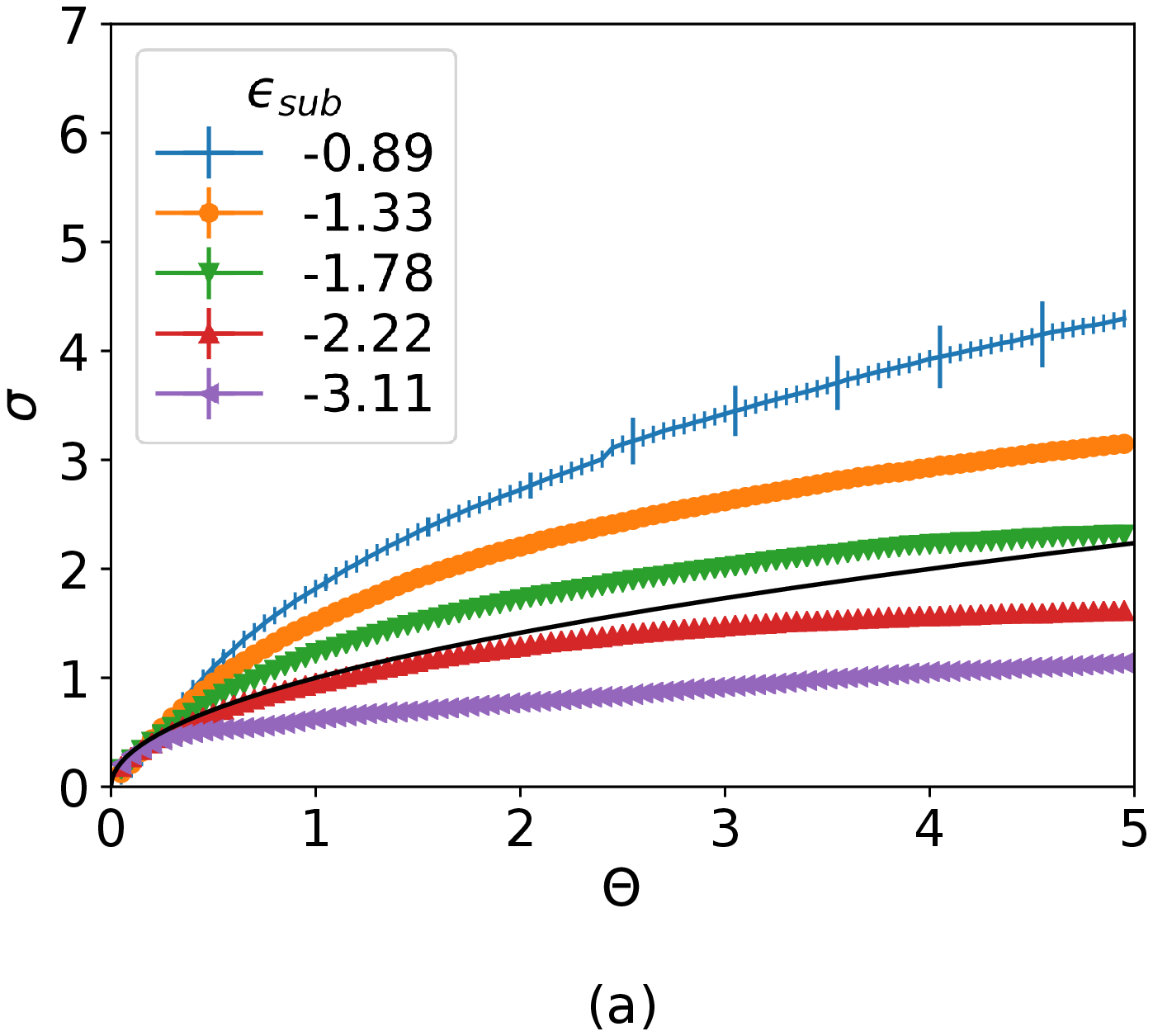}
		\includegraphics[width=0.49\linewidth]{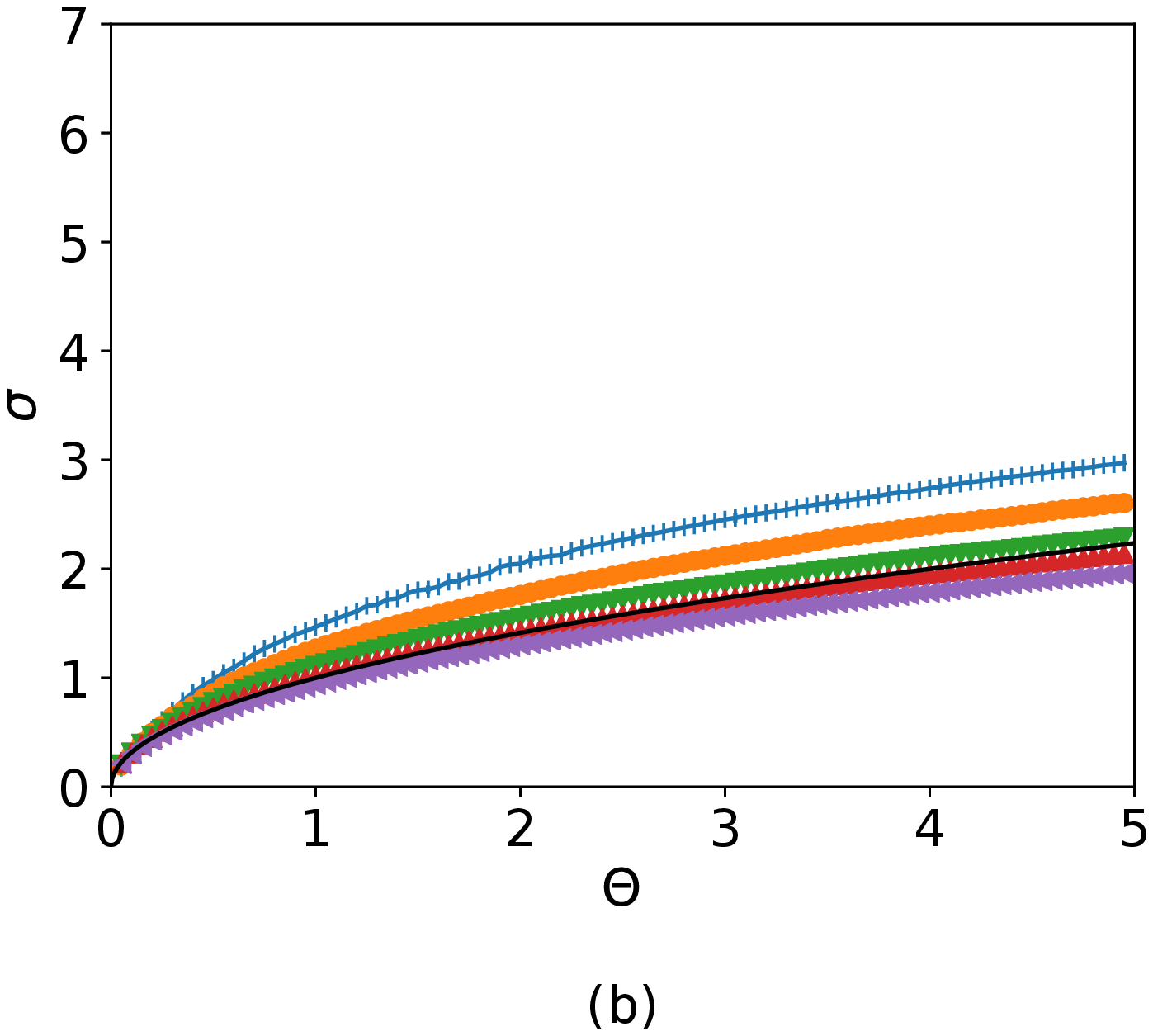}
	\caption{
Evolution of roughness in the CGM $\epsilon_\text{ES} = \infty$ at $\Gamma=10^4$ and for different $\esub$. 
(a) $\epsilon=-3$, (b) $\epsilon=-5$. The black line denotes the statistical roughness evolution $\sigma = \sqrt{\Theta}$.
}
	\label{fig:esb_0_roughness}
\end{figure*}

On the other hand, inter-layer diffusion is still possible in the CGM, albeit as a multi--step process 
in which particles will first detach from the film, perform diffusion moves inside the gas phase, and later 
reattach to the film, possibly in a different layer. This means that the ES barrier in this system is effectively 
lowered to a finite value, leading to a strong deviation from the behavior in the SOS model. 
In Fig. \ref{fig:esb_0_roughness} this is illustrated for two inter-particle attraction strengths $\epsilon=-3$
and $-5$ and for a range of substrate attractions $\esub$. For small $|\esub|$, the roughness grows faster  
than in stochastic growth, reflecting island formation. For larger $|\esub|$ the roughness decreases.
For lower $\epsilon$ it can go significantly below the roughness from stochastic growth, reaching 
$\sigma \lesssim 1$ as in LBL growth. For larger $\epsilon$ it saturates near the $\sigma \propto \Theta^{1/2}$ curve from 
stochastic growth.  
Other observables, such as the  filling of each layer vs. time also confirm the saturation in the stochastic growth mode.

To study the transition from LBL--like behavior to stochastic growth for very attractive substrates, 
we compute $\psid(|\epsilon|)$ after deposition of 1 ML (as in Sec. \ref{subs:dyn_layer}) and again find a tanh-like behavior of the observable. 
Here we set $\esub = -10^6$, i.e. quasi-infinite, so particles which reach the substrate will stay within the first layer.

\begin{figure}[h]
	\centering
	\includegraphics[width=\linewidth]{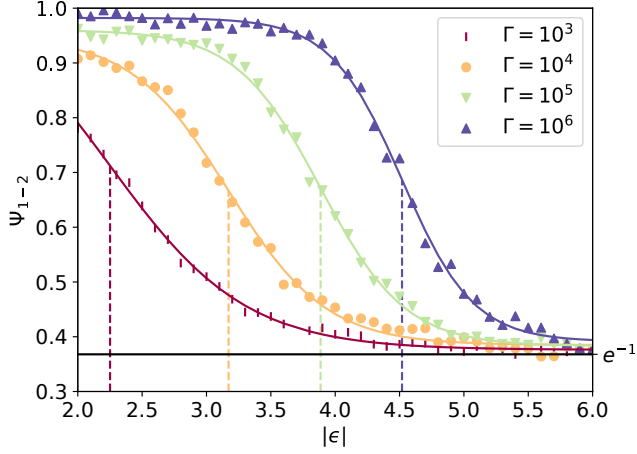}
	\caption{$\psid(|\epsilon|)$ after deposition of 1 ML in the CGM at $\esub=-10^6$. The dashed lines indicate the inflection points at the respective $\Gamma$.}
	\label{fig:stat_trans}
\end{figure}

At low $|\epsilon|$, $\psid$ is close to $1$, i.e. almost all particles are confined to the first monolayer. 
This is expected, since particles in the second layer may step down at these parameters via a multi--step process. 
Upon increasing $|\epsilon|$, $\psid$ goes to $\exp(-1)$ which is the value corresponding to stochastic growth 
when no inter--layer diffusion is possible. 
This indicates that here the inter-particle attraction is too strong for a 
significant amount of particles to desorb from the film.

The shape of $\psid(|\epsilon|)$ again allows to fit a tanh curve and identify the inflection point as the critical $\epsilon$ above which the system will 
grow via stochastic growth. 
This critical attraction strength increases with increasing $\Gamma$ and should disappear for $\Gamma \to \infty$.
In this limit (growth rate going to zero) the particles will always be able to desorb into the gas and then attach to the substrate in the first layer 
where they will be effectively trapped, i.e. the stochastic growth mode will never occur.

\end{document}